\documentclass[aps,prb,notitlepage,superscriptaddress,reprint]{revtex4-1}

\usepackage{amsfonts}
\usepackage{amsmath}
\usepackage{amssymb}
\usepackage{hyperref}
\usepackage{graphicx}
\usepackage{bm}
\usepackage{textcomp}
\usepackage{xcolor, soul}

\begin{document}

\title{Crossover from weakly indirect to direct excitons in atomically thin films of InSe}

\author{Adri\'{a}n Ceferino}
\email{adrian.ceferino@postgrad.manchester.ac.uk}
\affiliation{Department of Physics and Astronomy, University of Manchester, Oxford Road, Manchester, M13 9PL, UK}
\affiliation{National Graphene Institute, Booth Street East, Manchester, M13 9PL, United Kingdom}
\author{Kok Wee Song}
\email{kokwee.song@manchester.ac.uk}
\affiliation{National Graphene Institute, Booth Street East, Manchester, M13 9PL, United Kingdom}
\author{Samuel J. Magorrian}
\affiliation{National Graphene Institute, Booth Street East, Manchester, M13 9PL, United Kingdom}
\author{Viktor Z\'{o}lyomi}
\affiliation{STFC Hartree Centre, Daresbury Laboratory,
Daresbury, Warrington, WA4 4AD, United Kingdom}
\author{Vladimir I. Fal'ko}
\affiliation{Department of Physics and Astronomy, University of Manchester, Oxford Road, Manchester, M13 9PL, UK}
\affiliation{National Graphene Institute, Booth Street East, Manchester, M13 9PL, United Kingdom}
\affiliation{Henry Royce Institute for Advanced Materials, Manchester, M13 9PL, United Kingdom}

\begin{abstract}
We perform a $\mathbf{k \cdot p}$ theory analysis of the spectra of the lowest energy and excited states of the excitons in few-layer atomically thin films of InSe taking into account in plane electric polarizability of the film and the influence of the encapsulation environment. For the thinner films, the lowest-energy state of the exciton is weakly indirect in momentum space, with its dispersion showing minima at a layer-number-dependent wave number, due to an inverted edge of a relatively flat topmost valence band branch of the InSe film spectrum and we compute the activation energy from the momentum dark exciton ground state into the bright state. For the films with more than seven In$_2$Se$_2$ layers, the exciton dispersion minimum shifts to $\Gamma$-point.
\end{abstract}
\maketitle

Two dimensional (2D) materials create new opportunities for semiconductor optoelectronics\cite{Bhimanapati:ACSnano9-2015,Qian:2DMat2-2015,Novoselov:Science353-2016}. Among those new materials, post-transition metal chalcogenides (InSe and GaSe) occupy a special place, as they offer a flexibility to choose a desirable size of their bandgap (in the range from 3eV to 1.3eV) depending on the number of atomic planes in a thin film\cite{Mudd:AdvMat25-2013,Lei:ACSNano8-2014,Brotons-Gisbert:NanoLett16-2016,Bandurin:NatNano12-2016,Terry:2DMat5-2018}. While the experimental studies of the band gap and optical properties of few layer films of InSe\cite{Bandurin:NatNano12-2016,Sanchez-Royo:NanoResearch7-2014,Hamer:ACSnano13-2019} and GaSe\cite{Li:SciRep4-2014,Aziza:PRB96-2017,*Aziza:PRB98-2018,trushin_GaSe} have found a reasonably close quantitative interpretation at the single-particle level, based on density functional theory (DFT) \cite{Rybkovskiy:PRB84-2011,*Rybkovskiy:PRB90-2014,Zolyomi:PRB87-2013,Zolyomi:PRB89-2014,Zhou:PRB96-2017,*Zhou:PRB99-2019,Sun:Nanoscale10-2018} and the DFT-parameterized tight-binding model \cite{Magorrian:PRB94-2016}, the fine tuning of the theory requires taking into account excitonic effects in the system, which remains an open question for atomically thin InSe films.

\begin{figure}
\centering
\includegraphics[width=3.3in]{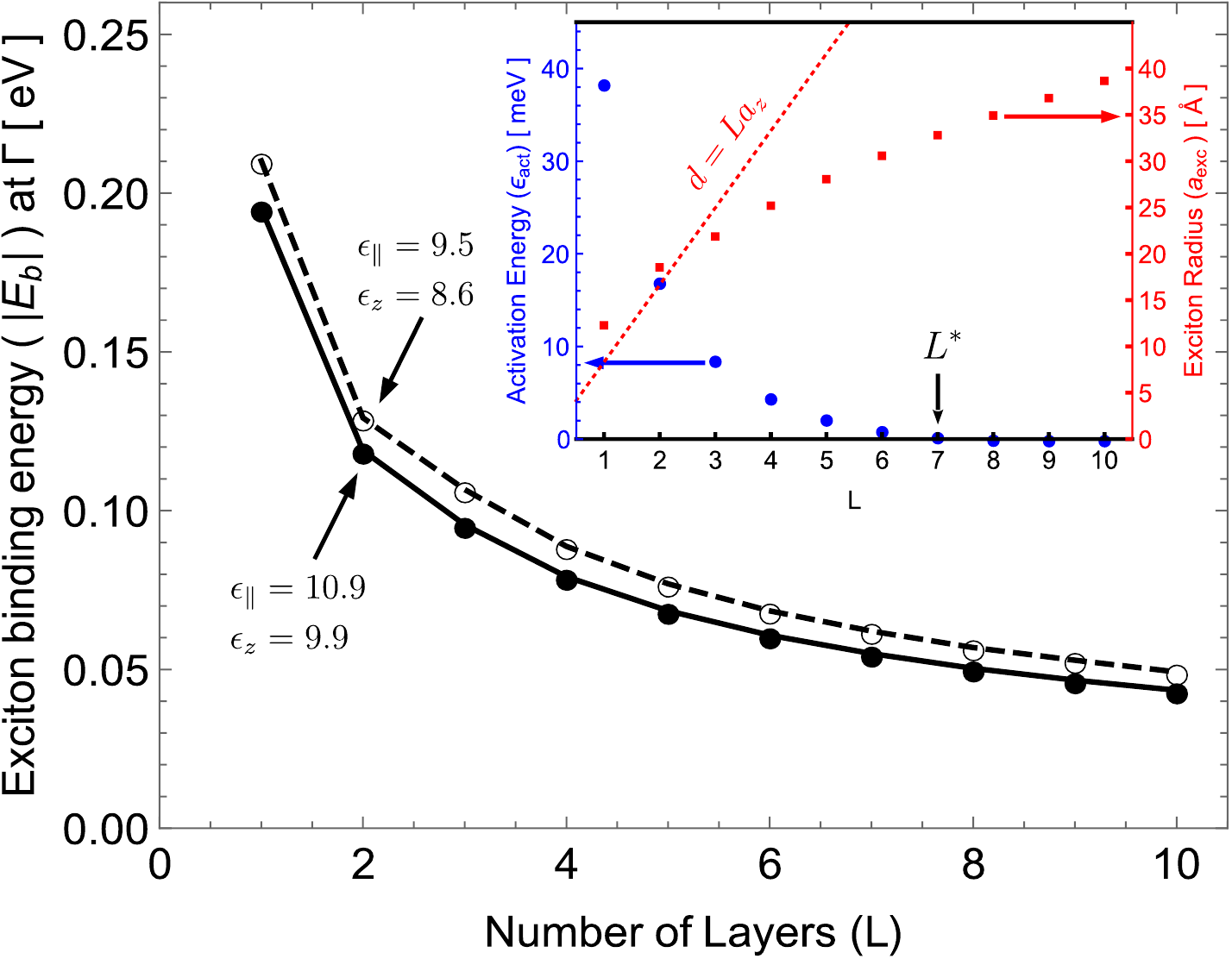}
\caption{The dependence of the exciton binding energy on the number of layers ($L$) for hBN encapsulated InSe films. Binding energies at $
\Gamma$-point, $E_{b}(0)$ are compared for two values of bulk InSe dielectric constants $\epsilon_{\parallel}$ and $\epsilon_{z}$. The inset shows the activation energy $\varepsilon_ {act} = E_\mathrm {b} (0) - E_\mathrm {b} (Q_ {\min}) $ (closed blue circles) where $Q_{\min} $ is approximately the wavevector between the $\Gamma$-point and the edge of the highest valence band. The radius of the exciton, $a_ {exc} =  \sqrt{\langle|\mathbf {r} _e - \mathbf {r}_h | ^2\rangle} $, (closed red square). The red dashed line shows the thickness of the InSe film.
}\label{fig:E0-L}
\end{figure}

Here, we develop a mesoscale theory for the binding energies, dispersions and excited state spectra of excitons in mono-, bi-, tri-, and few-layer InSe films ($\gamma$ polytype), taking into account the strongly non-parabolic features of the valence band dispersion in these 2D materials and the influence of various encapsulation environments. In particular, we study the role of a weak inversion of the hole dispersion near the top of the valence band \cite{Zolyomi:PRB89-2014,Zolyomi:PRB87-2013,Louieferro,trushin_excitons,Skinnerexcitons}, established in the thinnest InSe and GaSe films using angle-resolved photoemission spectroscopy\cite{Hamer:ACSnano13-2019} and high field magneto-optics studies\cite{Mudd:SciRep6-2016}, and analyze the crossover of the excitons from weakly indirect to direct in momentum space, as a function of the InSe film thickness. The crossover of the exciton dispersion from indirect [$\varepsilon(Q)$=min at $Q \neq 0$] to direct [$\varepsilon(Q)$=min at $Q=0$] exciton was found at $L=7$ layers. For films with $1 \leq L \leq 10$, we compute the binding energies of the excitons for hBN-encapsulated InSe films  and the activation energies from the momentum-dark excitonic bound states, with the results summarized in Fig. \ref{fig:E0-L}.

\begin{figure*}
\centering
\includegraphics[width=7in]{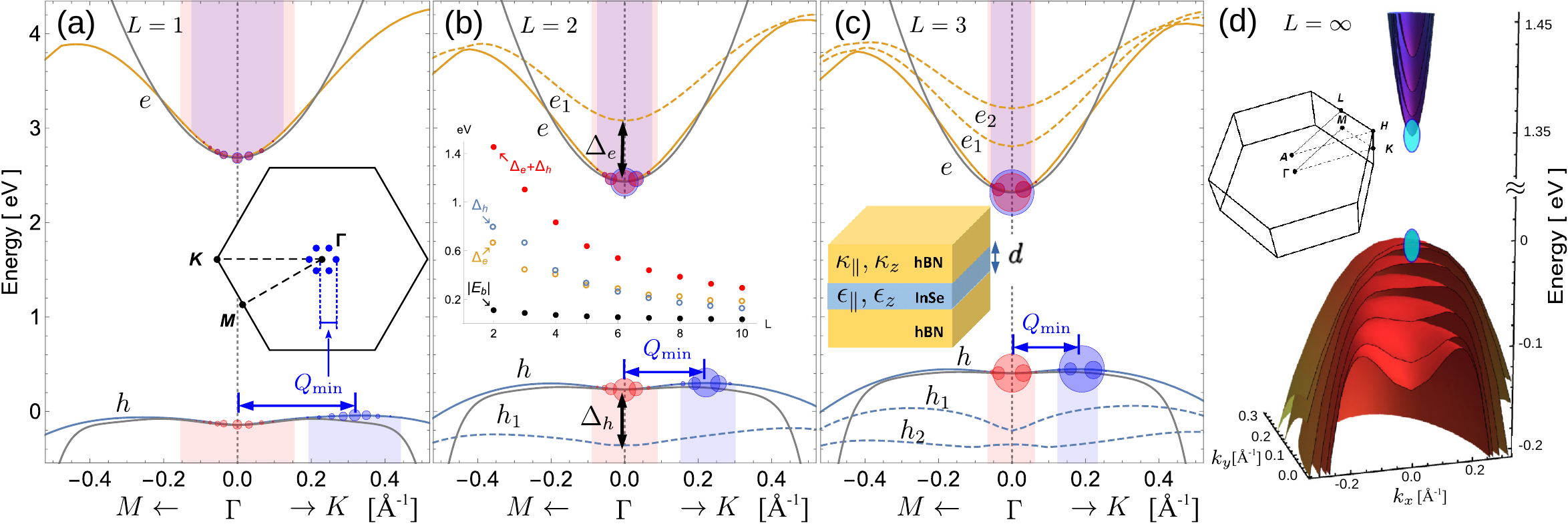}
\caption{The bound electron-hole states and the quasiparticle dispersion in momentum space: (a)--(c) The plots for $L=1-3$ layers film.  The solid (dashed) yellow and blue curves are conduction and valence (sub)band tight-binding dispersions\cite{Magorrian:PRB94-2016}. The gray curves are the $\mathbf{k}\cdot \mathbf{p}$ dispersion where $\varepsilon_c$ and $\varepsilon_v$ are expanded into a polynomial of $\mathbf{k}$. The sizes of the red (blue) circles are proportional to the probability density $|\psi_\mathbf{Q}(\mathbf{k})|^2$ with $\mathbf{Q}=0$ ($\mathbf{Q}=Q_\text{min}$ being the total momentum of the lowest energy exciton). (d) Plot for bulk InSe dispersion near A-point: Each of the bands plotted both in conduction and in valence band correspond to $k_{z}=0-0.06\AA^{-1}$ in steps of $0.01\AA^{-1}$. Blue shaded region indicates the region in in-plane momentum and in energy covered by the ground state exciton in the bulk limit as determined by the size of the Gaussian wavepacket and the exciton binding energy. Insets: (a) Brillouin zone of 2D-InSe, blue circles showing $C_6$-symmetric localization of holes. (b) A comparison of the $\mathbf{Q}=0$ exciton binding energy (closed black circles) with the subband energy splittings. $\Delta_e$ (open yellow circles) and $\Delta_h$ (open blue circles) correspond to the $e$-$e_1$ and $h$-$h_1$ splitting at the $\Gamma$ point, respectively. The closed black circles are the total splitting, $\Delta_e+\Delta_h$. (c) Schematic of hBN-encapsulated InSe, with their different dielectric constants. (d) Brillouin zone of bulk InSe.
}\label{fig:Ex-weight}
\end{figure*}
In the analysis presented below, we describe excitons using two-particle wavefunctions, $\Psi^{\dagger}_{\mathbf{Q}}=\sum_{nm} \int d^{2}k \psi_{\mathbf{Q},nm}(\mathbf{k})a^{\dagger}_{\mathbf{k}+\mathbf{Q},n}c_{\mathbf{k},m}$, written in the wavenumber representation for the constituent electrons and holes occupying states with wavenumbers $\mathbf{k}+\mathbf{Q}$ and $\mathbf{k}$ in subbands \cite{Mogorrian:PRB97-2018,Zultak:NatComm11-2020} $n$ and $m$ on the conduction ($a_{\mathbf{k}+\mathbf{Q},n}$) and valence ($c_{\mathbf{k},m}$) band side of few-layer InSe film spectrum. Below, we project all electron and hole states onto the lowest subbands ($n=1$) in the film, which is justified by the much larger inter-subband energies, as compared to the exciton binding energies in the thin films (with $L\lesssim 10$ see Fig. 2). As a result, the exciton creation operator takes the approximate form $\Psi^{\dagger}_{\mathbf{Q}}=\int d^{2}k \psi_{\mathbf{Q}}(\mathbf{k})a^{\dagger}_{\mathbf{k}+\mathbf{Q},1}c_{\mathbf{k},1}$ where $\psi_{\mathbf{Q}}\equiv \psi_{\mathbf{Q},11}$. This gives the Bethe-Salpeter equation\cite{Onida:RMP74-2002,Jiang:PRB75-2007,Trolle:PRB89-2014,Wu:PRB91-2015,Deilmann:2DMat6-2019} 
\begin{equation}
\int_\mathbf{q}
[(\varepsilon_c(\mathbf{k})\!-\!\varepsilon_v(\mathbf {k}\!-\!\mathbf{Q})\!-\Omega)\delta_{\mathbf{q},0}\!+\!V(\mathbf{q})]\psi_\mathbf{Q}(\mathbf {k}\!+\mathbf{q})\!=\!0,\label{eqn:TB-BSE0}
\end{equation}
for an effectively 2D exciton with momentum $\mathbf{Q}$ and energy $\Omega$ (the latter is a sum $\Omega=E_{g}+E_{b}$ of the gap $E_{g}$ and the binding energy $E_{b}$). Here, we use the notation $\int_\mathbf{q}\equiv\int\frac{\mathrm{d}^2q}{(2\pi)^2}$. The electron-hole (e-h) attraction is accounted for by the Fourier transform of the interaction potential,
 \begin{align}
 &
 V(\mathbf{q})=-\frac{4\pi e^{2}}{\epsilon_{0}}\iint |\phi_{e\mathbf{k}_e}(z)|^2W(\mathbf{q},z,z')|\phi_{h\mathbf{k}_h}(z')|^2dzdz',\notag \\
 &
 W(\mathbf{q},z,z')\!=\frac{\cosh[\tilde{q}(\tfrac{d}{2}\!-\!z)\!+\!\eta]\cosh[\tilde{q}(\tfrac{d}{2}\!+\!z')\!+\!\eta]}{\sqrt{\epsilon_\parallel\epsilon_z}q\sinh(\tilde{q}d+2\eta)};\notag \\ 
 &
 \tilde{q}=\sqrt{\epsilon_\parallel/\epsilon_z}q;\quad \eta =\frac{1}{2}\ln\frac{\sqrt{\epsilon_\parallel\epsilon_z}+\sqrt{\kappa_{\parallel}\kappa_{z}}}{\sqrt{\epsilon_\parallel\epsilon_z}-\sqrt{\kappa_{\parallel}\kappa_{z}}},\label{eqn:W}
 \end{align}
 designed to take into account both the dielectric polarizability of the 2DM and the dielectric environment\cite{Keldysh:JETP29-1979,Rytova:MUPhys3-1967,Latini:PRB92-2015,Trolle:SciRep7-2017} (e.g hBN\cite{Geick:PR146-1966,Laturia:njp2DMat2-2018}, with $\kappa_{\parallel}=6.9$ and $\kappa_{z}=3.7$). For $L$-layer InSe, film thickness is $d=La_z$, where $a_z=8.32$\AA\ is the interlayer distance and $\epsilon_\parallel$ and $\epsilon_z$ are the in- and out-of-plane permitivities of bulk InSe\cite{Kuroda:SSComm34-1980}. The above expression takes into account $z$-dependence of the lowest electron/hole subband wave functions $\phi_{e/h,\mathbf{k}}(z)$, and $W$ is quoted for $z\geq z'$ (for $z<z'$, $z$ should be interchanged with $z'$). We note that, for $L=1$ and $2$, the 2D potential $V(\mathbf{q})$ can be simplified to the Keldysh potential\cite{Keldysh:JETP29-1979,Rytova:MUPhys3-1967,Cudazzo:PRB84-2011,Trolle:SciRep7-2017},
\begin{equation*}
 V(\mathbf{q})\!\approx\!-\frac{2\pi e^2}{\sqrt{\kappa_{z}\kappa_{\parallel}} }\frac{1}{q(1\!+\!r_\ast q)},\quad
 r_\ast=\tfrac{\sqrt{\epsilon_z\epsilon_\parallel}-1}{2\sqrt{\kappa_{z}\kappa_{\parallel}}}d.
\end{equation*}
In the above equation, $r_{*}$ is the screening length\footnote{$r_\ast\approx  L \times 7.7$ \AA} indicating the region dominated by the logarithmically divergent potential at lengthscales smaller than $r_{*}$ and the region dominated by the Coulombic interaction potential at distances greater than $r_{*}$. However, for $L\geq3$, the exciton radius ($a_{exc}$) appears to be smaller than the film thickness, so that the electron/hole charge distribution along the $z$-axis in Eq.\eqref{eqn:W} needs to be taken into account in full details. To do that, we use the quantum-well approximation for the $z$-distribution of the lowest subband,\cite{Mogorrian:PRB97-2018,Florez:arXiv02-2020} $\phi_{e/h, \mathbf{k}}(z)\approx\sqrt{2/d}\cos(\pi z/d)$. We note that separating wavefunction variables and discarding higher energy subbands in Eq. \eqref{eqn:TB-BSE0} is applicable if the quantization energy due to confinement is much larger than the excitonic energy scale, which will be justified later by comparing the intersubband energies to the calculated exciton binding energies.

To implement numerical diagonalization of the Bethe-Salpeter equation \eqref{eqn:TB-BSE0}, we use a basis of harmonic oscillator functions for the bound electron-hole states
\footnote{The bound electron-hole states in an exciton resemble the two-body system in a hydrogen atom. Therefore, these bound states may be well described by an atomic orbital basis set. For the sake of numerical simplicity, we use harmonic oscillator basis function which is an orthogonal Gaussian-type orbital basis.}, $\psi_\mathbf{Q}(\mathbf{k})=\sum^{N_\mathrm{max}}_{0\leq n_{x}\!+n_{y}}\mathcal{A}^\mathbf{Q}_{n_x,n_y}\varphi_{n_x}(k_x)\varphi_{n_y}(k_y)$ where
$\varphi_{n}(k)\!=\!\sqrt{\frac{\lambda}{\pi^{1/2}2^{n}n! }}(-i)^{n}\mathrm{e}^{-k^2\lambda^2/2} H_{n}(k\lambda)$ and $H_n(x)$ is the $n$-th Hermite polynomial. In the above described basis, the choice of the length $\lambda$ and the cutoff $N_\mathrm{max}$ are optimized for speeding up a converging calculation(see Appendix \ref{app:Hermite} for details).
We also checked the performance of the developed code by comparing its results to the exact solution of the 2D hydrogen problem, aiming at $<2\%$ error as compared to the ground state energy of the Rydberg series.
A software package for the implementation of numerical diagonalization of Eq. \eqref{eqn:TB-BSE0} with arbitrary parameters for InSe films and encapsulation environment and instructions for interested users are included in the Supplementary Material\cite{SM}.

\begin{figure}
\centering
\includegraphics[width=3.2in]{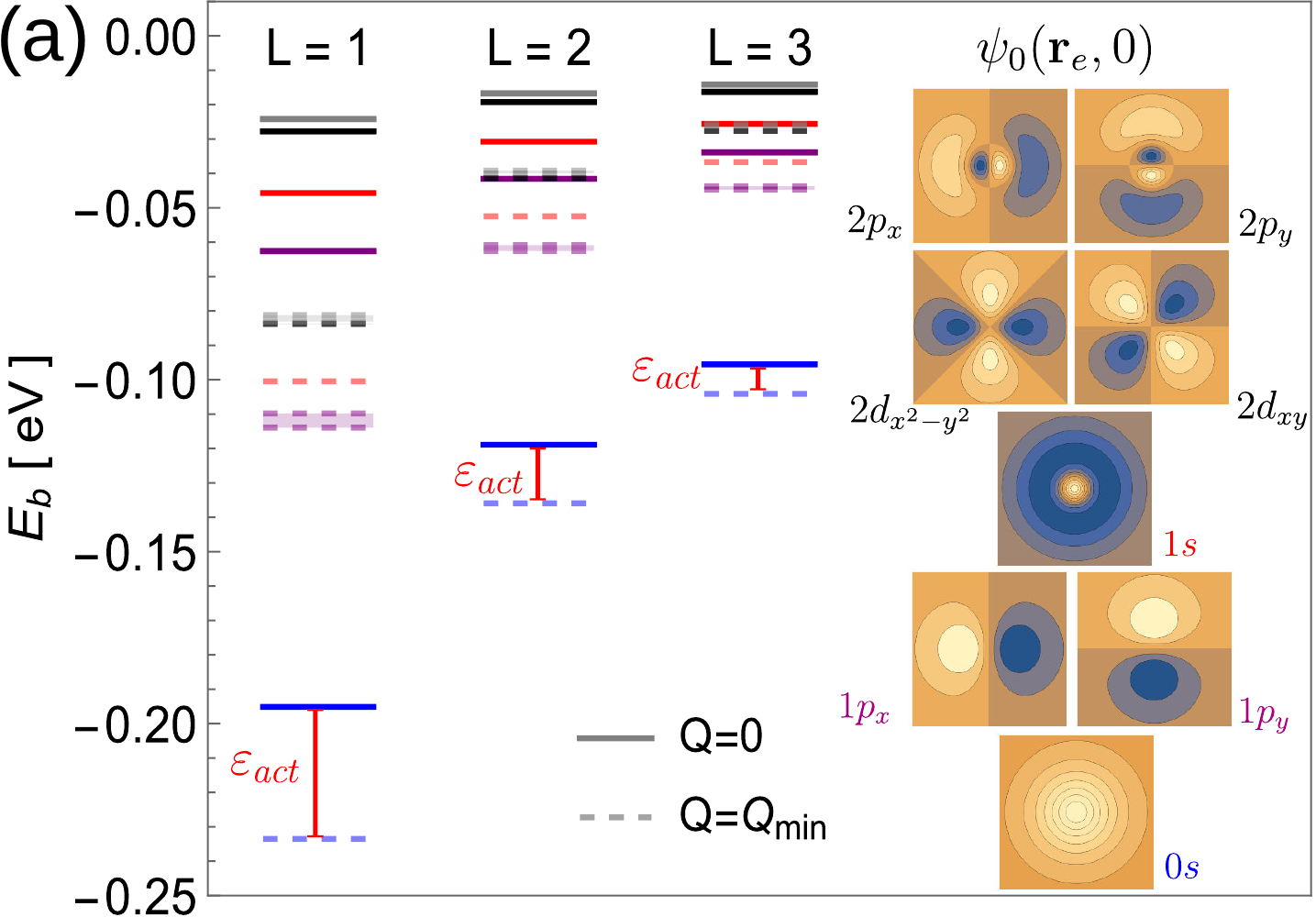}
\includegraphics[width=3.2in]{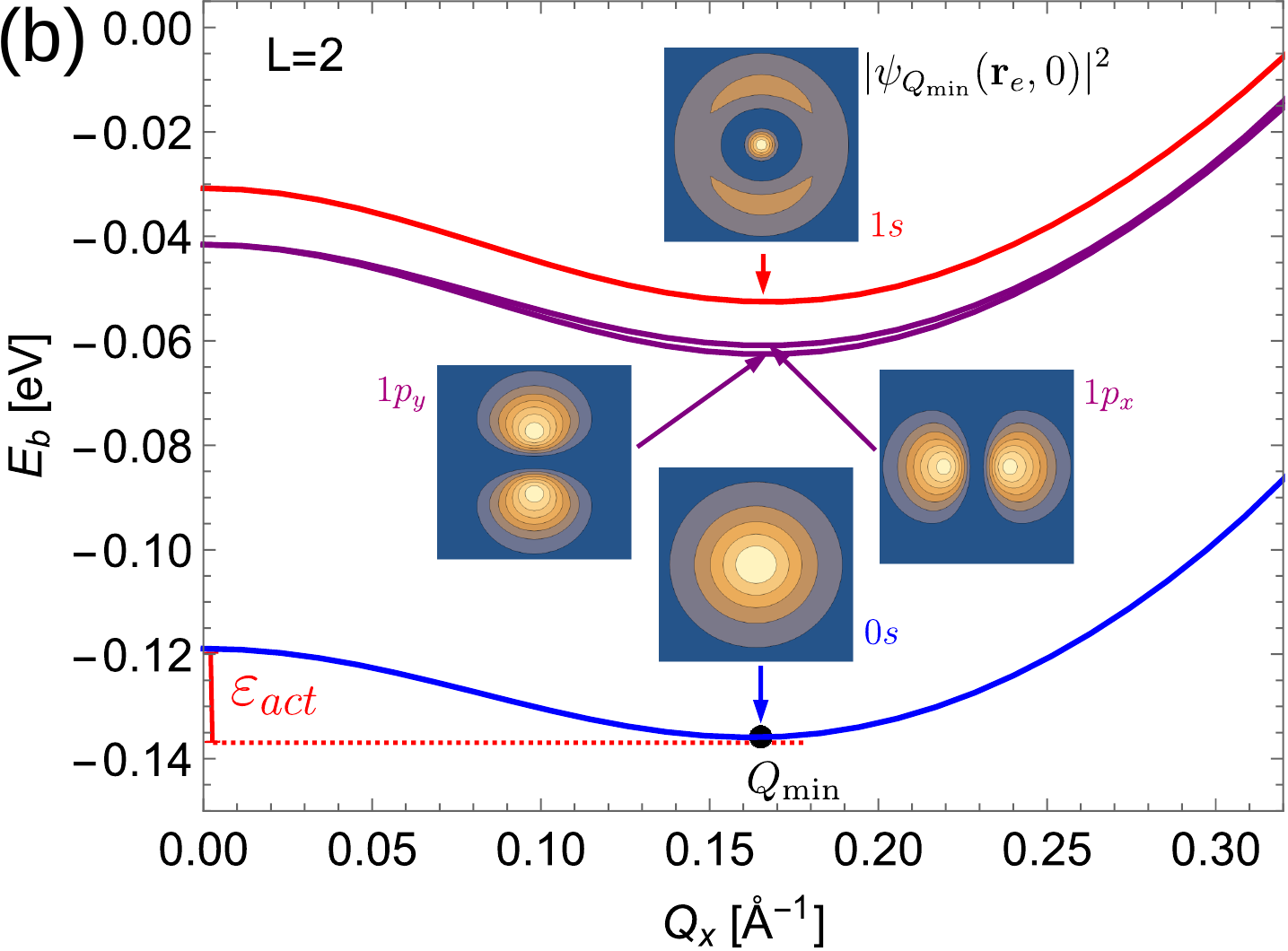}
\includegraphics[width=3.2in]{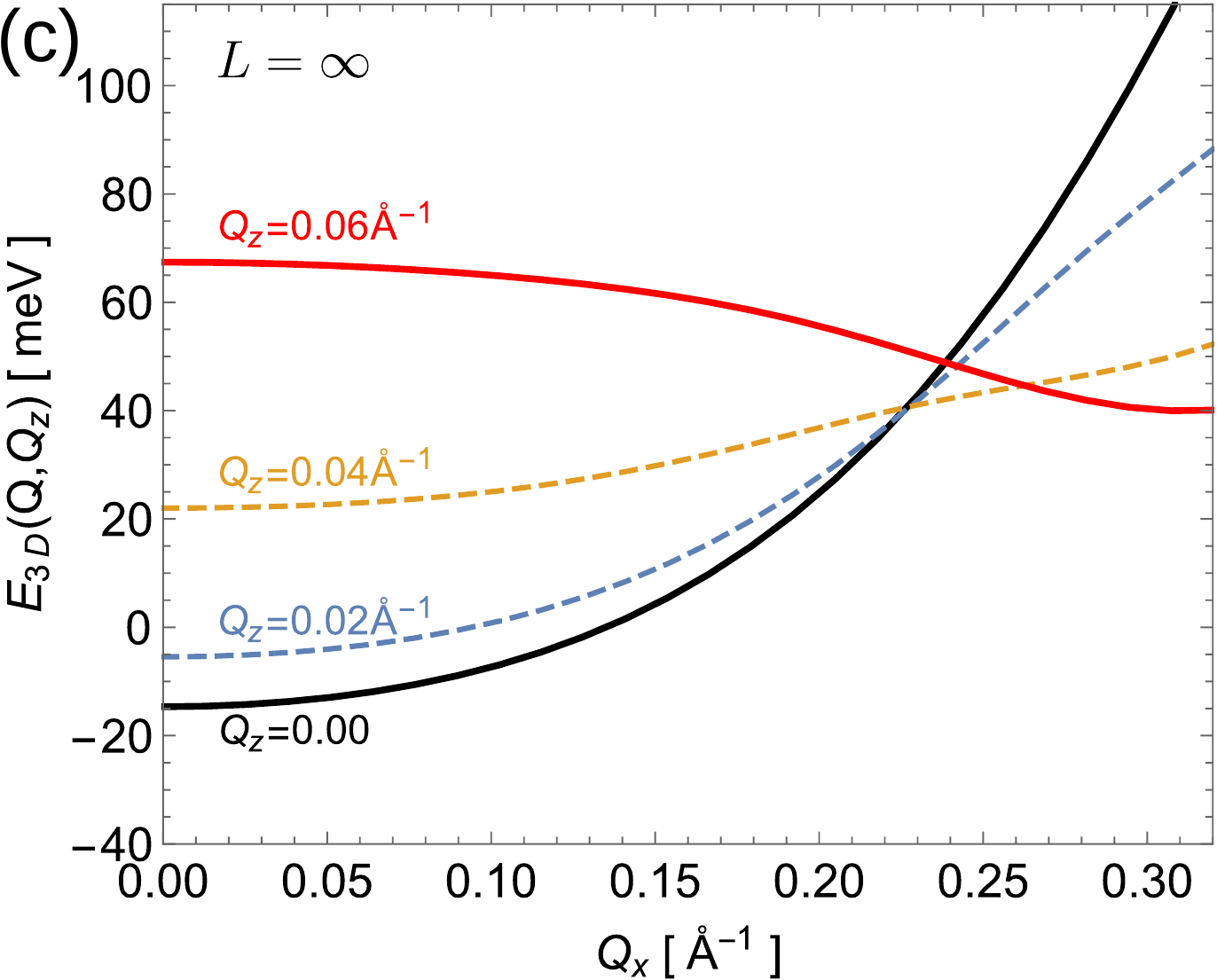}
\caption{(a)The first eight low energy states of the exciton at $Q=0, Q_{\min}$ in hbN-encapsulated InSe. Insets on the right are the schematic exciton wavefunctions $\psi_{0}(\mathbf{r}_e,0)$ in the real-space ($\psi_\mathbf{Q}(\mathbf{r}_e,\mathbf{r}_h)\equiv\sum_{\mathbf{k}}\psi_\mathbf{Q}(\mathbf{k})\mathrm{e}^{i\mathbf{k}\cdot(\mathbf{r}_e-\mathbf{r}_h)}$) which are sorted from higher to lower binding energies(bottom to top). The dark and bright region correspond to a negative and positive value for the wavefunctions amplitude. (b) The exciton dispersion for the first four low energy states. The schematic real-space probability distributions of a bound electron, $|\psi_{Q_{\min}}(\mathbf{r}_e,0)|^2$, are illustrated by the insets. (c) The energy dispersion of the exciton ground state in bulk InSe with $Q_y=0$. The different $Q_z$ values are indicated in the plot. We use the adjusted dielectric parameters in the plot which are $\epsilon_\parallel=9.5$ and $\epsilon_\parallel=8.6$.
}\label{fig:ExState}
\end{figure}

With this numerical setting, we solve Eq. \eqref{eqn:TB-BSE0} using the DFT-parameterized $\mathbf{k \cdot p}$ theory for InSe films with dispersions illustrated in Fig. \ref{fig:Ex-weight}. In particular, we used a polynomial expansion around $\Gamma$ point both for the conduction and valence bands \footnote{In the expansion, one has to ensure that the truncated polynomials must not allow $\varepsilon_c(\mathbf{k}_e)-\varepsilon_v(\mathbf{k}_h)$ smaller than the bandgap for all $\mathbf{k}_{e/h}$.} computed using the GW-parameterized hybrid $\mathbf{k}\cdot \mathbf{p}$ tight-binding model (see Appendix \ref{app:kptb}), $\varepsilon_{c/v}(\mathbf{k})=\sum_{i,j=0}A^{e/h}_{ij}k^i_{x}k^j_{y}$ also plotted in Figs. \ref{fig:Ex-weight} (a)--(c). For comparison, in Fig.  \ref{fig:Ex-weight}(d), we show the conduction and valence band dispersion of bulk InSe near the band edges (which are at A-point in the 3D Brillouin zone), where the inversion of $\varepsilon_{v}(\mathbf{k},k_{z})$ develops upon the increase of $z$-axis momentum, $k_{z}$ (counted from A-point).

In Fig. \ref{fig:ExState}(a), we show the first eight bound states energies of the $\Gamma$-point exciton with $Q=0$ (solid line)\footnote{The degeneracies for the non-$s$-wave states are lifted by less than $1$ meV in the tight-binding model which only has the 6-fold rotational symmetry.} and the lowest-energy momentum-dark state of the  exciton at $Q=Q_{\min}$ (dashed line) for $1\leq L\leq3$. A minimum at $Q=Q_{\min}$ in the exciton dispersion for each state [Fig. \ref{fig:ExState}(b)] is due to the sombrero of the $h$-band (see Fig. \ref{fig:Ex-weight}). 
The non-hydrogen-like energy sequence\cite{Ye:Nat513-2014,Zhang:JPhysCondMat31-2019} is due to the 2D screening of the e-h interaction in the film leading to a Keldysh-like potential for $L$=1 and 2.

To illustrate the layer-number dependence of the exciton dispersion, we compare the exciton binding energy at the $\Gamma$-point, $E_\mathrm{b}(0)$, and at its dispersion minimum, $E_\mathrm{b}(Q_{\min})$, for $L$ up to $10$ layers. In the inset in Fig. \ref{fig:E0-L}, we plot the activation energy, $\varepsilon_{act}$, from the dark exciton state (at $Q=Q_{\min}$) to the optically active state (at $Q=0$). We find that $\varepsilon_{act}\rightarrow0$ at $L^\ast=7$, which indicates that an indirect to direct crossover for the exciton occurs before the expectation based on a single particle valence band dispersion (at $L^\ast=10$, see Appendix \ref{app:kptb}). For completeness, we also analyzed excitons in bulk 3D InSe using bulk band dispersions shown in Fig. \ref{fig:Ex-weight}d and  $V(\mathbf{q},q_z)=-\frac{4\pi e^{2}/\epsilon_0}{\epsilon_\parallel q^2+\epsilon_{z}q^{2}_{z}}$.
We solve the Bethe-Salpeter equation for bulk InSe using the 3D harmonic oscillator basis, and use dielectric constants\cite{Kuroda:SSComm34-1980} $\epsilon_{\parallel}=10.9$, $\epsilon_{z}=9.9$ for InSe, together with the GW-computed valence band masses ($m_{v\parallel}=-5.35m_{0}$, $m_{vz}=-0.078m_{0}$) and the conduction band masses ($m_{c\parallel}/m_0=0.14$ and $m_{cz}/m_0=0.08$ where $m_0$ is the free electron mass) which are close to those measured in cyclotron resonance experiments\cite{cyclotron}. The examples of computed bulk exciton dispersions, $E_{3D}(\mathbf{Q},Q_{z})$ are shown in Fig. \ref{fig:ExState}c. Using ($Q_{z}\approx \frac{\pi}{La_{z}}$) for the quantization of the transverse exciton motion, we find that the crossover into indirect spectrum should be expected at $L\approx6-7$ layers, in agreement with the transition number of layers $L^\ast$ found in the layer dependence of the activation energy $\varepsilon_{act}$ (inset in Fig. \ref{fig:E0-L}). 
We note that the computed bulk (3D) exciton binding energy is about 30\% lower than the experimentally claimed \cite{Kuroda:JPSJ47-1979,Schindlmayr:EurJPhys18-1997} values of 13-15meV. Binding energy can be increased to 14.6meV by choosing $\epsilon_{\parallel}=9.5$, $\epsilon_{z}=8.6$ (with $\sqrt{\epsilon_{z}/\epsilon_{\parallel}}=0.95$ as in Ref. \onlinecite{Kuroda:SSComm34-1980}). For this reason we computed and compared the exciton spectra in the films using two choices of dielectric parameters $\epsilon_{\parallel}=10.9$,$\epsilon_{z}=9.9$ and $\epsilon_{\parallel}=9.5$,$\epsilon_{z}=8.6$. We find that in thin films $L\leq10$ such a variation of InSe dielectric parameters has a much weaker influence on the exciton bindings than in the bulk material. These calculated binding energies compare well with the values observed in the recent experiments \cite{Zultak:NatComm11-2020} on hBN-encapsulated thin InSe films.
In summary, we present a mesoscale theory which is particularly useful for investigating the energy spectrum of a Wannier-Mott exciton in large gap semiconductors ($E_g\gg |E_b|$). Most interestingly, this theoretical framework can also be applied to study direct and indirect excitons in complex van der Waals heterostructures \cite{Fang:PNAS111-2014,Rivera:NatComm6-2015,*Rivera:Science351-2016,Wang:NanoLett18-2017,Ciarrocchi:NatPhoto13-2018,Calman:NatComm9-2018,Ubrig:NatMat-2020,Mak:NatNano13-2018,KostyaSuperfluid}.

\begin{acknowledgments}
The authors would like to thank H. Deng, M. van Schilfgaarde, D. Pashov and V. Enaldiev for useful discussions. AC acknowledges support from EPSRC CDT Graphene NOWNANO. SJM, KWS and VF acknowledge support from ERC Synergy Grant Hetero2D, EC Quantum Technologies Flagship Project 2D-SIPC, EPSRC EP/N010345, and the Lloyd Register Foundation Nanotechnology grant.
\end{acknowledgments}

\appendix
\section{Parameterization of electron/hole dispersion in $L$-layer InSe}\label{app:k.p-InSe}

Here, we give the details on the parameterized electron and hole dispersion by using polynomial fit. The conduction and valence band dispersions near the $\Gamma$-point are approximated by:
\begin{align}
&\varepsilon_c(\mathbf{k})=\frac{1}{2m_c}\mathbf{k}^2\label{eqn:poly-Ec},\\
&\varepsilon_v(\mathbf{k})=A^h_2\mathbf{k}^2+A^h_4 \mathbf{k}^4+A^h_6\mathbf{k}^6 +A^h_8\mathbf{k}^8\label{eqn:poly-Ev}
\end{align}
where the hexagonal wraping terms are ignored, because the exciton wave function is strongly localized in the k-space. These polynomials are obtained by fitting to bands from the GW-parameterized hybrid $\mathbf{k}\cdot \mathbf{p}$ tight-binding (HkpTB) model, Appendix \ref{app:kptb}, and the fitted values for 1- to 10-layer InSe film are listed in Table \ref{tbl:Ecv}. We note for $L\leq9$ that the quadratic term in the valence band dispersion corresponds to negative effective hole masses. This yields a sombrero-shaped dispersion in the valence band and requires to retain higher-order terms in the expansion for fitting. The hole mass becomes positive at $L=10$. For the 3D bulk dispersion near the conduction and valence band edges of $\gamma$-InSe, we employ the following polynomial of the form,
\begin{align}
    &\varepsilon_c(\mathbf{k},k_{z})\!=\!\frac{k^2}{2m_{c\parallel}}+\frac{k_z^2}{2m_{cz}},\label{eqn:poly-Ec3D}\\
    &\varepsilon_v(\mathbf{k},k_{z})\!=\!\frac{\mathbf{k}^{2}}{2m_{v\parallel}}+\frac{k^{2}_{z}}{2m_{vz}}\!+\!\gamma \mathbf{k}^{4}\!+\!\alpha \mathbf{k}^{2}k^{2}_{z}\!+\!\gamma_{z}k^{4}_{z}.\label{eqn:poly-Ev3D}
\end{align}
Here, $k_z$ is measured from the A-point. In the fit, obtained using GW-DFT computed bands, the effective in-plane and out-of-plane masses for the electron are $m_{c\parallel}=0.16m_{0}$ and $m_{cz}=0.086m_{0}$, close to the experimentally measured \cite{cyclotron} values of $m_{c\parallel} \approx 0.14m_{0}$ and $m_{cz} \approx 0.08m_{0}$ respectively. For the valence band, the fitted-parameters are: $m_{v\parallel}=-5.35m_{0}$, $m_{vz}=-0.078m_{0}$, $\gamma=-10.84$eV\AA$^{4}$, $\alpha=1074$eV\AA$^{4}$ and $\gamma_{z}=1688$eV\AA$^{4}$. 
\begin{table}
\begin{ruledtabular}
\begin{tabular}{c|cccc||c||}
$L$ & $A^h_8$ (eV\AA$^{8}$) & $A^h_6$ (eV\AA$^{6}$) & $A^h_4$ (eV\AA$^{4}$)& $A^h_2$ (eV\AA$^{2}$) & $m_c/m_0$ \\
\hline
1 & -1188.591  & 471.809 &  -68.601 &  3.674 & 0.266\\
2 & -1210.270  & 388.158 &  -49.004 &  1.989 & 0.223\\
3 & -1308.626  & 371.401 &  -43.048 &  1.372 & 0.207\\
4 & -1411.696  & 364.846 & -39.437  &  0.985 & 0.198\\
5 & -1565.869  & 366.036 & -36.797  &  0.703 & 0.193\\
6 & -1745.505  & 368.254 & -34.556  &  0.487 & 0.189\\
7 & -1938.337  & 369.112 &  -32.543 &  0.316 & 0.187\\
8 & -2130.725  & 367.119 & -30.684  &  0.179 & 0.184\\
9 & -2302.573  & 361.073 &  -28.941 &  0.068 & 0.183\\
10& -2085.138  & 331.905 &  -27.004 & -0.026 & 0.181\\
\end{tabular}
\end{ruledtabular}
\caption{Polynomial fit of the sombrero dispersion as for the topmost valence band and parabolic dispersion of the lowest conduction band. $m_0$ is the free electron mass.}\label{tbl:Ecv}
\end{table}

\section{Hybrid multiband $\mathbf{k\cdot p}$ tight-binding theory with parameters from quasiparticle self-consistent GW calculations}\label{app:kptb}
\subsection{Hybrid multiband $\mathbf{k\cdot p}$ tight-binding model}
The model used in this study is built using two main components: a multiband $\mathbf{k\cdot p}$ model describing the monolayer bands (following
Refs. \onlinecite{Zhou:PRB96-2017} and  \onlinecite{Appelbaum}), and interlayer coupling in few-layer and bulk systems, described using a tight-binding approach based on the monolayer $\mathbf{k\cdot p}$ bands, (similar to the hybrid $\mathbf{k\cdot p}$ tight-binding approach taken in Refs. \onlinecite{Mogorrian:PRB97-2018} and \onlinecite{tmd_isb}).

In this description we model the bands of few-layer and bulk InSe near the $\Gamma$ point using a Hamiltonian with the form
\begin{equation}
\label{eq:all_fl}
H = \sum_{\mathbf{k},\sigma}\left[\sum_{n=1}^N H_{ML,\mathbf{k},\sigma}^{n} + \sum_{n=1}^{N-1}H_{IL,\mathbf{k},\sigma}^{n,n+1}+\mathrm{H.c.}\right],
\end{equation}
where $H_{ML,\mathbf{k},\sigma}^{n}$ is the monolayer $\mathbf{k\cdot p}$ Hamiltonian on layer $n$ of the $N$-layer crystal, at $\mathbf{k}$ with z-projection of spin $\sigma=\pm\frac{1}{2}$. $H_{I-L}^{n,n+1}$ includes the interlayer tight-binding hops between the monolayer bands.
\subsubsection{Monolayer $\mathbf{k\cdot p}$ Hamiltonian}

The monolayer Hamiltonian follows the multiband $\mathbf{k\cdot p}$ approaches of Refs. \onlinecite{Zhou:PRB96-2017} and  \onlinecite{Appelbaum}. While in our previous works\cite{Magorrian:PRB94-2016,Mogorrian:PRB97-2018} the basis of monolayer bands was a basis of single-band $\mathbf{k\cdot p}$ expansions, so that matrix elements such as couplings to electromagnetic fields and the interlayer hops mentioned above had to depend on $k$, here we follow the multiband approach and take as our basis the bands at $\Gamma$, and introduce $k$-dependent off-diagonal terms to account for the variation of the bands with $k$. At the expense of an increase in the dimensionality of the parameter space, this allows us to make the approximation that the interlayer hops are independent of $k$, and assists in the capture of higher-order effects, such as the offset valence band maximum, while keeping the $\mathbf{k\cdot p}$ expansions to order $k^2$. The monolayer Hamiltonian for layer $n$ of an $N$-layer crystal takes the form
\begin{widetext}
\begin{align}
\label{eq:ml}
H_{ML,\mathbf{k},\sigma}^{n} = &(\varepsilon_{c_1}+\alpha_{c_1}k^2)a_{n,c_1,\mathbf{k}}^{\sigma\dagger}a_{n,c_1,\mathbf{k}}^{\sigma}+(\varepsilon_{c}+\alpha_{c}k^2)a_{n,c,\mathbf{k}}^{\sigma\dagger}a_{n,c,\mathbf{k}}^{\sigma}+(\varepsilon_{v}+\alpha_{v}k^2)a_{n,v,\mathbf{k}}^{\sigma\dagger}a_{n,v,\mathbf{k}}^{\sigma}\\
+&(\varepsilon_{v_1}+(\alpha_{v_1}k^2+\alpha_{v_1}'(k_x^2-k_y^2))a_{n,v_{1x},\mathbf{k}}^{\sigma\dagger}a_{n,v_{1x},\mathbf{k}}^{\sigma}+(\varepsilon_{v_1}+\alpha_{v_1}k^2+\alpha_{v_1}'(k_y^2-k_x^2))a_{n,v_{1y},\mathbf{k}}^{\sigma\dagger}a_{n,v_{1y},\mathbf{k}}^{\sigma}\nonumber\\
+&(\varepsilon_{v_2}+(\alpha_{v_2}k^2+\alpha_{v_2}'(k_x^2-k_y^2))a_{n,v_{2x},\mathbf{k}}^{\sigma\dagger}a_{n,v_{2x},\mathbf{k}}^{\sigma}+(\varepsilon_{v_2}+\alpha_{v_2}k^2+\alpha_{v_2}'(k_y^2-k_x^2))a_{n,v_{2y},\mathbf{k}}^{\sigma\dagger}a_{n,v_{2y},\mathbf{k}}^{\sigma}\nonumber\\
+&2\alpha_{v_1}'k_xk_ya_{n,v_{1x},\mathbf{k}}^{\sigma\dagger}a_{n,v_{1y},\mathbf{k}}^{\sigma}+2\alpha_{v_2}'k_xk_ya_{n,v_{2x},\mathbf{k}}^{\sigma\dagger}a_{n,v_{2y},\mathbf{k}}^{\sigma}\nonumber\\
+&\beta_{c_1,v} k^2 a_{n,c_1,\mathbf{k}}^{\sigma\dagger}a_{n,v,\mathbf{k}}^{\sigma}+i\beta_{c_1,v_{2}}(k_xa_{n,c_1,\mathbf{k}}^{\sigma\dagger}a_{n,v_{2x},\mathbf{k}}^{\sigma}+k_ya_{n,c_1,\mathbf{k}}^{\sigma\dagger}a_{n,v_{2y},\mathbf{k}}^{\sigma})\nonumber\\
+&i\beta_{v,v_{2}}(k_xa_{n,v,\mathbf{k}}^{\sigma\dagger}a_{n,v_{2x},\mathbf{k}}^{\sigma}+k_ya_{n,v,\mathbf{k}}^{\sigma\dagger}a_{n,v_{2y},\mathbf{k}}^{\sigma})+i\beta_{c,v_{1}}(k_xa_{n,c,\mathbf{k}}^{\sigma\dagger}a_{n,v_{1x},\mathbf{k}}^{\sigma}+k_ya_{n,c,\mathbf{k}}^{\sigma\dagger}a_{n,v_{1y},\mathbf{k}}^{\sigma})\nonumber\\
-&2\lambda_{v_{1,2}}i\sigma(a_{n,v_{1x},\mathbf{k}}^{\sigma\dagger}a_{n,v_{1y},\mathbf{k}}^{\sigma}+a_{n,v_{2x},\mathbf{k}}^{\sigma\dagger}a_{n,v_{2y},\mathbf{k}}^{\sigma})+\lambda_{v,v_1}(-2\sigma a_{n,v,\mathbf{k}}^{\sigma\dagger}a_{n,v_{1x},\mathbf{k}}^{-\sigma}+ia_{n,v,\mathbf{k}}^{\sigma\dagger}a_{n,v_{1y},\mathbf{k}}^{-\sigma}).\nonumber
\end{align}
\end{widetext}
The bands which form the basis of the model are the monolayer $\Gamma$-point bands in the absence of spin-orbit coupling (SOC)\cite{Magorrian:PRB94-2016}. The operator $a_{n,j,\mathbf{k}}^{\sigma(\dagger)}$ annihilates (creates) an electron in layer $n$, band $j$, with spin $\sigma=\pm\frac{1}{2}$ and in-plane momentum $\mathbf{k}$. As singly-degenerate bands which are totally in-plane symmetric at $\Gamma$, bands $c_1,c,v$ are assigned $\Gamma$-point energies $\varepsilon_{c_1,c,v}$ with quadratic `onsite' dispersions with respective coefficients $\alpha_{c_1,c,v}$. In contrast, in the absence of SOC bands $v_1$ and $v_2$, being dominated by $p_x$ and $p_y$ orbitals, are twice-degenerate at $\Gamma$ with energies $\varepsilon_{v_1,v_2}$ . The dispersions of their two light- and heavy-hole branches are handled using two components corresponding to a basis of their $p_x$ and $p_y$ components, with quadratic intra- and inter-component contributions with coefficients $\alpha_{v_{1,2}}^{(\prime)}$. In the multiband $\mathbf{k\cdot p}$ picture away from $\Gamma$ the bands are modified by off-diagonal terms between them. These terms must preserve the $\sigma_h$ symmetry of the monolayer, so only involve the pairs $c_1,v$, $c,v_1$ and $v,v_2$. Of these, $c_1,v$ is between bands which are totally in-plane symmetric at $\Gamma$, so the off-diagonal term is quadratic, whilst terms involving the $x$ and $y$ components of $v_{1,2}$ are linear in $k_x$ and $k_y$, respectively. The coefficients of these terms are denoted as $\beta_{c_1,v}$, $\beta_{c,v_1}$, and $\beta_{v,v_2}$, respectively.
Finally, spin-orbit coupling (SOC) is included within the components of $v_1$ and $v_2$ ($l_zs_z$ with coupling strength $\lambda_{v_{1,2}}$) and between $v_1$ and $v$ (the `spin-flip' $l_xs_x+l_ys_y$ with coupling strength $\lambda_{v,v_1}$). Cross-gap `spin-flip' terms are neglected.
\subsubsection{Interlayer tight-binding hops}
The nonzero interlayer tight-binding hops between the monolayer bands, and their form, can be informed by the symmetries of the bands involved in the hop. The resulting interlayer contribution to the Hamiltonian takes the form,
\begin{align}
\label{eq:int_hops}
H_{IL,\mathbf{k},\sigma}^{n,n+1} =& \sum_{j=c_1,c,v}{t_ja_{n,j,\mathbf{k}}^{\sigma\dagger}a_{n+1,j,\mathbf{k}}^{\sigma}} \\
&+t_{c_1,c}(a_{n,c_{1},\mathbf{k}}^{\sigma\dagger}a_{n+1,c,\mathbf{k}}^{\sigma}-a_{n,c,\mathbf{k}}^{\sigma\dagger}a_{n+1,c_{1},\mathbf{k}}^{\sigma})\nonumber\\
&+t_{c,v}(a_{n,c,\mathbf{k}}^{\sigma\dagger}a_{n+1,v,\mathbf{k}}^{\sigma}-a_{n,v,\mathbf{k}}^{\sigma\dagger}a_{n+1,c,\mathbf{k}}^{\sigma})\nonumber\\
&+t_{v_{1,2}}\sum_{i=x,y}(a_{n,v_{1i},\mathbf{k}}^{\sigma\dagger}a_{n+1,v_{1i},\mathbf{k}}^{\sigma}-a_{n,v_{1i},\mathbf{k}}^{\sigma\dagger}a_{n+1,v_{2i},\mathbf{k}}^{\sigma}\nonumber\\
&-a_{n,v_{2i},\mathbf{k}}^{\sigma\dagger}a_{n+1,v_{2i},\mathbf{k}}^{\sigma}+a_{n,v_{2i},\mathbf{k}}^{\sigma\dagger}a_{n+1,v_{1i},\mathbf{k}}^{\sigma})\nonumber. 
\end{align}
Since the $\gamma$ stacking preserves the $C_3$ rotational symmetry of the monolayer, the bands may be divided into two groups, with no hopping between the singly- and doubly-degenerate basis bands, with the $x$ and $y$ components also not mixed by the interlayer hops. We have made the approximation that, since interlayer hops are dominated by interlayer Se-Se pairs\cite{Magorrian:PRB94-2016}, they may be taken as $z/-z$ symmetric. As a result, hops between $c_1$ and $c$, and between $c$ and $v$, which are pairs of bands with opposing symmetry under $z/-z$ reflection in the monolayer, are antisymmetric under exchange of layers. We neglect the hop $t_{c_1,v}$ as the bands are well separated in energy, and interlayer hops involving $c_1$ are expected to be weak owing to the dominance of the $c_1$ wavefunction by orbitals on the indium atoms in the center of each layer. Finally, using the domination of $v_1$ and $v_2$ by Se $p_x$ and $p_y$ orbitals, we assume that all hops within and between $v_1$ and $v_2$ are of the same magnitude, $t_{v_{1,2}}$.
\subsection{Parametrisation - bulk $\gamma$-InSe}
 Since DFT can often underestimate band gaps, significantly so in the case of thicker 2D and bulk InSe, a means by which one may obtain spectra of more use in comparision with experiments is the use of a `scissor operator' - a rigid shift upwards in energy of the unoccupied bands with respect to the occupied bands. In other words, one assumes that features of the DFT bands, such as effective masses, band widths, matrix elements, and so on, are all correct, other than the size of the gap itself. This has been shown to be a useful procedure in theoretical studies of semiconductors\cite{sc3,sc4,sc5}, and in 2D InSe\cite{Bandurin:NatNano12-2016,Magorrian:PRB94-2016}. However, the magnitude of the underestimation of the gap (approaching a factor of $\sim 4$ in the bulk limit) for InSe can make the procedure more complex. For example, a straight scissor correction without taking into account other effects of the underestimation of the gap can lead to an overestimation of the interband out-of-plane electric dipole matrix element\cite{Magorrian:PRB94-2016}, or an underestimation of the band-edge effective masses in the bulk case and hence an overestimation of the splitting of subbands in the few-layer case\cite{Mogorrian:PRB97-2018}. While there are means by which some of these problems may be overcome (for example, the out-of-plane effective mass was corrected in Ref. \onlinecite{Mogorrian:PRB97-2018} by applying a scissor correction to the monolayer bands \textit{after} parametrisation of the interlayer hops), the presence of cross-gap off-diagonal matrix elements in even the monolayer Hamiltonian presents challenges in the determination of the appropriate means of compensating for an underestimation of the band gap in a DFT reference.

In this case, therefore, we take as our first-principles reference a quasiparticle self-consistent GW (QSGW) calculation for the bulk crystal. For this we use the QUESTAAL package\cite{kotani2007quasiparticle,*Pashov:CompPhysComm149-2020}, using the Bethe-Salpeter equation (BSE) to determine the polarization in the calculation of W. Since the bands and gaps of InSe have been shown to be sensitive to strain\cite{Li2DMaterials2018,Song2018}, we use an experimental lattice with crystal structure parameters found using x-ray diffraction\cite{iucr_bulk}. The DFT part of the calculation sampled the Brillouin zone with a $24\times 24\times 24$ $\mathbf{k}$-point grid, while for the QSGW part a $6\times 6\times 6$ grid was used. In the calculation of W, nine occupied bands and 15 unoccupied bands were handled using the BSE, while the rest of the bands were handled at the random-phase approximation level. In the DFT part, the G-vector cutoff for the interstitial density mesh was 9.1 Ry$^{1/2}$, while in the QSGW part the cutoffs for the plane-wave expansions of the eigenfunctions and the Coulomb integrals were 3.4 a.u. and 2.9 a.u., respectively. The QSGW calculation of the self-energy is carried out without taking SOC into account, with the effects of SOC included at the DFT level afterwards. We choose a calculation of the bulk crystal as a reference for finding model parameters as a QSGW calculation for few-layer InSe would be prohibitively expensive given the number of atoms in a unit cell. The calculation gives a quasiparticle band gap of 1.367~eV for the bulk, close to the experimentally obtained 1.351~eV\cite{Camassel:PRB17-1978}.

In the case of the model, Eq. (\ref{eq:all_fl}) is amended to describe the bulk with a unit cell corresponding to a single layer as
\begin{equation}
H = H_{1L}^{1} + H_{IL}^{1,1}e^{ik_za_z}+\mathrm{H.c.},
\end{equation}
\begin{figure}
	\centering
	\includegraphics[width=0.9\linewidth]{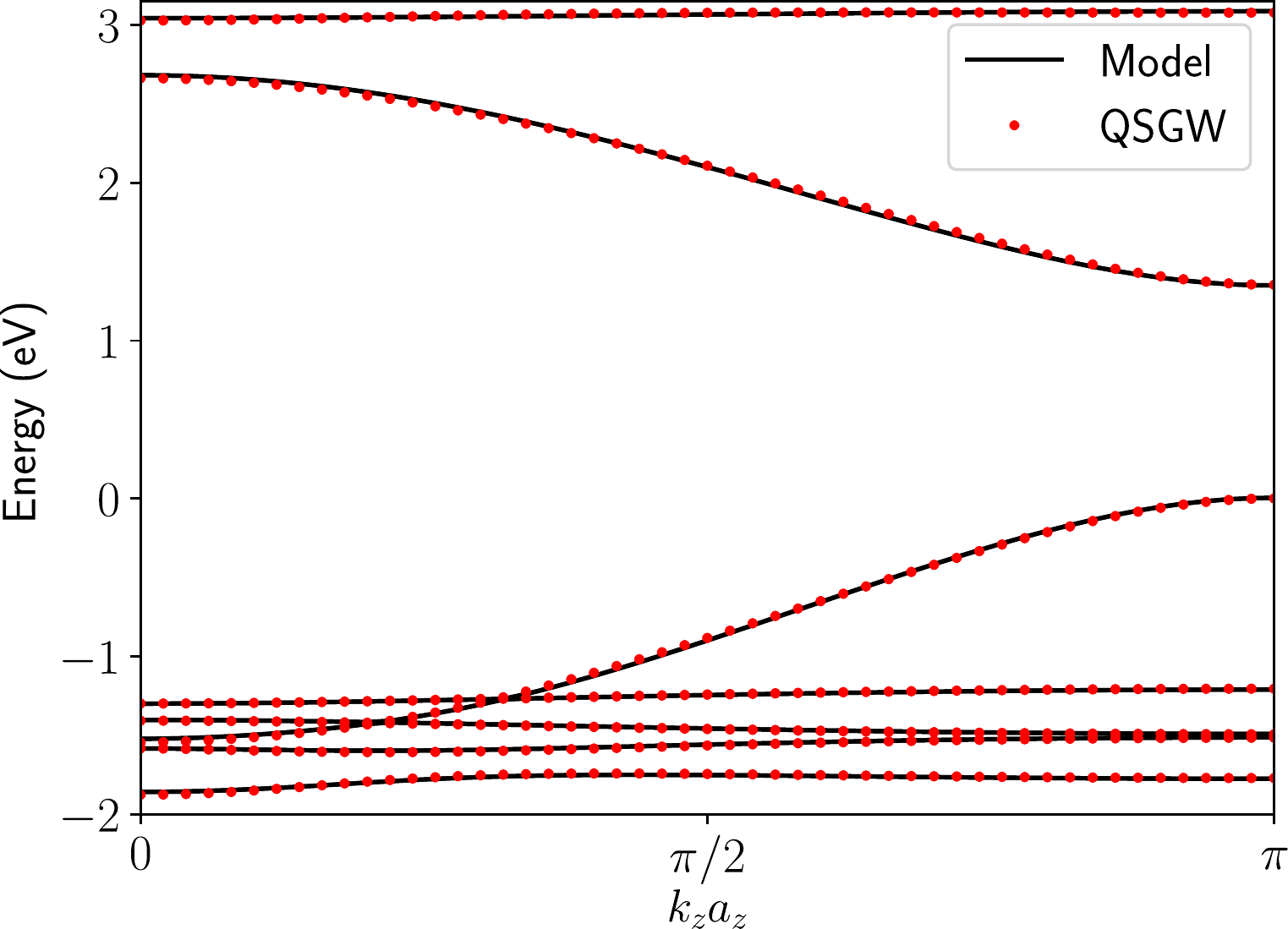}\vspace{0.3cm}
	\includegraphics[width=0.9\linewidth]{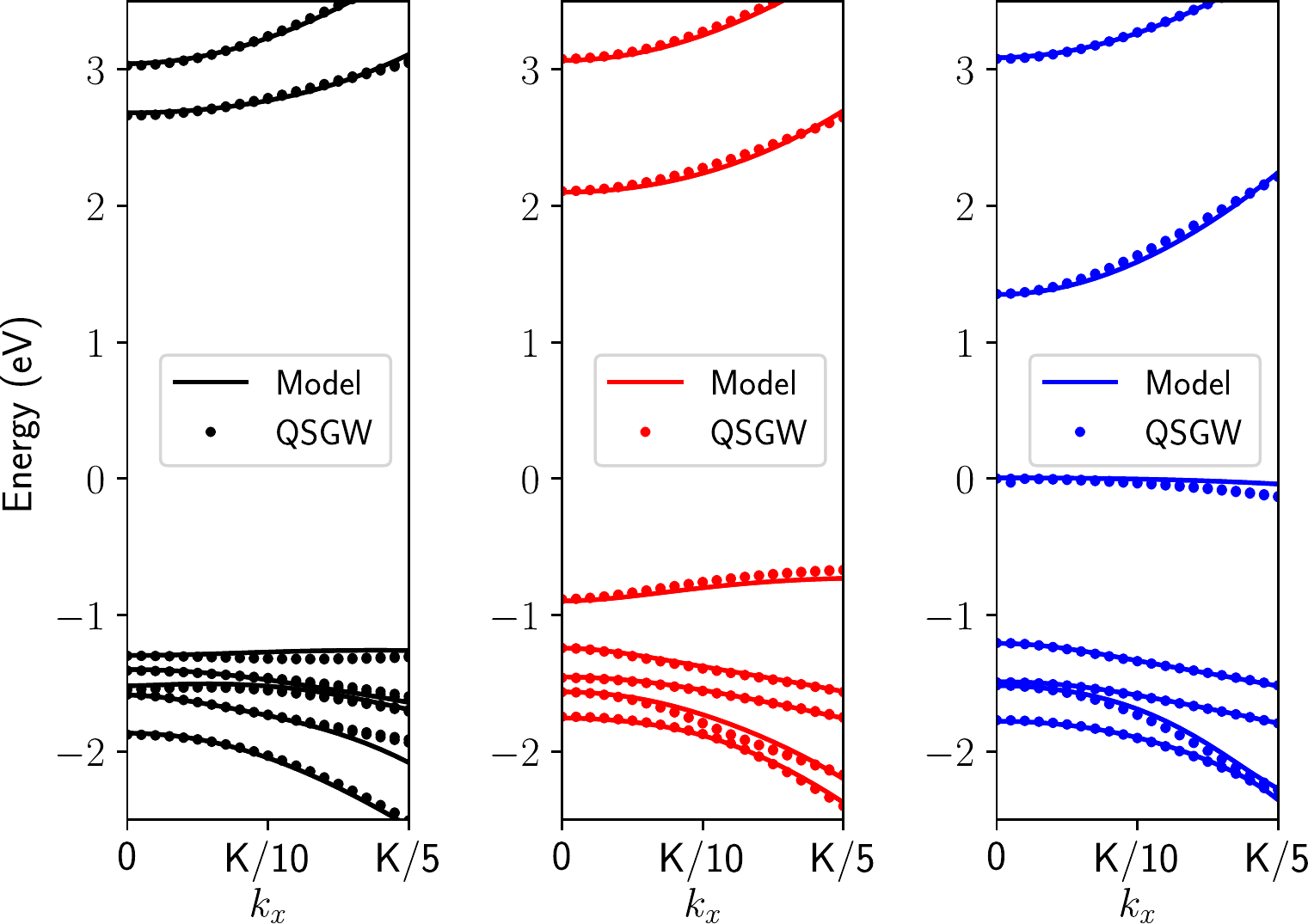}
	\caption{Upper panel: QSGW (dots) and fitted model (lines) out-of-plane dispersions for bulk $\gamma$-InSe, for in plane momentum $\mathbf{k}=0$. Lower panels: in-plane dispersions (along $k_x$) for (from left to right) $k_za_z=0,\pi/2,\pi$. 0~eV set to valence band edge in all cases.}
	\label{fig:gammafit}
\end{figure}
where $k_z$ is the out-of-plane momentum and $a_z=8.315$~\AA~is the distance between successive layers\cite{iucr_bulk}. The parametrization is carried out in two steps. Firstly, we fit bands for 50 $k_z$ points between $k_z=0$ and $k_z=\pi/a_z$ for $\mathbf{k}=0$, as we show in Fig. \ref{fig:gammafit}, then holding the 2D $\Gamma$-point parameters fixed, we fit the in-plane dispersions for small $\mathbf{k}$ near $\Gamma$ up to $k=\mathrm{K}/5$ for each $k_z$ used in the first stage of the fitting. In Fig. \ref{fig:gammafit} we show the in-plane QSGW and model dispersions for $k_za_z=0,\pi/2,\pi$. The model parameters are given in Table \ref{tab:vv1}.
\begin{table}[h]
	\caption{\label{tab:vv1}Model parameters for Eqs. (\ref{eq:ml}) \& (\ref{eq:int_hops}) fitted to QSGW bands for bulk InSe. 0~eV is set to the valence band edge in the bulk.}
	\begin{center}
		\begin{tabular}{cr}
			\hline
			\hline
			$\varepsilon_{c_1}$ & 3.064 eV\\
			$\varepsilon_{c}$ & 2.015 eV\\
			$\varepsilon_{v}$ & $-0.855$ eV\\
			$\varepsilon_{v_1}$ & $-1.449$ eV\\
			$\varepsilon_{v_2}$ & $-1.538$ eV\\
            \hline$\lambda_{v_{1,2}}$ &0.142 eV \\
			$\lambda_{v,v_1}$ & 0.119 eV\\
			\hline$t_{c_1}$ & $-0.011$ eV\\
			$t_{c}$ & 0.333 eV\\
			$t_{v}$ & $-0.420$ eV\\
			$t_{v_{1,2}}$ & $-0.048$ eV\\
			$t_{c_1,c}$ & 0.019 eV\\
			$t_{c,v}$ & 0.251 eV\\
			\hline
			\hline
		\end{tabular}\quad
		\begin{tabular}{cr}
	\hline
	\hline
	$\alpha_{c_1}$ & 1.54 eV\AA$^2$\\
	$\alpha_{c}$ & $-18.7$ eV \AA$^2$\\
	$\alpha_{v}$ & $-4.95$ eV\AA$^2$\\
	$\alpha_{v_1}$ & 6.48 eV\AA$^2$\\
	$\alpha_{v_1}'$ & $-10.51$ eV\AA$^2$\\
	$\alpha_{v_2}$ & $-0.28$ eV\AA$^2$\\
	$\alpha_{v_2}'$ & $-4.20$ eV\AA$^2$\\
	\hline$\beta_{c_1,v}$ & 3.77 eV\AA$^2$\\
	$\beta_{c_1,v_2}$ & 8.51 eV\AA\\
	$\beta_{c,v_1}$ &  10.54 eV\AA\\
	$\beta_{v,v_2}$ & $-2.78$ eV\AA\\
	\hline
	\hline
\end{tabular}
	\end{center}
\end{table}
\subsection{Few-layer bands}
\begin{figure}[h!]
	\centering
	\includegraphics[width=0.9\linewidth]{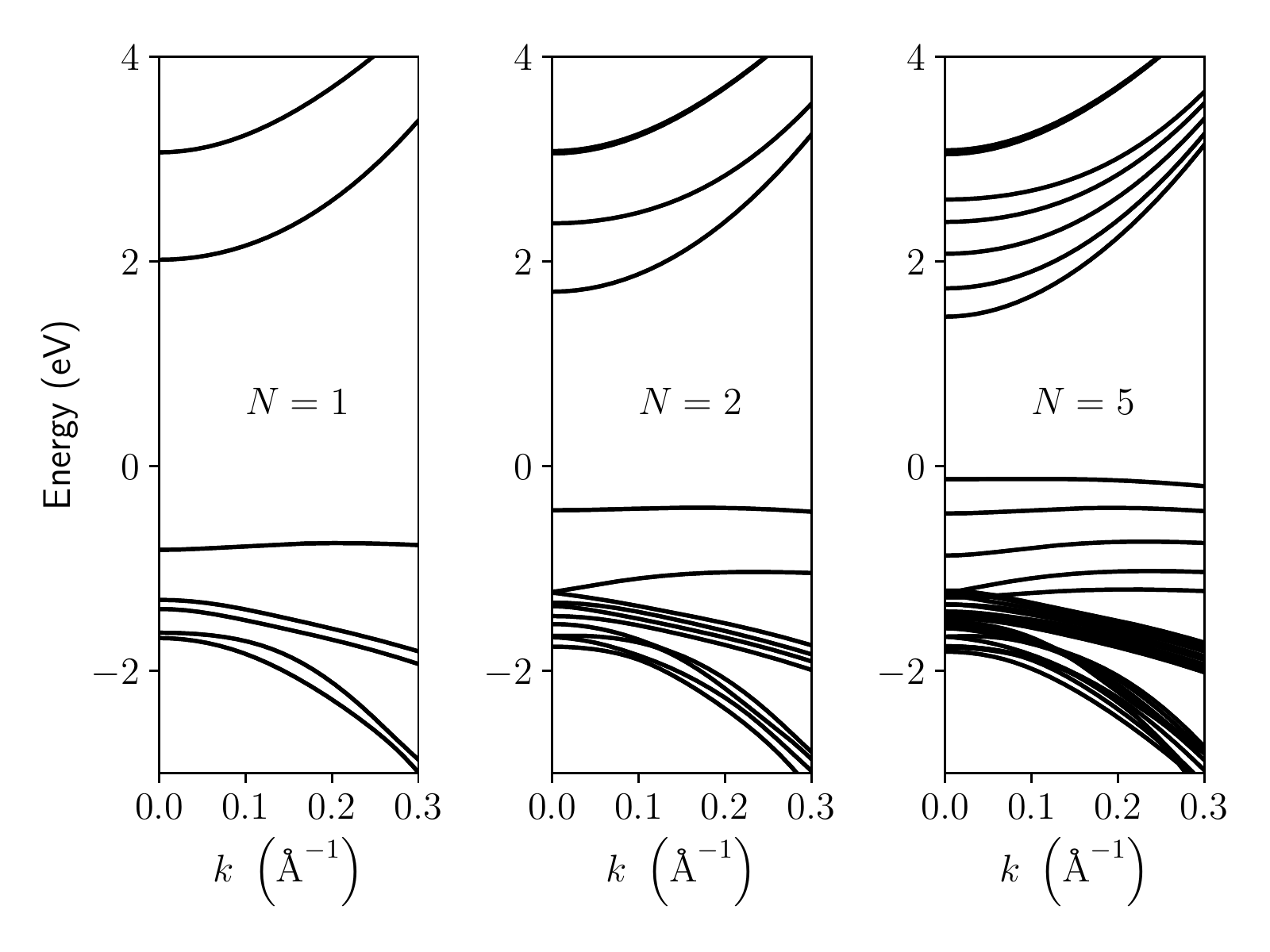}\vspace{0.3cm}
	\includegraphics[width=0.9\linewidth]{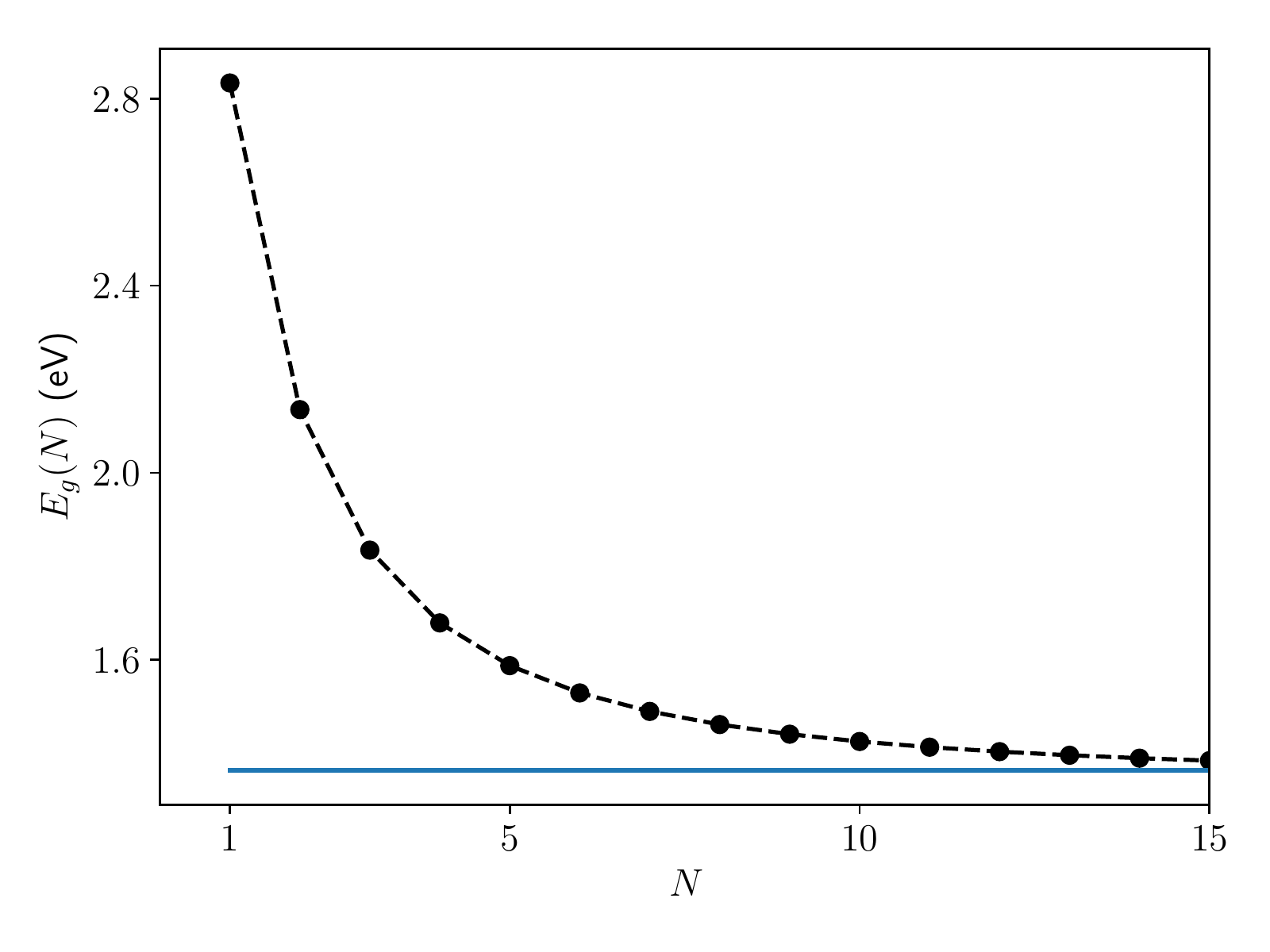}\vspace{0.3cm}
	\includegraphics[width=0.9\linewidth]{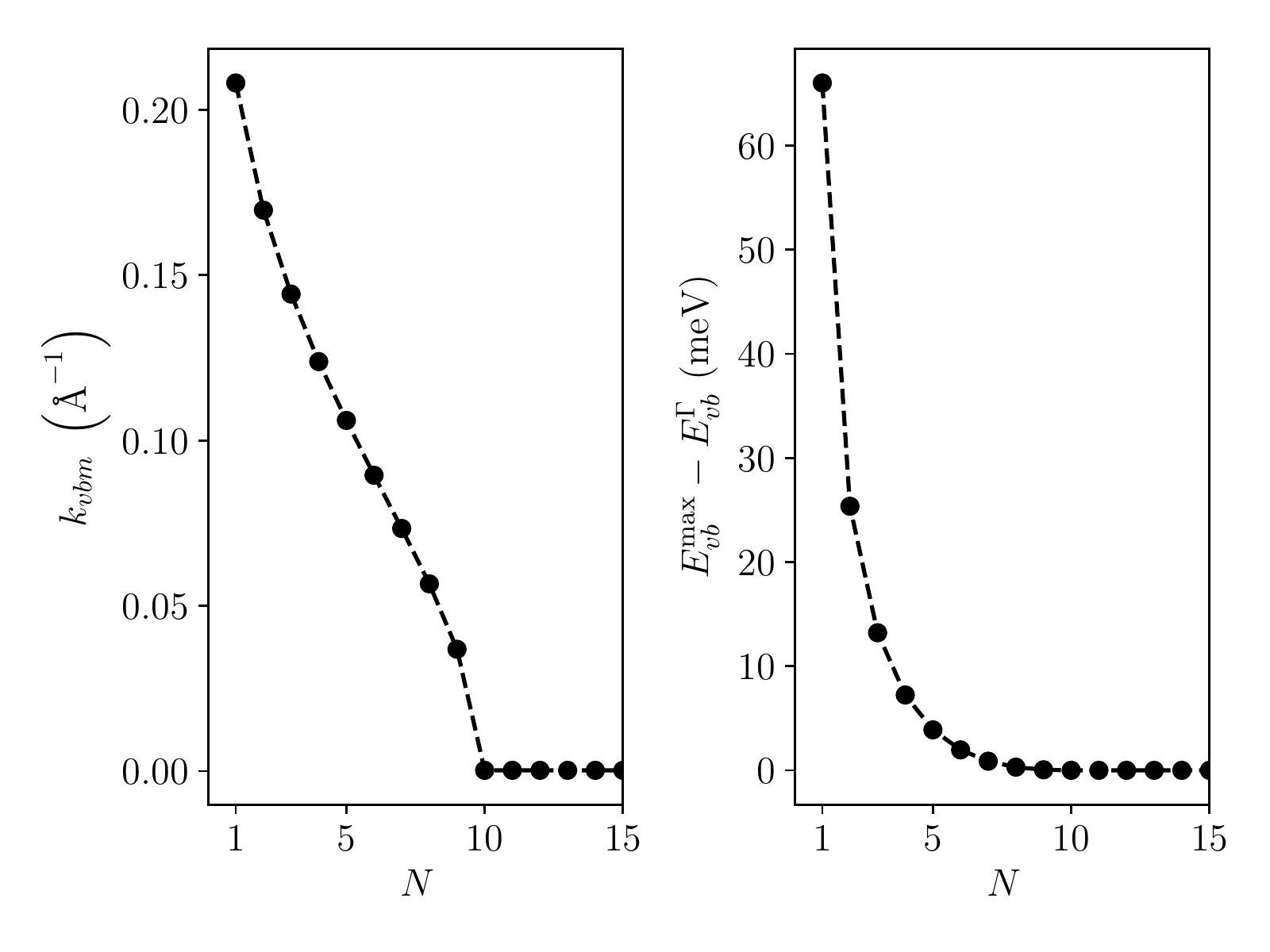}
	\caption{Upper panels: 2D model dispersions for monolayer, bilayer and 5-layer InSe.
		Middle panel: vertical band gaps at $\Gamma$ for $N=1-15$-layer InSe. Solid line is the bulk band gap. Lower panels: position (left) and magnitude (right) of offset of valence band maximum from $\Gamma$-point for $N=1-15$-layer InSe, showing indirect-direct gap transition at 10 layers.}
	\label{fig:fl_general}
\end{figure}

Having found a parameter set for the model, we now explore its behavior in the few-layer case, with an overview of some of the key features of the bands of few-layer InSe shown in Fig. \ref{fig:fl_general}. The dispersive nature of the bulk conduction and valence bands, arising from the strong interlayer hops $t_c, t_v, t_{cv}$ between bands with strong wavefunction contributions from selenium $p_z$ orbitals, translate to large splittings between subbands in the few-layer case. It is this strong interlayer hybridisation which is responsible for the large variation of band gap with crystal thickness \cite{Bandurin:NatNano12-2016,Magorrian:PRB94-2016}, reaching $>2.8$~eV for monolayer films. In contrast, $v_1$ and $v_2$, being dominated by $p_{x,y}$ orbitals which lie mostly in the 2D crystal plane, have weak interband hops and exhibit much weaker splitting. As a consequence when the conduction and valence bands acquire contributions from $v_1$ and $v_2$ (due to, in the model, interband $\mathbf{k\cdot p}$ mixing) away from $\Gamma$ their splitting becomes weaker. In the conduction band this manifests itself as a difference between the effective masses of successive subbands, which in Ref. \onlinecite{Mogorrian:PRB97-2018} was handled by a $k$-dependent $t_c$.

For the valence band the situation is more complex. As has been theoretically predicted\cite{Rybkovskiy:PRB90-2014,Zolyomi:PRB89-2014} and shown in ARPES experiments\cite{Hamer:ACSnano13-2019}, for the thinnest films an offset in the valence band maximum develops, leading to a slightly indirect band gap, in contrast to the direct gap found in thicker films and in the bulk crystal. In the multiband $\mathbf{k\cdot p}$ picture a key contribution to this phenomenon can be understood\cite{Appelbaum} as repulsion away from $\Gamma$ between bands $v$ and $v_2$. When in the few-layer case $v$ splits much more than $v_2$ this repulsion becomes much weaker. Coupled with a weaker splitting of $v$ itself at larger $k$ in a similar manner to that of the conduction band, this causes the depth and radius of the `Mexican hat' offest to decrease rapidly with increasing crystal thickness, ultimately leading to a direct gap in the model for $N\ge 10$ layers.

\section{Numerical implementation of harmonic oscillator basis}\label{app:Hermite}

In the harmonic oscillator basis described in the text, the BSE \eqref{eqn:TB-BSE0} takes the form:
\begin{equation}
\sum_{n'_xn'_y}\!\Big[\mathcal{H}^0_{n_xn_y;n'_xn'_y}\!-\!\mathcal{V}_{n_xn_y;n'_xn'_y}\Big]\mathcal{A}^\mathbf{Q}_{n'_xn'_y}
\!=\!\Omega\mathcal{A}^\mathbf{Q}_{n_xn_y},\label{eqn:H2p}
\end{equation}
with the kinetic energy matrix
\begin{align}
\mathcal{H}^0_{n_xn_y;n'_xn'_y}(\mathbf{Q})&=\int \mathrm{d}^2k\Big[\varepsilon_c(\mathbf{k})-\varepsilon_v(\mathbf{k}\!-\!\mathbf{Q})\Big]\notag\\
&\varphi^\ast_{n_x}(k_x)\varphi^\ast_{n_y}(k_y)\varphi_{n'_x}(k_x)\varphi_{n'_y}(k_y),\label{eqn:bandE}
\end{align}
and the interaction matrix
\begin{align}
&\mathcal{V}_{n_xn_y;n'_xn'_y}=\!\int\frac{\mathrm{d}^2k\mathrm{d}^2q}{(2\pi)^2}V(\mathbf{q})\notag\\
&\varphi^\ast_{n_x}(k_x)\varphi^\ast_{n_y}(k_y)\varphi_{n'_x}(k_x+q_x)\varphi_{n'_y}(k_y+q_y).\label{eqn:CoulE}
\end{align}
In the following, we explain how to choose an optimal harmonic oscillator basis set to speed up the convergence in a calculation. We also give the details for how to construct the matrix equation in Eq. \eqref{eqn:H2p}.

\begin{figure}
\includegraphics[width=3.2in]{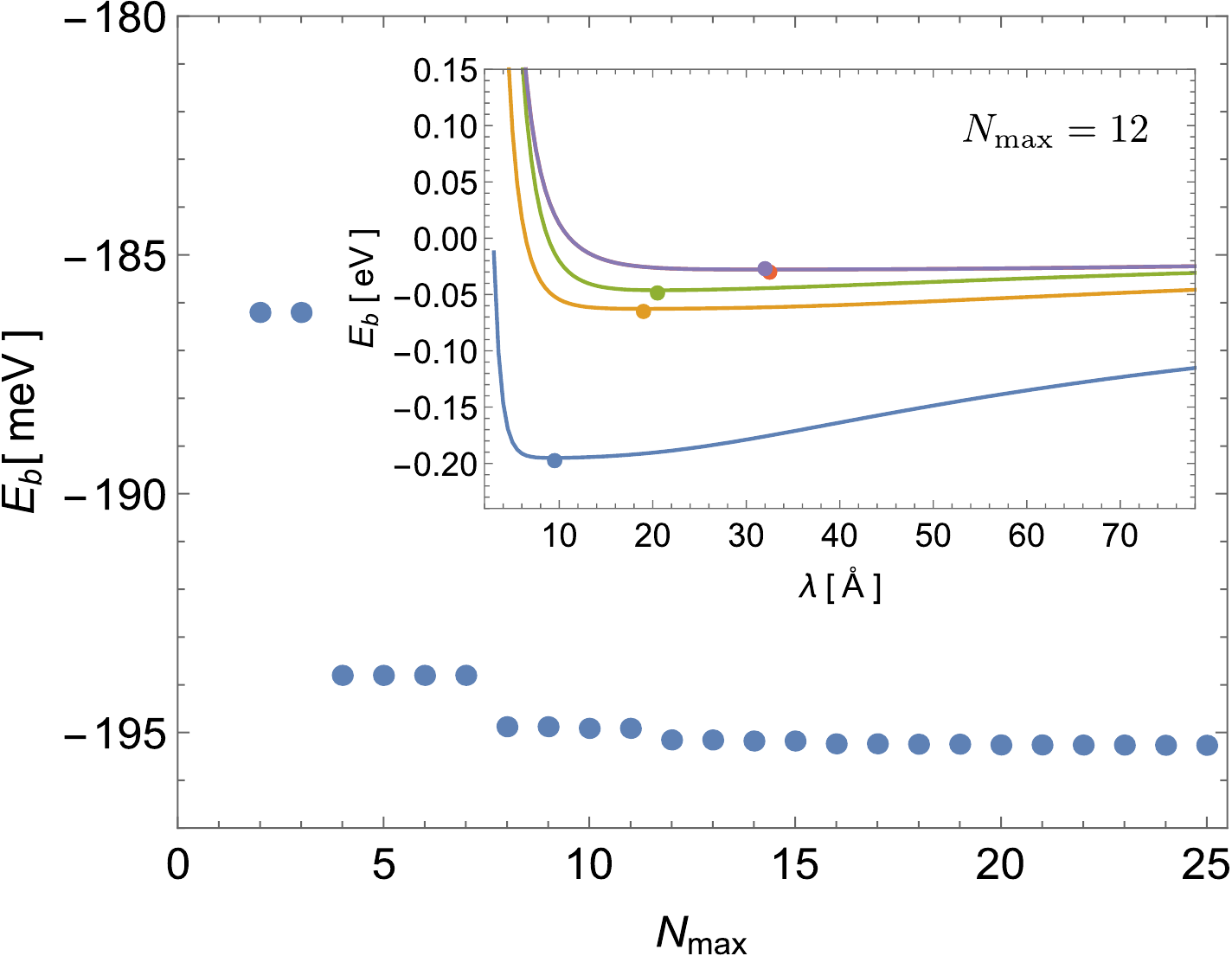}
\caption{The convergence in the calculation of exciton ground state binding energy by using the optimal $\lambda$ for a monolayer InSe with Keldysh potential. The optimal $\lambda$ for each states are determined by maximizing their corresponding binding energy which are marked by the closed circles in the inset. The optimization for the first four $\Gamma$-point exciton binding energy in the inset is performed with fixed $N_\mathrm{max}=12$.}\label{fig:ConvTest}
\end{figure}

\textit{Choice of basis set} -- To diagonalize the BSE in Eq. \eqref{eqn:H2p}, we first need to specify the harmonic oscillator basis set which is determined by the parameters, $\lambda$, the length scale of the oscillator, and, $N_\mathrm{max}$, the cutoff of the oscillator modes with $n_x+n_y\leq N_\mathrm{max}$. In principle, $\lambda$ can be arbitrary since a unique result can be obtained provided that $N_\mathrm{max}$ is large enough. In practice, working with a large basis set is undesirable because large matrix diagonalization is a very demanding computational task. In the following, we show that a good convergent result can be obtained with a relatively small basis set if a proper choice of $\lambda$ is used.

The procedure for obtaining the optimal $\lambda$ is to maximize the exciton binding energy against $\lambda$ (inset of Fig. \ref{fig:ConvTest}). This $\lambda$ corresponds to the optimal coverage of the exciton by the basis set in the momentum/real space. 
In Fig. \ref{fig:ConvTest}, we demonstrate how the binding energy depends on $\lambda$ of a finite basis set with $n_x+n_y\leq N_\text{max}=12$. We note that the optimal $\lambda$ for each different states need to be determined separately since each states have a very different characteristic localized length scale. In Fig. \ref{fig:ConvTest}, one can see that once the optimal $\lambda $ is determined, we obtain a good convergent result for the binding energy at $N_\mathrm{max}\sim12$. Increasing the number of basis beyond $N_\mathrm{max}=12$ only leads to no more than 2 meV correction.

\textit{Kinetic energy matrix} -- For a general band dispersion such as those in tight-binding model, analytical expression may not be available and the query of the band energy may be computationally expansive. Therefore, a straightforward numerical integration in Eq. \eqref{eqn:bandE} is not a practical approach. A feasible numerical method is to expand the band dispersion (periodic function) into a fast convergent Fourier series. Namely,
\begin{equation}
\varepsilon_c(\mathbf{k})-\varepsilon_v(\mathbf{k}-\mathbf{Q})\!=\!\sum_{\mathbf{s}=-\infty}^{\infty}\!\left[C_{\mathbf{s}}-V_{\mathbf{s}}\mathrm{e}^{i2\pi \mathbf{s}\cdot\bar{\mathbf{Q}}}\right]\mathrm{e}^{-i2\pi \mathbf{s}\cdot\bar{\mathbf{k}}},
\end{equation}
where $\mathbf{s}=(s_x,s_y)$, $\bar{\mathbf{k}}=(k_x/T_x,k_y/T_y)$, and $\bar{\mathbf{Q}}=(Q_x/T_x,Q_y/T_y)$ with the $(T_x,T_y)$ are the periodicity of the dispersion in each dimensions. The Fourier coefficients are therefore defined as
\begin{align}\label{eqn:FT-C}
\begin{bmatrix}
C_{\mathbf{s}}\\
V_{\mathbf{s}}
\end{bmatrix}=\int^{\frac{T_x}{2}}_{-\frac{T_x}{2}}\frac{\mathrm{d}k_x}{T_x}\int^{\frac{T_y}{2}}_{-\frac{T_y}{2}}\frac{\mathrm{d}k_y}{T_y}\mathrm{e}^{i2\pi \mathbf{s}\cdot \bar{\mathbf{k}}}
\begin{bmatrix}
\varepsilon_c(\mathbf{k})\\
\varepsilon_v(\mathbf{k})
\end{bmatrix}.
\end{align}
With this expansion, we can integrate out the momentum explicitly. Hence, the band energy matrix in Eq. \eqref{eqn:bandE} become
\begin{align}
&\mathcal{H}^0_{n_xn_y;n_x'n_y'}(\mathbf{Q})
=\sum_{s_x,s_y=-\infty}^{\infty}\left[C_{\mathbf{s}}-V_{\mathbf{s}}\mathrm{e}^{i2\pi \mathbf{s}\cdot\bar{\mathbf{Q}}}\right]\times\notag\\
&\prod_{j}^{x,y}\frac{2^{\zeta_j-\frac{1}{2}(n_j+n'_j)}\zeta_j!(\bar{a}_js_j)^{\Delta_j}}{i^{n_j'-n_j+\Delta_j}\sqrt{ n_j! n'_j! }}\mathrm{e}^{-\tfrac{1}{4}\bar{a}_j^2s_j^2}L_{\zeta_j}^{\Delta_j}(\tfrac{1}{2}\bar{a}_j^2s_j^2),\label{eqn:H0-TB}
\end{align}
with $\zeta_j=\min[n_j,n'_j]$, $\Delta_j=|n_j'-n_j|$, $\bar{a}_j=2\pi/(T_j \lambda)$ and $L^\alpha_n(x)$ is the associated Laguerre polynomial. Since the band dispersion is periodic, only a few of the Fourier modes are relevant to the series. Moreover, we note that the higher order term in the sum are exponentially suppressed. This implies that we have transformed the numerical integration problem into a fast convergent summation.

\begin{figure}
\centering
\includegraphics[width=3.3in]{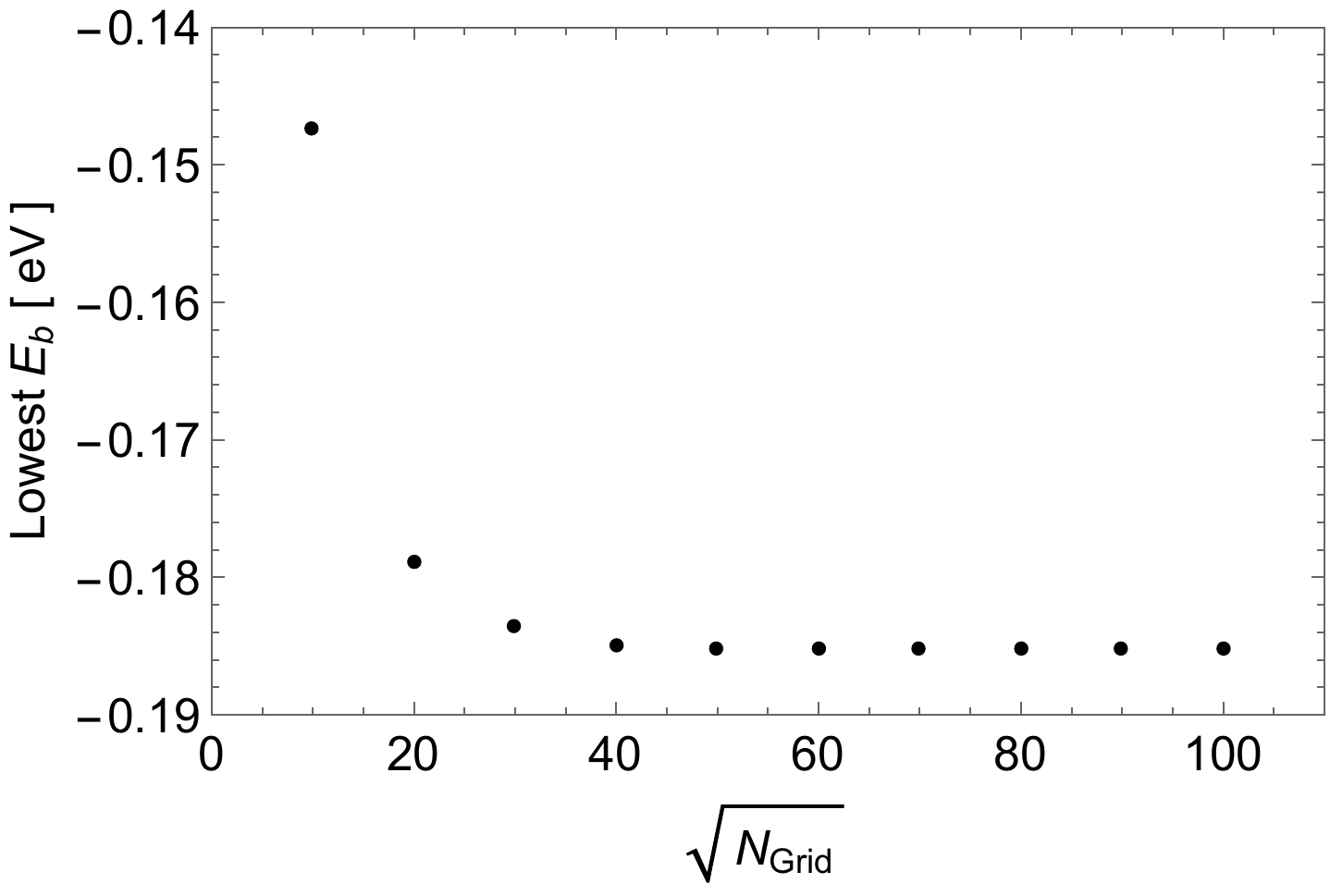}
\caption{The convergence of the $\mathbf{Q}=0$ exciton ground state energy with different grid size for constructing the Fourier series of $\varepsilon_c$ and $\varepsilon_v$.}\label{fig:Grid-test}
\end{figure}

To calculate the Fourier coefficients, we can approximate the integral in Eq. \eqref{eqn:FT-C} as a Riemann sum by discretizing the momentum space into a uniform grid. The calculation of Riemann sum is the same as calculating the discrete Fourier transformation which can be very efficiently evaluated by the fast-Fourier transformation. In this numerical approach, the tight-binding Hamiltonian only needs to be diagonalized once in constructing the uniform grid. Depending on the smoothness of the band structure, typically, the grid size greater than $50\times50$ points is good enough for a desirable convergent result(see Fig. \ref{fig:Grid-test}). In this paper, we use $100\times100$ grid points for the calculation.

Although we have used a straightforward method with Fast-Fourier transformation. The idea of our method is essentially the same as K-points sampling in Ref. \onlinecite{Chadi:PRB8-1973,Cunningham:PRB10-1974,Monkhorst:PRB13-1976}. The K-points sampling method is much more efficient since it utilizes all the symmetry in the function and regrouping the Fourier series into a faster convergent series. The Fourier coefficient in the series can be very efficiently calculated by the Monkhorst-Pack grid in the reduced Brillouin zone. This method was originally discussed in Ref. \onlinecite{Monkhorst:PRB13-1976} as a `hybrid method'.

We can further simplify the calculation in Eq. \eqref{eqn:bandE} if only the low-energy exciton is in our interest. As indicated in Fig. \ref{fig:Ex-weight}, only the low-energy electronic modes (red/blue shaded region) which are well described by the $\mathbf{k}\cdot \mathbf{p}$ model are relevant for exciton binding. In this low-energy regime, one may approximate $\varepsilon_{c/v}$ by expanding it into polynomial. Thus, in this approach, we can use the following identity to calculate Eq. \eqref{eqn:bandE} analytically
\begin{align}\label{eqn:ladder-identity}
\int\mathrm{d}kk^l &\frac{\mathrm{e}^{-k^2}H_m(k)H_n(k)}{\sqrt{\pi m! n! 2^{m+n}}}=\sqrt{\frac{n!}{m!}}\sum_{r=0}^{\lfloor l/2\rfloor}\sum_{s=0}^{\mathrm{min}[m,l-2r]}\binom{m}{s}\notag\\
&\times\frac{2^{s-l-\frac{1}{2}(m-n)}l!}{r!(l-2r-s)!}\delta_{l+m-2r-2s,n},
\end{align}
where $\lfloor l/2\rfloor$ is the largest integer that is equal or smaller than $l/2$.

\textit{Interaction matrix} -- In this paper, we assume in-plane rotational symmetry in the e-h interaction. Hence, the $k$-integration in Eq. \eqref{eqn:CoulE} can be carried out explicitly by using $H_n(x+y)=\sum_{s=0}^{n}\binom{n}{s}H_{s}(x)(2y)^{n-s}$ and this yields
\begin{align}
    &\mathcal{V}_{n_xn_y;n_x'n_y'}
\!=\!\int\!\frac{q\mathrm{d}q}{(2\pi)^2}V(q)\mathrm{e}^{-\frac{1}{4}q^2\lambda^2}\prod^{x,y}_j\sum_{s_j=0}^{\min[m_j,n_j]}(\lambda q)^{\sigma_j}\notag\\
&\binom{n_j}{s_j}\binom{m_j}{s_j}\frac{(-\tfrac{1}{2})^{\frac{1}{2}\sigma_j}s_j!}{\sqrt{m_j!n_j!}}2\text{B}(\tfrac{\sigma_x+1}{2},\tfrac{\sigma_y+1}{2})
\end{align}
where $\sigma_j=n_j+m_j-2s_j$ and $\mathrm{B}(x,y)$ is the beta function. For Keldysh potential, $V(q)=-\frac{2\pi e^2}{\sqrt{\kappa_\parallel\kappa_z}q(1+r_\ast q)}$, we have the following analytical expression for
\begin{align*}
    &\int\frac{q\mathrm{d}q}{(2\pi)^2}V(q)(\lambda q)^{\sigma_x+\sigma_y}\mathrm{e}^{-\frac{1}{4}q^2\lambda^2}=\frac{-e^2}{2\pi\sqrt{\kappa_\parallel\kappa_z}}(-\frac{\lambda}{r_\ast})^{\sigma_x+\sigma_y}\\
    &\Bigg\{\frac{\mathrm{e}^{-\frac{\lambda ^2}{4r^{2}_\ast}}}{2r_\ast/\lambda}\Big[\pi\mathrm{erf}(\tfrac{\lambda }{2r_{*}})\!-\!\mathrm{Ei} (\tfrac{\lambda ^2}{4r^{2}_{*}})\Big]\!-\!\sum^{\sigma_x\!+\!\sigma_y\!-\!1}_{j=0}\!\Gamma(\tfrac{j+1}{2})(-\tfrac{2r_{\ast}}{\lambda})^{j}\Bigg\}
\end{align*}
where $\Gamma(x)$ is the gamma function, $\mathrm{Ei}(x)$ is the expontential-integral, and $\mathrm{erf}(x)$ is error function.

\begin{table}
\begin{ruledtabular}
\begin{tabular}{l|cc|cc|cc|cc}
 & \small{HO basis} & & \small{2D Hydrogen} & & \small{HO basis} & & \small{suspended MoS$_{2}$}\cite{Zhang:JPhysCondMat31-2019} & \\
\hline
$0s$ & 92.5 & & 94.2 & &  554 &  & 555 &\\
$1p_{x,y}$ & 10.38 & & 10.47 & & 315 & & 316 &\\
$1s$  & 9.6 & & 10.47 & & 257 & & 258 &\\
$2d_{xy}$  & 3.76 & & 3.77 & & 209 & & 209 &\\
$2p_{x,y}$  & 3.69 & & 3.77 & & 184 & & 185 &\\
\hline
\end{tabular}
\end{ruledtabular}
\caption{Comparison of binding energies in meV as obtained from the harmonic oscillator basis against analytical and calculated results\cite{Zhang:JPhysCondMat31-2019} for suspended MoS$_{2}$. Basis size used in the comparison with MoS$_{2}$ monolayer corresponded to $N_{max}=12$ (basis size=91 states) and $\lambda$ was optimised. For the 2D hydrogen atom with a reduced effective mass of $\mu=0.14$ and $\epsilon=9$ the basis size used for the comparison was $N_{max}=20$ for every states in the table except the $1s$. As for the $1s$ state a greater basis size of $N_{max}=24$ as used.}\label{tbl:HO-test}
\end{table}

\textit{Comparison with hydrogen-like exciton levels for $V \propto -1/r$ and Keldysh interaction} -- As shown in Table \ref{tbl:HO-test}, in comparing the binding energy as obtained from the harmonic oscillator basis with the analytically obtained 2D hydrogen atom energy levels, the discrepancy between the two was found smaller than 2$\%$ as for the ground state energy and even lower for the states with $l\neq0$. The higher excited states with $l=0$ required a very large basis size in order to accurately calculate the binding energy due to the very sharp singularity of the wavefunction appearing at $r=0$ (Kato cusp). This situation is similar to the well-known problem in the Slater-type versus Gaussian-type orbitals in quantum chemistry\cite{Magalhaes:JChemEdu91-2014}, since the harmonic oscillator (Hermite function) is essentially a Gaussian basis. Such a sharp feature in the excitonic wavefunction is mitigated in the Keldysh potential as the $1/r$-divergence becomes logarithmic. In this case, the harmonic oscillator basis yields better accuracy for each binding state in the spectrum. In comparing our binding energy calculation with the calculated bindings for MoS$_{2}$, the error was significantly reduced for the same basis size with $<$0.3$\%$ as for the ground state and lower for the $l\neq0$ states.

\textit{Connection to the real-space formalism}
It is also instructive to describe the excitonic problem in term of real-space. To do this, we can Fourier transform the Bethe-Salpeter equation in \eqref{eqn:TB-BSE0} by using
\begin{equation}
\psi(\mathbf{r}_e,\mathbf{r}_h)=\sum_{\mathbf{k}_e,\mathbf{k}_h}\psi_\mathbf{Q}(\mathbf{k})\mathrm{e}^{i( \mathbf{k}_e\cdot\mathbf{r}_e-\mathbf{k}_h\cdot\mathbf{r}_h)}.
\end{equation}
We remind that $\mathbf{k}_e=\mathbf{k}$ and $\mathbf{k}_h=\mathbf{k}-\mathbf{Q}$. This transformation turn all momentum in the dispersion in Eq. \eqref{eqn:TB-BSE0} into derivative operators 
and yields
\begin{equation}\label{eqn:GMW-eqn}
    [\varepsilon_{c}(-i\nabla_{\mathbf{r}_{e}})-\varepsilon_{v}(i\nabla_{\mathbf{r}_{h}})-\Omega+V(\mathbf{r}_e-\mathbf{r}_h)]\psi(\mathbf{r}_e,\mathbf{r}_h)=0
\end{equation}
where $V(\mathbf{r})=\int_{\mathbf{q}}\mathrm{e}^{i\mathbf{r}\cdot\mathbf{q}}V(\mathbf{q})$ is the Fourier transformation of the potential. The above equation yields Mott-Wannier model if only the quadratic mass term in $\varepsilon_{c/v}$ is kept. However, in our model, we need to retain higher-order terms in the hole dispersion. 

Similar to Mott-Wannier model, Eq. \eqref{eqn:GMW-eqn} can be reduced to a one-body problem by using the canonical transformation. Since the hole effective mass is not well-defined due to the sombrero-shaped dispersion, instead of using the center-of-mass frame coordinate system, we choose
\begin{align*}
\begin{bmatrix}
\bm{X}\\
\bm{x}
\end{bmatrix}=\frac{1}{\sqrt{2}}
\begin{bmatrix}
\mathbf{r}_{e}+\mathbf{r}_{h}\\
\mathbf{r}_{e}-\mathbf{r}_{h}
\end{bmatrix}, \quad
\begin{bmatrix}
\hat{\mathbf{P}}\\
\hat{\mathbf{p}}
\end{bmatrix}=-\frac{i}{\sqrt{2}}
\begin{bmatrix}
\nabla_{\mathbf{r}_{e}}+\nabla_{\mathbf{r}_{h}}\\
\nabla_{\mathbf{r}_{e}}-\nabla_{\mathbf{r}_{h}}
\end{bmatrix},
\end{align*}
The crucial requirement for this transformation is that the new coordinate system satisfies $[x_j,\hat{p}_j]=[X_j,\hat{P}_j]=i$ such that the physical phase space volume is preserved. Using the $(\bm{X},\bm{x})$ coodinate, Eq. \eqref{eqn:GMW-eqn} in the real-space reads
\begin{equation}
  [\varepsilon_c(\tfrac{\hat{\mathbf{p}}+\hat{\mathbf{P}}}{\sqrt{2}})-\varepsilon_v(\tfrac{\hat{\mathbf{p}}-\hat{\mathbf{P}}}{\sqrt{2}})-\Omega-V(\sqrt{2}\bm{x})]\psi(\bm{X},\bm{x})=0.\label{eqn:kp-BSE}  
\end{equation}
We note that, in this coordinate system, the correspondence between momentum and real space representation of the exciton momentum is $\mathbf{Q}=\mathbf{k}_e-\mathbf{k}_h\leftrightarrow -i\nabla_{\mathbf{r}_e}-i\nabla_{\mathbf{r}_h}=\sqrt{2} \hat{\mathbf{P}}$.

Firstly, it is noted that $[\hat{\mathbf{P}},H]=0$ where $H$ (independent of $\bm{X}$) is the electron-hole two-particle Hamiltonian in \eqref{eqn:GMW-eqn} indicating that $\mathbf{P}$ is a well-defined quantum number which give the exciton momentum $\mathbf{Q}=\sqrt{2}\mathbf{P}$. Therefore, the wavefunction is uniquely dependent on $\bm{x}$
\begin{align}
\psi(\bm{X},\bm{x})=e^{i\mathbf{P}\cdot\bm{X}}\chi(\bm{x}),
\end{align}
which is the eigenfunction of $H$. Substituting the above ansatz wavefunction into Eq. \eqref{eqn:GMW-eqn}, we reduce the equation into a one-body Shr\"odinger equation as
\begin{equation}
[\varepsilon_c(\tfrac{\sqrt{2}\hat{\mathbf{p}}+\mathbf{Q}}{2})-\varepsilon_v(\tfrac{\sqrt{2}\hat{\mathbf{p}}-\mathbf{Q}}{2})-\Omega+V(\sqrt{2}\bm{x})]
\chi(\bm{x})=0.\label{eqn:kp-BSETrans}
\end{equation}

Expanding $\chi(\bm{x})$ into harmonic oscilltor basis as
$\chi(\bm{x})=\sum_{\mathbf{n}}\mathcal{C}^\mathbf{Q}_{n_xn_y}\varphi_{n_x}(\rho_x)\varphi_{n_y}(\rho_y)$
where $\bm{\rho}=\mathbf{r}_e-\mathbf{r}_h=\sqrt{2}\bm{x}$ is the relative coordinate of electron and hole. The real-space basis function, $\varphi_{n}(\rho)$, is the Fourier transformation of $\varphi_{n}(k)$
which is also a harmonic oscillator.
Therefore, the matrix representation for Eq. \eqref{eqn:kp-BSETrans} is
\begin{equation}\label{eqn:H2p-real}
\sum_{n_x'n_y'}\!\Big[\mathcal{H}^0_{n_xn_y;n_x'n_y'}\!+\!\mathcal{V}_{n_xn_y;n_x'n_y'}\Big]\mathcal{C}^\mathbf{Q}_{n_x'n_y'}\!=\!\Omega\mathcal{C}^\mathbf{Q}_{n_xn_y}
\end{equation}
with the kinetic Hamiltonian
\begin{align}
&\mathcal{H}^0_{n_xn_y;n_x'n_y'}(\mathbf{Q})\!=\int \mathrm{d}^2\rho\varphi_{n_x}(\rho_x)\varphi_{n_y}(\rho_y)\notag\\
&\times\Big[\varepsilon_c(\tfrac{\sqrt{2}\hat{\mathbf{p}}+\mathbf{Q}}{2})-\varepsilon_v(\tfrac{\sqrt{2}\hat{\mathbf{p}}-\mathbf{Q}}{2})\Big]\varphi_{n_x'}(\rho_x)\varphi_{n_y'}(\rho_y)\label{eqn:H0-x}
\end{align}
and the Coulomb interaction matrix
\begin{align*}
\mathcal{V}_{n_xn_y;n_x'n_y'}=&\!\int\mathrm{d}^2\rho V(\bm{\rho})\varphi_{n_x}(\rho_x)\varphi_{n_y}(\rho_y)\varphi_{n_x'}(\rho_x)\varphi_{n_y'}(\rho_y)
\end{align*}

The integration in Eq. \eqref{eqn:H0-x} can be carried out exactly by using chain rule to rewrite $\hat{\mathbf{p}}=-i\nabla_{\bm{x}}=\sqrt{2}(-i\nabla_{\bm{\rho}})$ and then using the recursive relation of the Hermite functions. Alternatively, one may also calculate it by turning $-i\nabla_{\bm{\rho}}$ into the simple harmonic ladder operators and carrying out the commutation algebra. Nevertheless, the calculated result from both methods is identical to the momentum space calculation in Eq. \eqref{eqn:ladder-identity}.

\bibliography{InSeEx}

\begin{thebibliography}{76}%
\makeatletter
\providecommand \@ifxundefined [1]{%
 \@ifx{#1\undefined}
}%
\providecommand \@ifnum [1]{%
 \ifnum #1\expandafter \@firstoftwo
 \else \expandafter \@secondoftwo
 \fi
}%
\providecommand \@ifx [1]{%
 \ifx #1\expandafter \@firstoftwo
 \else \expandafter \@secondoftwo
 \fi
}%
\providecommand \natexlab [1]{#1}%
\providecommand \enquote  [1]{``#1''}%
\providecommand \bibnamefont  [1]{#1}%
\providecommand \bibfnamefont [1]{#1}%
\providecommand \citenamefont [1]{#1}%
\providecommand \href@noop [0]{\@secondoftwo}%
\providecommand \href [0]{\begingroup \@sanitize@url \@href}%
\providecommand \@href[1]{\@@startlink{#1}\@@href}%
\providecommand \@@href[1]{\endgroup#1\@@endlink}%
\providecommand \@sanitize@url [0]{\catcode `\\12\catcode `\$12\catcode
  `\&12\catcode `\#12\catcode `\^12\catcode `\_12\catcode `\%12\relax}%
\providecommand \@@startlink[1]{}%
\providecommand \@@endlink[0]{}%
\providecommand \url  [0]{\begingroup\@sanitize@url \@url }%
\providecommand \@url [1]{\endgroup\@href {#1}{\urlprefix }}%
\providecommand \urlprefix  [0]{URL }%
\providecommand \Eprint [0]{\href }%
\providecommand \doibase [0]{http://dx.doi.org/}%
\providecommand \selectlanguage [0]{\@gobble}%
\providecommand \bibinfo  [0]{\@secondoftwo}%
\providecommand \bibfield  [0]{\@secondoftwo}%
\providecommand \translation [1]{[#1]}%
\providecommand \BibitemOpen [0]{}%
\providecommand \bibitemStop [0]{}%
\providecommand \bibitemNoStop [0]{.\EOS\space}%
\providecommand \EOS [0]{\spacefactor3000\relax}%
\providecommand \BibitemShut  [1]{\csname bibitem#1\endcsname}%
\let\auto@bib@innerbib\@empty
%</preamble>
\bibitem [{\citenamefont {Bhimanapati}\ \emph {et~al.}(2015)\citenamefont
  {Bhimanapati}, \citenamefont {Lin}, \citenamefont {Meunier}, \citenamefont
  {Jung}, \citenamefont {Cha}, \citenamefont {Das}, \citenamefont {Xiao},
  \citenamefont {Son}, \citenamefont {Strano}, \citenamefont {Cooper},
  \citenamefont {Liang}, \citenamefont {Louie}, \citenamefont {Ringe},
  \citenamefont {Zhou}, \citenamefont {Kim}, \citenamefont {Naik},
  \citenamefont {Sumpter}, \citenamefont {Terrones}, \citenamefont {Xia},
  \citenamefont {Wang}, \citenamefont {Zhu}, \citenamefont {Akinwande},
  \citenamefont {Alem}, \citenamefont {Schuller}, \citenamefont {Schaak},
  \citenamefont {Terrones},\ and\ \citenamefont
  {Robinson}}]{Bhimanapati:ACSnano9-2015}%
  \BibitemOpen
  \bibfield  {author} {\bibinfo {author} {\bibfnamefont {G.~R.}\ \bibnamefont
  {Bhimanapati}}, \bibinfo {author} {\bibfnamefont {Z.}~\bibnamefont {Lin}},
  \bibinfo {author} {\bibfnamefont {V.}~\bibnamefont {Meunier}}, \bibinfo
  {author} {\bibfnamefont {Y.}~\bibnamefont {Jung}}, \bibinfo {author}
  {\bibfnamefont {J.}~\bibnamefont {Cha}}, \bibinfo {author} {\bibfnamefont
  {S.}~\bibnamefont {Das}}, \bibinfo {author} {\bibfnamefont {D.}~\bibnamefont
  {Xiao}}, \bibinfo {author} {\bibfnamefont {Y.}~\bibnamefont {Son}}, \bibinfo
  {author} {\bibfnamefont {M.~S.}\ \bibnamefont {Strano}}, \bibinfo {author}
  {\bibfnamefont {V.~R.}\ \bibnamefont {Cooper}}, \bibinfo {author}
  {\bibfnamefont {L.}~\bibnamefont {Liang}}, \bibinfo {author} {\bibfnamefont
  {S.~G.}\ \bibnamefont {Louie}}, \bibinfo {author} {\bibfnamefont
  {E.}~\bibnamefont {Ringe}}, \bibinfo {author} {\bibfnamefont
  {W.}~\bibnamefont {Zhou}}, \bibinfo {author} {\bibfnamefont {S.~S.}\
  \bibnamefont {Kim}}, \bibinfo {author} {\bibfnamefont {R.~R.}\ \bibnamefont
  {Naik}}, \bibinfo {author} {\bibfnamefont {B.~G.}\ \bibnamefont {Sumpter}},
  \bibinfo {author} {\bibfnamefont {H.}~\bibnamefont {Terrones}}, \bibinfo
  {author} {\bibfnamefont {F.}~\bibnamefont {Xia}}, \bibinfo {author}
  {\bibfnamefont {Y.}~\bibnamefont {Wang}}, \bibinfo {author} {\bibfnamefont
  {J.}~\bibnamefont {Zhu}}, \bibinfo {author} {\bibfnamefont {D.}~\bibnamefont
  {Akinwande}}, \bibinfo {author} {\bibfnamefont {N.}~\bibnamefont {Alem}},
  \bibinfo {author} {\bibfnamefont {J.~A.}\ \bibnamefont {Schuller}}, \bibinfo
  {author} {\bibfnamefont {R.~E.}\ \bibnamefont {Schaak}}, \bibinfo {author}
  {\bibfnamefont {M.}~\bibnamefont {Terrones}}, \ and\ \bibinfo {author}
  {\bibfnamefont {J.~A.}\ \bibnamefont {Robinson}},\ }\href {\doibase
  10.1021/acsnano.5b05556} {\bibfield  {journal} {\bibinfo  {journal} {{ACS}
  Nano}\ }\textbf {\bibinfo {volume} {9}},\ \bibinfo {pages} {11509} (\bibinfo
  {year} {2015})}\BibitemShut {NoStop}%
\bibitem [{\citenamefont {Qian}\ \emph {et~al.}(2015)\citenamefont {Qian},
  \citenamefont {Wang}, \citenamefont {Li}, \citenamefont {Lu},\ and\
  \citenamefont {Li}}]{Qian:2DMat2-2015}%
  \BibitemOpen
  \bibfield  {author} {\bibinfo {author} {\bibfnamefont {X.}~\bibnamefont
  {Qian}}, \bibinfo {author} {\bibfnamefont {Y.}~\bibnamefont {Wang}}, \bibinfo
  {author} {\bibfnamefont {W.}~\bibnamefont {Li}}, \bibinfo {author}
  {\bibfnamefont {J.}~\bibnamefont {Lu}}, \ and\ \bibinfo {author}
  {\bibfnamefont {J.}~\bibnamefont {Li}},\ }\href {\doibase
  10.1088/2053-1583/2/3/032003} {\bibfield  {journal} {\bibinfo  {journal} {2D
  Materials}\ }\textbf {\bibinfo {volume} {2}},\ \bibinfo {pages} {032003}
  (\bibinfo {year} {2015})}\BibitemShut {NoStop}%
\bibitem [{\citenamefont {Novoselov}\ \emph {et~al.}(2016)\citenamefont
  {Novoselov}, \citenamefont {Mishchenko}, \citenamefont {Carvalho},\ and\
  \citenamefont {Neto}}]{Novoselov:Science353-2016}%
  \BibitemOpen
  \bibfield  {author} {\bibinfo {author} {\bibfnamefont {K.~S.}\ \bibnamefont
  {Novoselov}}, \bibinfo {author} {\bibfnamefont {A.}~\bibnamefont
  {Mishchenko}}, \bibinfo {author} {\bibfnamefont {A.}~\bibnamefont
  {Carvalho}}, \ and\ \bibinfo {author} {\bibfnamefont {A.~H.~C.}\ \bibnamefont
  {Neto}},\ }\href {\doibase 10.1126/science.aac9439} {\bibfield  {journal}
  {\bibinfo  {journal} {Science}\ }\textbf {\bibinfo {volume} {353}},\ \bibinfo
  {pages} {aac9439} (\bibinfo {year} {2016})}\BibitemShut {NoStop}%
\bibitem [{\citenamefont {Mudd}\ \emph {et~al.}(2013)\citenamefont {Mudd},
  \citenamefont {Svatek}, \citenamefont {Ren}, \citenamefont {Patan{\`e}},
  \citenamefont {Makarovsky}, \citenamefont {Eaves}, \citenamefont {Beton},
  \citenamefont {Kovalyuk}, \citenamefont {Lashkarev}, \citenamefont
  {Kudrynskyi} \emph {et~al.}}]{Mudd:AdvMat25-2013}%
  \BibitemOpen
  \bibfield  {author} {\bibinfo {author} {\bibfnamefont {G.~W.}\ \bibnamefont
  {Mudd}}, \bibinfo {author} {\bibfnamefont {S.~A.}\ \bibnamefont {Svatek}},
  \bibinfo {author} {\bibfnamefont {T.}~\bibnamefont {Ren}}, \bibinfo {author}
  {\bibfnamefont {A.}~\bibnamefont {Patan{\`e}}}, \bibinfo {author}
  {\bibfnamefont {O.}~\bibnamefont {Makarovsky}}, \bibinfo {author}
  {\bibfnamefont {L.}~\bibnamefont {Eaves}}, \bibinfo {author} {\bibfnamefont
  {P.~H.}\ \bibnamefont {Beton}}, \bibinfo {author} {\bibfnamefont {Z.~D.}\
  \bibnamefont {Kovalyuk}}, \bibinfo {author} {\bibfnamefont {G.~V.}\
  \bibnamefont {Lashkarev}}, \bibinfo {author} {\bibfnamefont {Z.~R.}\
  \bibnamefont {Kudrynskyi}},  \emph {et~al.},\ }\href@noop {} {\bibfield
  {journal} {\bibinfo  {journal} {Advanced Materials}\ }\textbf {\bibinfo
  {volume} {25}},\ \bibinfo {pages} {5714} (\bibinfo {year}
  {2013})}\BibitemShut {NoStop}%
\bibitem [{\citenamefont {Lei}\ \emph {et~al.}(2014)\citenamefont {Lei},
  \citenamefont {Ge}, \citenamefont {Najmaei}, \citenamefont {George},
  \citenamefont {Kappera}, \citenamefont {Lou}, \citenamefont {Chhowalla},
  \citenamefont {Yamaguchi}, \citenamefont {Gupta}, \citenamefont {Vajtai},
  \citenamefont {Mohite},\ and\ \citenamefont {Ajayan}}]{Lei:ACSNano8-2014}%
  \BibitemOpen
  \bibfield  {author} {\bibinfo {author} {\bibfnamefont {S.}~\bibnamefont
  {Lei}}, \bibinfo {author} {\bibfnamefont {L.}~\bibnamefont {Ge}}, \bibinfo
  {author} {\bibfnamefont {S.}~\bibnamefont {Najmaei}}, \bibinfo {author}
  {\bibfnamefont {A.}~\bibnamefont {George}}, \bibinfo {author} {\bibfnamefont
  {R.}~\bibnamefont {Kappera}}, \bibinfo {author} {\bibfnamefont
  {J.}~\bibnamefont {Lou}}, \bibinfo {author} {\bibfnamefont {M.}~\bibnamefont
  {Chhowalla}}, \bibinfo {author} {\bibfnamefont {H.}~\bibnamefont
  {Yamaguchi}}, \bibinfo {author} {\bibfnamefont {G.}~\bibnamefont {Gupta}},
  \bibinfo {author} {\bibfnamefont {R.}~\bibnamefont {Vajtai}}, \bibinfo
  {author} {\bibfnamefont {A.~D.}\ \bibnamefont {Mohite}}, \ and\ \bibinfo
  {author} {\bibfnamefont {P.~M.}\ \bibnamefont {Ajayan}},\ }\href {\doibase
  10.1021/nn405036u} {\bibfield  {journal} {\bibinfo  {journal} {{ACS} Nano}\
  }\textbf {\bibinfo {volume} {8}},\ \bibinfo {pages} {1263} (\bibinfo {year}
  {2014})}\BibitemShut {NoStop}%
\bibitem [{\citenamefont {Brotons-Gisbert}\ \emph {et~al.}(2016)\citenamefont
  {Brotons-Gisbert}, \citenamefont {Andres-Penares}, \citenamefont {Suh},
  \citenamefont {Hidalgo}, \citenamefont {Abargues}, \citenamefont
  {Rodr{\'{\i}}guez-Cant{\'{o}}}, \citenamefont {Segura}, \citenamefont {Cros},
  \citenamefont {Tobias}, \citenamefont {Canadell}, \citenamefont
  {Ordej{\'{o}}n}, \citenamefont {Wu}, \citenamefont {Mart{\'{\i}}nez-Pastor},\
  and\ \citenamefont {S{\'{a}}nchez-Royo}}]{Brotons-Gisbert:NanoLett16-2016}%
  \BibitemOpen
  \bibfield  {author} {\bibinfo {author} {\bibfnamefont {M.}~\bibnamefont
  {Brotons-Gisbert}}, \bibinfo {author} {\bibfnamefont {D.}~\bibnamefont
  {Andres-Penares}}, \bibinfo {author} {\bibfnamefont {J.}~\bibnamefont {Suh}},
  \bibinfo {author} {\bibfnamefont {F.}~\bibnamefont {Hidalgo}}, \bibinfo
  {author} {\bibfnamefont {R.}~\bibnamefont {Abargues}}, \bibinfo {author}
  {\bibfnamefont {P.~J.}\ \bibnamefont {Rodr{\'{\i}}guez-Cant{\'{o}}}},
  \bibinfo {author} {\bibfnamefont {A.}~\bibnamefont {Segura}}, \bibinfo
  {author} {\bibfnamefont {A.}~\bibnamefont {Cros}}, \bibinfo {author}
  {\bibfnamefont {G.}~\bibnamefont {Tobias}}, \bibinfo {author} {\bibfnamefont
  {E.}~\bibnamefont {Canadell}}, \bibinfo {author} {\bibfnamefont
  {P.}~\bibnamefont {Ordej{\'{o}}n}}, \bibinfo {author} {\bibfnamefont
  {J.}~\bibnamefont {Wu}}, \bibinfo {author} {\bibfnamefont {J.~P.}\
  \bibnamefont {Mart{\'{\i}}nez-Pastor}}, \ and\ \bibinfo {author}
  {\bibfnamefont {J.~F.}\ \bibnamefont {S{\'{a}}nchez-Royo}},\ }\href {\doibase
  10.1021/acs.nanolett.6b00689} {\bibfield  {journal} {\bibinfo  {journal}
  {Nano Letters}\ }\textbf {\bibinfo {volume} {16}},\ \bibinfo {pages} {3221}
  (\bibinfo {year} {2016})}\BibitemShut {NoStop}%
\bibitem [{\citenamefont {Bandurin}\ \emph {et~al.}(2016)\citenamefont
  {Bandurin}, \citenamefont {Tyurnina}, \citenamefont {Yu}, \citenamefont
  {Mishchenko}, \citenamefont {Z{\'{o}}lyomi}, \citenamefont {Morozov},
  \citenamefont {Kumar}, \citenamefont {Gorbachev}, \citenamefont {Kudrynskyi},
  \citenamefont {Pezzini}, \citenamefont {Kovalyuk}, \citenamefont {Zeitler},
  \citenamefont {Novoselov}, \citenamefont {Patan{\`{e}}}, \citenamefont
  {Eaves}, \citenamefont {Grigorieva}, \citenamefont {Fal{\textquotesingle}ko},
  \citenamefont {Geim},\ and\ \citenamefont {Cao}}]{Bandurin:NatNano12-2016}%
  \BibitemOpen
  \bibfield  {author} {\bibinfo {author} {\bibfnamefont {D.~A.}\ \bibnamefont
  {Bandurin}}, \bibinfo {author} {\bibfnamefont {A.~V.}\ \bibnamefont
  {Tyurnina}}, \bibinfo {author} {\bibfnamefont {G.~L.}\ \bibnamefont {Yu}},
  \bibinfo {author} {\bibfnamefont {A.}~\bibnamefont {Mishchenko}}, \bibinfo
  {author} {\bibfnamefont {V.}~\bibnamefont {Z{\'{o}}lyomi}}, \bibinfo {author}
  {\bibfnamefont {S.~V.}\ \bibnamefont {Morozov}}, \bibinfo {author}
  {\bibfnamefont {R.~K.}\ \bibnamefont {Kumar}}, \bibinfo {author}
  {\bibfnamefont {R.~V.}\ \bibnamefont {Gorbachev}}, \bibinfo {author}
  {\bibfnamefont {Z.~R.}\ \bibnamefont {Kudrynskyi}}, \bibinfo {author}
  {\bibfnamefont {S.}~\bibnamefont {Pezzini}}, \bibinfo {author} {\bibfnamefont
  {Z.~D.}\ \bibnamefont {Kovalyuk}}, \bibinfo {author} {\bibfnamefont
  {U.}~\bibnamefont {Zeitler}}, \bibinfo {author} {\bibfnamefont {K.~S.}\
  \bibnamefont {Novoselov}}, \bibinfo {author} {\bibfnamefont {A.}~\bibnamefont
  {Patan{\`{e}}}}, \bibinfo {author} {\bibfnamefont {L.}~\bibnamefont {Eaves}},
  \bibinfo {author} {\bibfnamefont {I.~V.}\ \bibnamefont {Grigorieva}},
  \bibinfo {author} {\bibfnamefont {V.~I.}\ \bibnamefont
  {Fal{\textquotesingle}ko}}, \bibinfo {author} {\bibfnamefont {A.~K.}\
  \bibnamefont {Geim}}, \ and\ \bibinfo {author} {\bibfnamefont
  {Y.}~\bibnamefont {Cao}},\ }\href {\doibase 10.1038/nnano.2016.242}
  {\bibfield  {journal} {\bibinfo  {journal} {Nature Nanotechnology}\ }\textbf
  {\bibinfo {volume} {12}},\ \bibinfo {pages} {223} (\bibinfo {year}
  {2016})}\BibitemShut {NoStop}%
\bibitem [{\citenamefont {Terry}\ \emph {et~al.}(2018)\citenamefont {Terry},
  \citenamefont {Z{\'{o}}lyomi}, \citenamefont {Hamer}, \citenamefont
  {Tyurnina}, \citenamefont {Hopkinson}, \citenamefont {Rakowski},
  \citenamefont {Magorrian}, \citenamefont {Clark}, \citenamefont {Andreev},
  \citenamefont {Kazakova}, \citenamefont {Novoselov}, \citenamefont {Haigh},
  \citenamefont {Fal'ko},\ and\ \citenamefont {Gorbachev}}]{Terry:2DMat5-2018}%
  \BibitemOpen
  \bibfield  {author} {\bibinfo {author} {\bibfnamefont {D.~J.}\ \bibnamefont
  {Terry}}, \bibinfo {author} {\bibfnamefont {V.}~\bibnamefont
  {Z{\'{o}}lyomi}}, \bibinfo {author} {\bibfnamefont {M.}~\bibnamefont
  {Hamer}}, \bibinfo {author} {\bibfnamefont {A.~V.}\ \bibnamefont {Tyurnina}},
  \bibinfo {author} {\bibfnamefont {D.~G.}\ \bibnamefont {Hopkinson}}, \bibinfo
  {author} {\bibfnamefont {A.~M.}\ \bibnamefont {Rakowski}}, \bibinfo {author}
  {\bibfnamefont {S.~J.}\ \bibnamefont {Magorrian}}, \bibinfo {author}
  {\bibfnamefont {N.}~\bibnamefont {Clark}}, \bibinfo {author} {\bibfnamefont
  {Y.~M.}\ \bibnamefont {Andreev}}, \bibinfo {author} {\bibfnamefont
  {O.}~\bibnamefont {Kazakova}}, \bibinfo {author} {\bibfnamefont
  {K.}~\bibnamefont {Novoselov}}, \bibinfo {author} {\bibfnamefont {S.~J.}\
  \bibnamefont {Haigh}}, \bibinfo {author} {\bibfnamefont {V.~I.}\ \bibnamefont
  {Fal'ko}}, \ and\ \bibinfo {author} {\bibfnamefont {R.}~\bibnamefont
  {Gorbachev}},\ }\href {\doibase 10.1088/2053-1583/aadfc3} {\bibfield
  {journal} {\bibinfo  {journal} {2D Materials}\ }\textbf {\bibinfo {volume}
  {5}},\ \bibinfo {pages} {041009} (\bibinfo {year} {2018})}\BibitemShut
  {NoStop}%
\bibitem [{\citenamefont {S{\'{a}}nchez-Royo}\ \emph
  {et~al.}(2014)\citenamefont {S{\'{a}}nchez-Royo}, \citenamefont
  {Mu{\~{n}}oz-Matutano}, \citenamefont {Brotons-Gisbert}, \citenamefont
  {Mart{\'{\i}}nez-Pastor}, \citenamefont {Segura}, \citenamefont {Cantarero},
  \citenamefont {Mata}, \citenamefont {Canet-Ferrer}, \citenamefont {Tobias},
  \citenamefont {Canadell}, \citenamefont {Marqu{\'{e}}s-Hueso},\ and\
  \citenamefont {Gerardot}}]{Sanchez-Royo:NanoResearch7-2014}%
  \BibitemOpen
  \bibfield  {author} {\bibinfo {author} {\bibfnamefont {J.~F.}\ \bibnamefont
  {S{\'{a}}nchez-Royo}}, \bibinfo {author} {\bibfnamefont {G.}~\bibnamefont
  {Mu{\~{n}}oz-Matutano}}, \bibinfo {author} {\bibfnamefont {M.}~\bibnamefont
  {Brotons-Gisbert}}, \bibinfo {author} {\bibfnamefont {J.~P.}\ \bibnamefont
  {Mart{\'{\i}}nez-Pastor}}, \bibinfo {author} {\bibfnamefont {A.}~\bibnamefont
  {Segura}}, \bibinfo {author} {\bibfnamefont {A.}~\bibnamefont {Cantarero}},
  \bibinfo {author} {\bibfnamefont {R.}~\bibnamefont {Mata}}, \bibinfo {author}
  {\bibfnamefont {J.}~\bibnamefont {Canet-Ferrer}}, \bibinfo {author}
  {\bibfnamefont {G.}~\bibnamefont {Tobias}}, \bibinfo {author} {\bibfnamefont
  {E.}~\bibnamefont {Canadell}}, \bibinfo {author} {\bibfnamefont
  {J.}~\bibnamefont {Marqu{\'{e}}s-Hueso}}, \ and\ \bibinfo {author}
  {\bibfnamefont {B.~D.}\ \bibnamefont {Gerardot}},\ }\href {\doibase
  10.1007/s12274-014-0516-x} {\bibfield  {journal} {\bibinfo  {journal} {Nano
  Research}\ }\textbf {\bibinfo {volume} {7}},\ \bibinfo {pages} {1556}
  (\bibinfo {year} {2014})}\BibitemShut {NoStop}%
\bibitem [{\citenamefont {Hamer}\ \emph {et~al.}(2019)\citenamefont {Hamer},
  \citenamefont {Zultak}, \citenamefont {Tyurnina}, \citenamefont
  {Z{\'{o}}lyomi}, \citenamefont {Terry}, \citenamefont {Barinov},
  \citenamefont {Garner}, \citenamefont {Donoghue}, \citenamefont {Rooney},
  \citenamefont {Kandyba}, \citenamefont {Giampietri}, \citenamefont {Graham},
  \citenamefont {Teutsch}, \citenamefont {Xia}, \citenamefont {Koperski},
  \citenamefont {Haigh}, \citenamefont {Fal'ko}, \citenamefont {Gorbachev},\
  and\ \citenamefont {Wilson}}]{Hamer:ACSnano13-2019}%
  \BibitemOpen
  \bibfield  {author} {\bibinfo {author} {\bibfnamefont {M.~J.}\ \bibnamefont
  {Hamer}}, \bibinfo {author} {\bibfnamefont {J.}~\bibnamefont {Zultak}},
  \bibinfo {author} {\bibfnamefont {A.~V.}\ \bibnamefont {Tyurnina}}, \bibinfo
  {author} {\bibfnamefont {V.}~\bibnamefont {Z{\'{o}}lyomi}}, \bibinfo {author}
  {\bibfnamefont {D.}~\bibnamefont {Terry}}, \bibinfo {author} {\bibfnamefont
  {A.}~\bibnamefont {Barinov}}, \bibinfo {author} {\bibfnamefont
  {A.}~\bibnamefont {Garner}}, \bibinfo {author} {\bibfnamefont
  {J.}~\bibnamefont {Donoghue}}, \bibinfo {author} {\bibfnamefont {A.~P.}\
  \bibnamefont {Rooney}}, \bibinfo {author} {\bibfnamefont {V.}~\bibnamefont
  {Kandyba}}, \bibinfo {author} {\bibfnamefont {A.}~\bibnamefont {Giampietri}},
  \bibinfo {author} {\bibfnamefont {A.}~\bibnamefont {Graham}}, \bibinfo
  {author} {\bibfnamefont {N.}~\bibnamefont {Teutsch}}, \bibinfo {author}
  {\bibfnamefont {X.}~\bibnamefont {Xia}}, \bibinfo {author} {\bibfnamefont
  {M.}~\bibnamefont {Koperski}}, \bibinfo {author} {\bibfnamefont {S.~J.}\
  \bibnamefont {Haigh}}, \bibinfo {author} {\bibfnamefont {V.~I.}\ \bibnamefont
  {Fal'ko}}, \bibinfo {author} {\bibfnamefont {R.~V.}\ \bibnamefont
  {Gorbachev}}, \ and\ \bibinfo {author} {\bibfnamefont {N.~R.}\ \bibnamefont
  {Wilson}},\ }\href {\doibase 10.1021/acsnano.8b08726} {\bibfield  {journal}
  {\bibinfo  {journal} {{ACS} Nano}\ } (\bibinfo {year} {2019}),\
  10.1021/acsnano.8b08726}\BibitemShut {NoStop}%
\bibitem [{\citenamefont {Li}\ \emph {et~al.}(2014)\citenamefont {Li},
  \citenamefont {Lin}, \citenamefont {Puretzky}, \citenamefont {Idrobo},
  \citenamefont {Ma}, \citenamefont {Chi}, \citenamefont {Yoon}, \citenamefont
  {Rouleau}, \citenamefont {Kravchenko}, \citenamefont {Geohegan},\ and\
  \citenamefont {Xiao}}]{Li:SciRep4-2014}%
  \BibitemOpen
  \bibfield  {author} {\bibinfo {author} {\bibfnamefont {X.}~\bibnamefont
  {Li}}, \bibinfo {author} {\bibfnamefont {M.-W.}\ \bibnamefont {Lin}},
  \bibinfo {author} {\bibfnamefont {A.~A.}\ \bibnamefont {Puretzky}}, \bibinfo
  {author} {\bibfnamefont {J.~C.}\ \bibnamefont {Idrobo}}, \bibinfo {author}
  {\bibfnamefont {C.}~\bibnamefont {Ma}}, \bibinfo {author} {\bibfnamefont
  {M.}~\bibnamefont {Chi}}, \bibinfo {author} {\bibfnamefont {M.}~\bibnamefont
  {Yoon}}, \bibinfo {author} {\bibfnamefont {C.~M.}\ \bibnamefont {Rouleau}},
  \bibinfo {author} {\bibfnamefont {I.~I.}\ \bibnamefont {Kravchenko}},
  \bibinfo {author} {\bibfnamefont {D.~B.}\ \bibnamefont {Geohegan}}, \ and\
  \bibinfo {author} {\bibfnamefont {K.}~\bibnamefont {Xiao}},\ }\href {\doibase
  10.1038/srep05497} {\bibfield  {journal} {\bibinfo  {journal} {Scientific
  Reports}\ }\textbf {\bibinfo {volume} {4}} (\bibinfo {year} {2014}),\
  10.1038/srep05497}\BibitemShut {NoStop}%
\bibitem [{\citenamefont {Aziza}\ \emph {et~al.}(2017)\citenamefont {Aziza},
  \citenamefont {Pierucci}, \citenamefont {Henck}, \citenamefont {Silly},
  \citenamefont {David}, \citenamefont {Yoon}, \citenamefont {Sirotti},
  \citenamefont {Xiao}, \citenamefont {Eddrief}, \citenamefont {Girard},\ and\
  \citenamefont {Ouerghi}}]{Aziza:PRB96-2017}%
  \BibitemOpen
  \bibfield  {author} {\bibinfo {author} {\bibfnamefont {Z.~B.}\ \bibnamefont
  {Aziza}}, \bibinfo {author} {\bibfnamefont {D.}~\bibnamefont {Pierucci}},
  \bibinfo {author} {\bibfnamefont {H.}~\bibnamefont {Henck}}, \bibinfo
  {author} {\bibfnamefont {M.~G.}\ \bibnamefont {Silly}}, \bibinfo {author}
  {\bibfnamefont {C.}~\bibnamefont {David}}, \bibinfo {author} {\bibfnamefont
  {M.}~\bibnamefont {Yoon}}, \bibinfo {author} {\bibfnamefont {F.}~\bibnamefont
  {Sirotti}}, \bibinfo {author} {\bibfnamefont {K.}~\bibnamefont {Xiao}},
  \bibinfo {author} {\bibfnamefont {M.}~\bibnamefont {Eddrief}}, \bibinfo
  {author} {\bibfnamefont {J.-C.}\ \bibnamefont {Girard}}, \ and\ \bibinfo
  {author} {\bibfnamefont {A.}~\bibnamefont {Ouerghi}},\ }\href {\doibase
  10.1103/physrevb.96.035407} {\bibfield  {journal} {\bibinfo  {journal}
  {Physical Review B}\ }\textbf {\bibinfo {volume} {96}} (\bibinfo {year}
  {2017}),\ 10.1103/physrevb.96.035407}\BibitemShut {NoStop}%
\bibitem [{\citenamefont {Aziza}\ \emph {et~al.}(2018)\citenamefont {Aziza},
  \citenamefont {Z{\'{o}}lyomi}, \citenamefont {Henck}, \citenamefont
  {Pierucci}, \citenamefont {Silly}, \citenamefont {Avila}, \citenamefont
  {Magorrian}, \citenamefont {Chaste}, \citenamefont {Chen}, \citenamefont
  {Yoon}, \citenamefont {Xiao}, \citenamefont {Sirotti}, \citenamefont
  {Asensio}, \citenamefont {Lhuillier}, \citenamefont {Eddrief}, \citenamefont
  {Fal{\textquotesingle}ko},\ and\ \citenamefont {Ouerghi}}]{Aziza:PRB98-2018}%
  \BibitemOpen
  \bibfield  {author} {\bibinfo {author} {\bibfnamefont {Z.~B.}\ \bibnamefont
  {Aziza}}, \bibinfo {author} {\bibfnamefont {V.}~\bibnamefont
  {Z{\'{o}}lyomi}}, \bibinfo {author} {\bibfnamefont {H.}~\bibnamefont
  {Henck}}, \bibinfo {author} {\bibfnamefont {D.}~\bibnamefont {Pierucci}},
  \bibinfo {author} {\bibfnamefont {M.~G.}\ \bibnamefont {Silly}}, \bibinfo
  {author} {\bibfnamefont {J.}~\bibnamefont {Avila}}, \bibinfo {author}
  {\bibfnamefont {S.~J.}\ \bibnamefont {Magorrian}}, \bibinfo {author}
  {\bibfnamefont {J.}~\bibnamefont {Chaste}}, \bibinfo {author} {\bibfnamefont
  {C.}~\bibnamefont {Chen}}, \bibinfo {author} {\bibfnamefont {M.}~\bibnamefont
  {Yoon}}, \bibinfo {author} {\bibfnamefont {K.}~\bibnamefont {Xiao}}, \bibinfo
  {author} {\bibfnamefont {F.}~\bibnamefont {Sirotti}}, \bibinfo {author}
  {\bibfnamefont {M.~C.}\ \bibnamefont {Asensio}}, \bibinfo {author}
  {\bibfnamefont {E.}~\bibnamefont {Lhuillier}}, \bibinfo {author}
  {\bibfnamefont {M.}~\bibnamefont {Eddrief}}, \bibinfo {author} {\bibfnamefont
  {V.~I.}\ \bibnamefont {Fal{\textquotesingle}ko}}, \ and\ \bibinfo {author}
  {\bibfnamefont {A.}~\bibnamefont {Ouerghi}},\ }\href {\doibase
  10.1103/physrevb.98.115405} {\bibfield  {journal} {\bibinfo  {journal}
  {Physical Review B}\ }\textbf {\bibinfo {volume} {98}} (\bibinfo {year}
  {2018}),\ 10.1103/physrevb.98.115405}\BibitemShut {NoStop}%
\bibitem [{\citenamefont {Budweg}\ \emph {et~al.}(2019)\citenamefont {Budweg},
  \citenamefont {Yadav}, \citenamefont {Grupp}, \citenamefont {Leitenstorfer},
  \citenamefont {Trushin}, \citenamefont {Pauly},\ and\ \citenamefont
  {Brida}}]{trushin_GaSe}%
  \BibitemOpen
  \bibfield  {author} {\bibinfo {author} {\bibfnamefont {A.}~\bibnamefont
  {Budweg}}, \bibinfo {author} {\bibfnamefont {D.}~\bibnamefont {Yadav}},
  \bibinfo {author} {\bibfnamefont {A.}~\bibnamefont {Grupp}}, \bibinfo
  {author} {\bibfnamefont {A.}~\bibnamefont {Leitenstorfer}}, \bibinfo {author}
  {\bibfnamefont {M.}~\bibnamefont {Trushin}}, \bibinfo {author} {\bibfnamefont
  {F.}~\bibnamefont {Pauly}}, \ and\ \bibinfo {author} {\bibfnamefont
  {D.}~\bibnamefont {Brida}},\ }\href {\doibase 10.1103/PhysRevB.100.045404}
  {\bibfield  {journal} {\bibinfo  {journal} {Phys. Rev. B}\ }\textbf {\bibinfo
  {volume} {100}},\ \bibinfo {pages} {045404} (\bibinfo {year}
  {2019})}\BibitemShut {NoStop}%
\bibitem [{\citenamefont {Rybkovskiy}\ \emph {et~al.}(2011)\citenamefont
  {Rybkovskiy}, \citenamefont {Arutyunyan}, \citenamefont {Orekhov},
  \citenamefont {Gromchenko}, \citenamefont {Vorobiev}, \citenamefont
  {Osadchy}, \citenamefont {Salaev}, \citenamefont {Baykara}, \citenamefont
  {Allakhverdiev},\ and\ \citenamefont {Obraztsova}}]{Rybkovskiy:PRB84-2011}%
  \BibitemOpen
  \bibfield  {author} {\bibinfo {author} {\bibfnamefont {D.~V.}\ \bibnamefont
  {Rybkovskiy}}, \bibinfo {author} {\bibfnamefont {N.~R.}\ \bibnamefont
  {Arutyunyan}}, \bibinfo {author} {\bibfnamefont {A.~S.}\ \bibnamefont
  {Orekhov}}, \bibinfo {author} {\bibfnamefont {I.~A.}\ \bibnamefont
  {Gromchenko}}, \bibinfo {author} {\bibfnamefont {I.~V.}\ \bibnamefont
  {Vorobiev}}, \bibinfo {author} {\bibfnamefont {A.~V.}\ \bibnamefont
  {Osadchy}}, \bibinfo {author} {\bibfnamefont {E.~Y.}\ \bibnamefont {Salaev}},
  \bibinfo {author} {\bibfnamefont {T.~K.}\ \bibnamefont {Baykara}}, \bibinfo
  {author} {\bibfnamefont {K.~R.}\ \bibnamefont {Allakhverdiev}}, \ and\
  \bibinfo {author} {\bibfnamefont {E.~D.}\ \bibnamefont {Obraztsova}},\ }\href
  {\doibase 10.1103/physrevb.84.085314} {\bibfield  {journal} {\bibinfo
  {journal} {Physical Review B}\ }\textbf {\bibinfo {volume} {84}} (\bibinfo
  {year} {2011}),\ 10.1103/physrevb.84.085314}\BibitemShut {NoStop}%
\bibitem [{\citenamefont {Rybkovskiy}\ \emph {et~al.}(2014)\citenamefont
  {Rybkovskiy}, \citenamefont {Osadchy},\ and\ \citenamefont
  {Obraztsova}}]{Rybkovskiy:PRB90-2014}%
  \BibitemOpen
  \bibfield  {author} {\bibinfo {author} {\bibfnamefont {D.~V.}\ \bibnamefont
  {Rybkovskiy}}, \bibinfo {author} {\bibfnamefont {A.~V.}\ \bibnamefont
  {Osadchy}}, \ and\ \bibinfo {author} {\bibfnamefont {E.~D.}\ \bibnamefont
  {Obraztsova}},\ }\href {\doibase 10.1103/physrevb.90.235302} {\bibfield
  {journal} {\bibinfo  {journal} {Phys. Rev. B}\ }\textbf {\bibinfo {volume}
  {90}} (\bibinfo {year} {2014}),\ 10.1103/physrevb.90.235302}\BibitemShut
  {NoStop}%
\bibitem [{\citenamefont {Z{\'{o}}lyomi}\ \emph {et~al.}(2013)\citenamefont
  {Z{\'{o}}lyomi}, \citenamefont {Drummond},\ and\ \citenamefont
  {Fal{\textquotesingle}ko}}]{Zolyomi:PRB87-2013}%
  \BibitemOpen
  \bibfield  {author} {\bibinfo {author} {\bibfnamefont {V.}~\bibnamefont
  {Z{\'{o}}lyomi}}, \bibinfo {author} {\bibfnamefont {N.~D.}\ \bibnamefont
  {Drummond}}, \ and\ \bibinfo {author} {\bibfnamefont {V.~I.}\ \bibnamefont
  {Fal{\textquotesingle}ko}},\ }\href {\doibase 10.1103/physrevb.87.195403}
  {\bibfield  {journal} {\bibinfo  {journal} {Physical Review B}\ }\textbf
  {\bibinfo {volume} {87}} (\bibinfo {year} {2013}),\
  10.1103/physrevb.87.195403}\BibitemShut {NoStop}%
\bibitem [{\citenamefont {Z\'olyomi}\ \emph {et~al.}(2014)\citenamefont
  {Z\'olyomi}, \citenamefont {Drummond},\ and\ \citenamefont
  {Fal'ko}}]{Zolyomi:PRB89-2014}%
  \BibitemOpen
  \bibfield  {author} {\bibinfo {author} {\bibfnamefont {V.}~\bibnamefont
  {Z\'olyomi}}, \bibinfo {author} {\bibfnamefont {N.~D.}\ \bibnamefont
  {Drummond}}, \ and\ \bibinfo {author} {\bibfnamefont {V.~I.}\ \bibnamefont
  {Fal'ko}},\ }\href {\doibase 10.1103/PhysRevB.89.205416} {\bibfield
  {journal} {\bibinfo  {journal} {Phys. Rev. B}\ }\textbf {\bibinfo {volume}
  {89}},\ \bibinfo {pages} {205416} (\bibinfo {year} {2014})}\BibitemShut
  {NoStop}%
\bibitem [{\citenamefont {Zhou}\ \emph {et~al.}(2017)\citenamefont {Zhou},
  \citenamefont {Zhang}, \citenamefont {Sun}, \citenamefont {Lou},
  \citenamefont {Zhang}, \citenamefont {Yang},\ and\ \citenamefont
  {Chang}}]{Zhou:PRB96-2017}%
  \BibitemOpen
  \bibfield  {author} {\bibinfo {author} {\bibfnamefont {M.}~\bibnamefont
  {Zhou}}, \bibinfo {author} {\bibfnamefont {R.}~\bibnamefont {Zhang}},
  \bibinfo {author} {\bibfnamefont {J.}~\bibnamefont {Sun}}, \bibinfo {author}
  {\bibfnamefont {W.-K.}\ \bibnamefont {Lou}}, \bibinfo {author} {\bibfnamefont
  {D.}~\bibnamefont {Zhang}}, \bibinfo {author} {\bibfnamefont
  {W.}~\bibnamefont {Yang}}, \ and\ \bibinfo {author} {\bibfnamefont
  {K.}~\bibnamefont {Chang}},\ }\href {\doibase 10.1103/physrevb.96.155430}
  {\bibfield  {journal} {\bibinfo  {journal} {Phys. Rev. B}\ }\textbf {\bibinfo
  {volume} {96}} (\bibinfo {year} {2017}),\
  10.1103/physrevb.96.155430}\BibitemShut {NoStop}%
\bibitem [{\citenamefont {Zhou}\ \emph {et~al.}(2019)\citenamefont {Zhou},
  \citenamefont {Zhang}, \citenamefont {Yu}, \citenamefont {Huang},
  \citenamefont {Chen}, \citenamefont {Yang},\ and\ \citenamefont
  {Chang}}]{Zhou:PRB99-2019}%
  \BibitemOpen
  \bibfield  {author} {\bibinfo {author} {\bibfnamefont {M.}~\bibnamefont
  {Zhou}}, \bibinfo {author} {\bibfnamefont {D.}~\bibnamefont {Zhang}},
  \bibinfo {author} {\bibfnamefont {S.}~\bibnamefont {Yu}}, \bibinfo {author}
  {\bibfnamefont {Z.}~\bibnamefont {Huang}}, \bibinfo {author} {\bibfnamefont
  {Y.}~\bibnamefont {Chen}}, \bibinfo {author} {\bibfnamefont {W.}~\bibnamefont
  {Yang}}, \ and\ \bibinfo {author} {\bibfnamefont {K.}~\bibnamefont {Chang}},\
  }\href {\doibase 10.1103/PhysRevB.99.155402} {\bibfield  {journal} {\bibinfo
  {journal} {Phys. Rev. B}\ }\textbf {\bibinfo {volume} {99}},\ \bibinfo
  {pages} {155402} (\bibinfo {year} {2019})}\BibitemShut {NoStop}%
\bibitem [{\citenamefont {Sun}\ \emph {et~al.}(2018)\citenamefont {Sun},
  \citenamefont {Luo}, \citenamefont {Zhao}, \citenamefont {Biswas},
  \citenamefont {Li},\ and\ \citenamefont {Zhang}}]{Sun:Nanoscale10-2018}%
  \BibitemOpen
  \bibfield  {author} {\bibinfo {author} {\bibfnamefont {Y.}~\bibnamefont
  {Sun}}, \bibinfo {author} {\bibfnamefont {S.}~\bibnamefont {Luo}}, \bibinfo
  {author} {\bibfnamefont {X.-G.}\ \bibnamefont {Zhao}}, \bibinfo {author}
  {\bibfnamefont {K.}~\bibnamefont {Biswas}}, \bibinfo {author} {\bibfnamefont
  {S.-L.}\ \bibnamefont {Li}}, \ and\ \bibinfo {author} {\bibfnamefont
  {L.}~\bibnamefont {Zhang}},\ }\href {\doibase 10.1039/c7nr09486h} {\bibfield
  {journal} {\bibinfo  {journal} {Nanoscale}\ }\textbf {\bibinfo {volume}
  {10}},\ \bibinfo {pages} {7991} (\bibinfo {year} {2018})}\BibitemShut
  {NoStop}%
\bibitem [{\citenamefont {Magorrian}\ \emph {et~al.}(2016)\citenamefont
  {Magorrian}, \citenamefont {Z{\'{o}}lyomi},\ and\ \citenamefont
  {Fal{\textquotesingle}ko}}]{Magorrian:PRB94-2016}%
  \BibitemOpen
  \bibfield  {author} {\bibinfo {author} {\bibfnamefont {S.~J.}\ \bibnamefont
  {Magorrian}}, \bibinfo {author} {\bibfnamefont {V.}~\bibnamefont
  {Z{\'{o}}lyomi}}, \ and\ \bibinfo {author} {\bibfnamefont {V.~I.}\
  \bibnamefont {Fal{\textquotesingle}ko}},\ }\href {\doibase
  10.1103/physrevb.94.245431} {\bibfield  {journal} {\bibinfo  {journal} {Phys.
  Rev. B}\ }\textbf {\bibinfo {volume} {94}} (\bibinfo {year} {2016}),\
  10.1103/physrevb.94.245431}\BibitemShut {NoStop}%
\bibitem [{\citenamefont {Cao}\ \emph {et~al.}(2015)\citenamefont {Cao},
  \citenamefont {Li},\ and\ \citenamefont {Louie}}]{Louieferro}%
  \BibitemOpen
  \bibfield  {author} {\bibinfo {author} {\bibfnamefont {T.}~\bibnamefont
  {Cao}}, \bibinfo {author} {\bibfnamefont {Z.}~\bibnamefont {Li}}, \ and\
  \bibinfo {author} {\bibfnamefont {S.~G.}\ \bibnamefont {Louie}},\ }\href
  {\doibase 10.1103/PhysRevLett.114.236602} {\bibfield  {journal} {\bibinfo
  {journal} {Phys. Rev. Lett.}\ }\textbf {\bibinfo {volume} {114}},\ \bibinfo
  {pages} {236602} (\bibinfo {year} {2015})}\BibitemShut {NoStop}%
\bibitem [{\citenamefont {Trushin}(2019)}]{trushin_excitons}%
  \BibitemOpen
  \bibfield  {author} {\bibinfo {author} {\bibfnamefont {M.}~\bibnamefont
  {Trushin}},\ }\href {\doibase 10.1103/PhysRevB.99.205307} {\bibfield
  {journal} {\bibinfo  {journal} {Phys. Rev. B}\ }\textbf {\bibinfo {volume}
  {99}},\ \bibinfo {pages} {205307} (\bibinfo {year} {2019})}\BibitemShut
  {NoStop}%
\bibitem [{\citenamefont {Skinner}(2016)}]{Skinnerexcitons}%
  \BibitemOpen
  \bibfield  {author} {\bibinfo {author} {\bibfnamefont {B.}~\bibnamefont
  {Skinner}},\ }\href {\doibase 10.1103/PhysRevB.93.235110} {\bibfield
  {journal} {\bibinfo  {journal} {Phys. Rev. B}\ }\textbf {\bibinfo {volume}
  {93}},\ \bibinfo {pages} {235110} (\bibinfo {year} {2016})}\BibitemShut
  {NoStop}%
\bibitem [{\citenamefont {Mudd}\ \emph {et~al.}(2016)\citenamefont {Mudd},
  \citenamefont {Molas}, \citenamefont {Chen}, \citenamefont {Z{\'{o}}lyomi},
  \citenamefont {Nogajewski}, \citenamefont {Kudrynskyi}, \citenamefont
  {Kovalyuk}, \citenamefont {Yusa}, \citenamefont {Makarovsky}, \citenamefont
  {Eaves}, \citenamefont {Potemski}, \citenamefont {Fal'ko},\ and\
  \citenamefont {Patan{\`{e}}}}]{Mudd:SciRep6-2016}%
  \BibitemOpen
  \bibfield  {author} {\bibinfo {author} {\bibfnamefont {G.~W.}\ \bibnamefont
  {Mudd}}, \bibinfo {author} {\bibfnamefont {M.~R.}\ \bibnamefont {Molas}},
  \bibinfo {author} {\bibfnamefont {X.}~\bibnamefont {Chen}}, \bibinfo {author}
  {\bibfnamefont {V.}~\bibnamefont {Z{\'{o}}lyomi}}, \bibinfo {author}
  {\bibfnamefont {K.}~\bibnamefont {Nogajewski}}, \bibinfo {author}
  {\bibfnamefont {Z.~R.}\ \bibnamefont {Kudrynskyi}}, \bibinfo {author}
  {\bibfnamefont {Z.~D.}\ \bibnamefont {Kovalyuk}}, \bibinfo {author}
  {\bibfnamefont {G.}~\bibnamefont {Yusa}}, \bibinfo {author} {\bibfnamefont
  {O.}~\bibnamefont {Makarovsky}}, \bibinfo {author} {\bibfnamefont
  {L.}~\bibnamefont {Eaves}}, \bibinfo {author} {\bibfnamefont
  {M.}~\bibnamefont {Potemski}}, \bibinfo {author} {\bibfnamefont {V.~I.}\
  \bibnamefont {Fal'ko}}, \ and\ \bibinfo {author} {\bibfnamefont
  {A.}~\bibnamefont {Patan{\`{e}}}},\ }\href {\doibase 10.1038/srep39619}
  {\bibfield  {journal} {\bibinfo  {journal} {Sci. Rep.}\ }\textbf {\bibinfo
  {volume} {6}} (\bibinfo {year} {2016}),\ 10.1038/srep39619}\BibitemShut
  {NoStop}%
\bibitem [{\citenamefont {Magorrian}\ \emph {et~al.}(2018)\citenamefont
  {Magorrian}, \citenamefont {Ceferino}, \citenamefont {Z\'olyomi},\ and\
  \citenamefont {Fal'ko}}]{Mogorrian:PRB97-2018}%
  \BibitemOpen
  \bibfield  {author} {\bibinfo {author} {\bibfnamefont {S.~J.}\ \bibnamefont
  {Magorrian}}, \bibinfo {author} {\bibfnamefont {A.}~\bibnamefont {Ceferino}},
  \bibinfo {author} {\bibfnamefont {V.}~\bibnamefont {Z\'olyomi}}, \ and\
  \bibinfo {author} {\bibfnamefont {V.~I.}\ \bibnamefont {Fal'ko}},\ }\href
  {\doibase 10.1103/PhysRevB.97.165304} {\bibfield  {journal} {\bibinfo
  {journal} {Phys. Rev. B}\ }\textbf {\bibinfo {volume} {97}},\ \bibinfo
  {pages} {165304} (\bibinfo {year} {2018})}\BibitemShut {NoStop}%
\bibitem [{\citenamefont {Zultak}\ \emph {et~al.}(2020)\citenamefont {Zultak},
  \citenamefont {Magorrian}, \citenamefont {Koperski}, \citenamefont {Garner},
  \citenamefont {Hamer}, \citenamefont {T{\'{o}}v{\'{a}}ri}, \citenamefont
  {Novoselov}, \citenamefont {Zhukov}, \citenamefont {Zou}, \citenamefont
  {Wilson}, \citenamefont {Haigh}, \citenamefont {Kretinin}, \citenamefont
  {Fal'ko},\ and\ \citenamefont {Gorbachev}}]{Zultak:NatComm11-2020}%
  \BibitemOpen
  \bibfield  {author} {\bibinfo {author} {\bibfnamefont {J.}~\bibnamefont
  {Zultak}}, \bibinfo {author} {\bibfnamefont {S.~J.}\ \bibnamefont
  {Magorrian}}, \bibinfo {author} {\bibfnamefont {M.}~\bibnamefont {Koperski}},
  \bibinfo {author} {\bibfnamefont {A.}~\bibnamefont {Garner}}, \bibinfo
  {author} {\bibfnamefont {M.~J.}\ \bibnamefont {Hamer}}, \bibinfo {author}
  {\bibfnamefont {E.}~\bibnamefont {T{\'{o}}v{\'{a}}ri}}, \bibinfo {author}
  {\bibfnamefont {K.~S.}\ \bibnamefont {Novoselov}}, \bibinfo {author}
  {\bibfnamefont {A.~A.}\ \bibnamefont {Zhukov}}, \bibinfo {author}
  {\bibfnamefont {Y.}~\bibnamefont {Zou}}, \bibinfo {author} {\bibfnamefont
  {N.~R.}\ \bibnamefont {Wilson}}, \bibinfo {author} {\bibfnamefont {S.~J.}\
  \bibnamefont {Haigh}}, \bibinfo {author} {\bibfnamefont {A.~V.}\ \bibnamefont
  {Kretinin}}, \bibinfo {author} {\bibfnamefont {V.~I.}\ \bibnamefont
  {Fal'ko}}, \ and\ \bibinfo {author} {\bibfnamefont {R.}~\bibnamefont
  {Gorbachev}},\ }\href {\doibase 10.1038/s41467-019-13893-w} {\bibfield
  {journal} {\bibinfo  {journal} {Nature Communications}\ }\textbf {\bibinfo
  {volume} {11}} (\bibinfo {year} {2020}),\
  10.1038/s41467-019-13893-w}\BibitemShut {NoStop}%
\bibitem [{\citenamefont {Onida}\ \emph {et~al.}(2002)\citenamefont {Onida},
  \citenamefont {Reining},\ and\ \citenamefont {Rubio}}]{Onida:RMP74-2002}%
  \BibitemOpen
  \bibfield  {author} {\bibinfo {author} {\bibfnamefont {G.}~\bibnamefont
  {Onida}}, \bibinfo {author} {\bibfnamefont {L.}~\bibnamefont {Reining}}, \
  and\ \bibinfo {author} {\bibfnamefont {A.}~\bibnamefont {Rubio}},\ }\href
  {\doibase 10.1103/revmodphys.74.601} {\bibfield  {journal} {\bibinfo
  {journal} {Rev. Mod. Phys.}\ }\textbf {\bibinfo {volume} {74}},\ \bibinfo
  {pages} {601} (\bibinfo {year} {2002})}\BibitemShut {NoStop}%
\bibitem [{\citenamefont {Jiang}\ \emph {et~al.}(2007)\citenamefont {Jiang},
  \citenamefont {Saito}, \citenamefont {Samsonidze}, \citenamefont {Jorio},
  \citenamefont {Chou}, \citenamefont {Dresselhaus},\ and\ \citenamefont
  {Dresselhaus}}]{Jiang:PRB75-2007}%
  \BibitemOpen
  \bibfield  {author} {\bibinfo {author} {\bibfnamefont {J.}~\bibnamefont
  {Jiang}}, \bibinfo {author} {\bibfnamefont {R.}~\bibnamefont {Saito}},
  \bibinfo {author} {\bibfnamefont {G.~G.}\ \bibnamefont {Samsonidze}},
  \bibinfo {author} {\bibfnamefont {A.}~\bibnamefont {Jorio}}, \bibinfo
  {author} {\bibfnamefont {S.~G.}\ \bibnamefont {Chou}}, \bibinfo {author}
  {\bibfnamefont {G.}~\bibnamefont {Dresselhaus}}, \ and\ \bibinfo {author}
  {\bibfnamefont {M.~S.}\ \bibnamefont {Dresselhaus}},\ }\href {\doibase
  10.1103/physrevb.75.035407} {\bibfield  {journal} {\bibinfo  {journal} {Phys.
  Rev. B}\ }\textbf {\bibinfo {volume} {75}} (\bibinfo {year} {2007}),\
  10.1103/physrevb.75.035407}\BibitemShut {NoStop}%
\bibitem [{\citenamefont {Trolle}\ \emph {et~al.}(2014)\citenamefont {Trolle},
  \citenamefont {Seifert},\ and\ \citenamefont {Pedersen}}]{Trolle:PRB89-2014}%
  \BibitemOpen
  \bibfield  {author} {\bibinfo {author} {\bibfnamefont {M.~L.}\ \bibnamefont
  {Trolle}}, \bibinfo {author} {\bibfnamefont {G.}~\bibnamefont {Seifert}}, \
  and\ \bibinfo {author} {\bibfnamefont {T.~G.}\ \bibnamefont {Pedersen}},\
  }\href {\doibase 10.1103/PhysRevB.89.235410} {\bibfield  {journal} {\bibinfo
  {journal} {Phys. Rev. B}\ }\textbf {\bibinfo {volume} {89}},\ \bibinfo
  {pages} {235410} (\bibinfo {year} {2014})}\BibitemShut {NoStop}%
\bibitem [{\citenamefont {Wu}\ \emph {et~al.}(2015)\citenamefont {Wu},
  \citenamefont {Qu},\ and\ \citenamefont {MacDonald}}]{Wu:PRB91-2015}%
  \BibitemOpen
  \bibfield  {author} {\bibinfo {author} {\bibfnamefont {F.}~\bibnamefont
  {Wu}}, \bibinfo {author} {\bibfnamefont {F.}~\bibnamefont {Qu}}, \ and\
  \bibinfo {author} {\bibfnamefont {A.~H.}\ \bibnamefont {MacDonald}},\ }\href
  {\doibase 10.1103/PhysRevB.91.075310} {\bibfield  {journal} {\bibinfo
  {journal} {Phys. Rev. B}\ }\textbf {\bibinfo {volume} {91}},\ \bibinfo
  {pages} {075310} (\bibinfo {year} {2015})}\BibitemShut {NoStop}%
\bibitem [{\citenamefont {Deilmann}\ and\ \citenamefont
  {Thygesen}(2019)}]{Deilmann:2DMat6-2019}%
  \BibitemOpen
  \bibfield  {author} {\bibinfo {author} {\bibfnamefont {T.}~\bibnamefont
  {Deilmann}}\ and\ \bibinfo {author} {\bibfnamefont {K.~S.}\ \bibnamefont
  {Thygesen}},\ }\href {\doibase 10.1088/2053-1583/ab0e1d} {\bibfield
  {journal} {\bibinfo  {journal} {2D Mater.}\ }\textbf {\bibinfo {volume}
  {6}},\ \bibinfo {pages} {035003} (\bibinfo {year} {2019})}\BibitemShut
  {NoStop}%
\bibitem [{\citenamefont {{Keldysh}}(1979)}]{Keldysh:JETP29-1979}%
  \BibitemOpen
  \bibfield  {author} {\bibinfo {author} {\bibfnamefont {L.~V.}\ \bibnamefont
  {{Keldysh}}},\ }\href@noop {} {\bibfield  {journal} {\bibinfo  {journal}
  {Soviet Journal of Experimental and Theoretical Physics Letters}\ }\textbf
  {\bibinfo {volume} {29}},\ \bibinfo {pages} {658} (\bibinfo {year}
  {1979})}\BibitemShut {NoStop}%
\bibitem [{\citenamefont {Rytova}(1967)}]{Rytova:MUPhys3-1967}%
  \BibitemOpen
  \bibfield  {author} {\bibinfo {author} {\bibfnamefont {N.~S.}\ \bibnamefont
  {Rytova}},\ }\href@noop {} {\bibfield  {journal} {\bibinfo  {journal} {Moscow
  Univ. Phys. Bull.}\ }\textbf {\bibinfo {volume} {3}},\ \bibinfo {pages} {18}
  (\bibinfo {year} {1967})}\BibitemShut {NoStop}%
\bibitem [{\citenamefont {Latini}\ \emph {et~al.}(2015)\citenamefont {Latini},
  \citenamefont {Olsen},\ and\ \citenamefont {Thygesen}}]{Latini:PRB92-2015}%
  \BibitemOpen
  \bibfield  {author} {\bibinfo {author} {\bibfnamefont {S.}~\bibnamefont
  {Latini}}, \bibinfo {author} {\bibfnamefont {T.}~\bibnamefont {Olsen}}, \
  and\ \bibinfo {author} {\bibfnamefont {K.~S.}\ \bibnamefont {Thygesen}},\
  }\href {\doibase 10.1103/PhysRevB.92.245123} {\bibfield  {journal} {\bibinfo
  {journal} {Phys. Rev. B}\ }\textbf {\bibinfo {volume} {92}},\ \bibinfo
  {pages} {245123} (\bibinfo {year} {2015})}\BibitemShut {NoStop}%
\bibitem [{\citenamefont {Trolle}\ \emph {et~al.}(2017)\citenamefont {Trolle},
  \citenamefont {Pedersen},\ and\ \citenamefont
  {Véniard}}]{Trolle:SciRep7-2017}%
  \BibitemOpen
  \bibfield  {author} {\bibinfo {author} {\bibfnamefont {M.~L.}\ \bibnamefont
  {Trolle}}, \bibinfo {author} {\bibfnamefont {T.~G.}\ \bibnamefont
  {Pedersen}}, \ and\ \bibinfo {author} {\bibfnamefont {V.}~\bibnamefont
  {Véniard}},\ }\href {https://doi.org/10.1038/srep39844} {\bibfield
  {journal} {\bibinfo  {journal} {Sci. Rep.}\ }\textbf {\bibinfo {volume}
  {7}},\ \bibinfo {pages} {39844} (\bibinfo {year} {2017})}\BibitemShut
  {NoStop}%
\bibitem [{\citenamefont {Geick}\ \emph {et~al.}(1966)\citenamefont {Geick},
  \citenamefont {Perry},\ and\ \citenamefont {Rupprecht}}]{Geick:PR146-1966}%
  \BibitemOpen
  \bibfield  {author} {\bibinfo {author} {\bibfnamefont {R.}~\bibnamefont
  {Geick}}, \bibinfo {author} {\bibfnamefont {C.~H.}\ \bibnamefont {Perry}}, \
  and\ \bibinfo {author} {\bibfnamefont {G.}~\bibnamefont {Rupprecht}},\ }\href
  {\doibase 10.1103/physrev.146.543} {\bibfield  {journal} {\bibinfo  {journal}
  {Phys. Rev.}\ }\textbf {\bibinfo {volume} {146}},\ \bibinfo {pages} {543}
  (\bibinfo {year} {1966})}\BibitemShut {NoStop}%
\bibitem [{\citenamefont {Laturia}\ \emph {et~al.}(2018)\citenamefont
  {Laturia}, \citenamefont {de~Put},\ and\ \citenamefont
  {Vandenberghe}}]{Laturia:njp2DMat2-2018}%
  \BibitemOpen
  \bibfield  {author} {\bibinfo {author} {\bibfnamefont {A.}~\bibnamefont
  {Laturia}}, \bibinfo {author} {\bibfnamefont {M.~L.~V.}\ \bibnamefont
  {de~Put}}, \ and\ \bibinfo {author} {\bibfnamefont {W.~G.}\ \bibnamefont
  {Vandenberghe}},\ }\href {\doibase 10.1038/s41699-018-0050-x} {\bibfield
  {journal} {\bibinfo  {journal} {npj 2D Mater. Appl.}\ }\textbf {\bibinfo
  {volume} {2}} (\bibinfo {year} {2018}),\
  10.1038/s41699-018-0050-x}\BibitemShut {NoStop}%
\bibitem [{\citenamefont {Kuroda}\ and\ \citenamefont
  {Nishina}(1980)}]{Kuroda:SSComm34-1980}%
  \BibitemOpen
  \bibfield  {author} {\bibinfo {author} {\bibfnamefont {N.}~\bibnamefont
  {Kuroda}}\ and\ \bibinfo {author} {\bibfnamefont {Y.}~\bibnamefont
  {Nishina}},\ }\href {\doibase 10.1016/0038-1098(80)90656-0} {\bibfield
  {journal} {\bibinfo  {journal} {Solid State Commun.}\ }\textbf {\bibinfo
  {volume} {34}},\ \bibinfo {pages} {481} (\bibinfo {year} {1980})}\BibitemShut
  {NoStop}%
\bibitem [{\citenamefont {Cudazzo}\ \emph {et~al.}(2011)\citenamefont
  {Cudazzo}, \citenamefont {Tokatly},\ and\ \citenamefont
  {Rubio}}]{Cudazzo:PRB84-2011}%
  \BibitemOpen
  \bibfield  {author} {\bibinfo {author} {\bibfnamefont {P.}~\bibnamefont
  {Cudazzo}}, \bibinfo {author} {\bibfnamefont {I.~V.}\ \bibnamefont
  {Tokatly}}, \ and\ \bibinfo {author} {\bibfnamefont {A.}~\bibnamefont
  {Rubio}},\ }\href {\doibase 10.1103/physrevb.84.085406} {\bibfield  {journal}
  {\bibinfo  {journal} {Phys. Rev. B}\ }\textbf {\bibinfo {volume} {84}}
  (\bibinfo {year} {2011}),\ 10.1103/physrevb.84.085406}\BibitemShut {NoStop}%
\bibitem [{Note1()}]{Note1}%
  \BibitemOpen
  \bibinfo {note} {$r_\ast \approx L \times 7.7$ \r A}\BibitemShut {NoStop}%
\bibitem [{\citenamefont {{Garc{\'\i}a Fl{\'o}rez}}\ \emph
  {et~al.}(2020)\citenamefont {{Garc{\'\i}a Fl{\'o}rez}}, \citenamefont
  {{Siebbeles}},\ and\ \citenamefont {{Stoof}}}]{Florez:arXiv02-2020}%
  \BibitemOpen
  \bibfield  {author} {\bibinfo {author} {\bibfnamefont {F.}~\bibnamefont
  {{Garc{\'\i}a Fl{\'o}rez}}}, \bibinfo {author} {\bibfnamefont {L.~D.~A.}\
  \bibnamefont {{Siebbeles}}}, \ and\ \bibinfo {author} {\bibfnamefont
  {H.~T.~C.}\ \bibnamefont {{Stoof}}},\ }\href@noop {} {\bibfield  {journal}
  {\bibinfo  {journal} {arXiv e-prints}\ ,\ \bibinfo {eid} {arXiv:2002.05921}}
  (\bibinfo {year} {2020})},\ \Eprint {http://arxiv.org/abs/2002.05921}
  {arXiv:2002.05921 [cond-mat.mes-hall]} \BibitemShut {NoStop}%
\bibitem [{Note2()}]{Note2}%
  \BibitemOpen
  \bibinfo {note} {The bound electron-hole states in an exciton resemble the
  two-body system in a hydrogen atom. Therefore, these bound states may be well
  described by an atomic orbital basis set. For the sake of numerical
  simplicity, we use harmonic oscillator basis function which is an orthogonal
  Gaussian-type orbital basis.}\BibitemShut {Stop}%
\bibitem [{SM()}]{SM}%
  \BibitemOpen
  \href@noop {} {}\bibinfo {note} {See Supplemental Material at [...] for the
  software package. The instruction is included in the file.}\BibitemShut
  {Stop}%
\bibitem [{Note3()}]{Note3}%
  \BibitemOpen
  \bibinfo {note} {In the expansion, one has to ensure that the truncated
  polynomials must not allow $\varepsilon _c(\protect \mathbf
  {k}_e)-\varepsilon _v(\protect \mathbf {k}_h)$ smaller than the bandgap for
  all $\protect \mathbf {k}_{e/h}$.}\BibitemShut {Stop}%
\bibitem [{Note4()}]{Note4}%
  \BibitemOpen
  \bibinfo {note} {The degeneracies for the non-$s$-wave states are lifted by
  less than $1$ meV in the tight-binding model which only has the 6-fold
  rotational symmetry.}\BibitemShut {Stop}%
\bibitem [{\citenamefont {Ye}\ \emph {et~al.}(2014)\citenamefont {Ye},
  \citenamefont {Cao}, \citenamefont {O'Brien}, \citenamefont {Zhu},
  \citenamefont {Yin}, \citenamefont {Wang}, \citenamefont {Louie},\ and\
  \citenamefont {Zhang}}]{Ye:Nat513-2014}%
  \BibitemOpen
  \bibfield  {author} {\bibinfo {author} {\bibfnamefont {Z.}~\bibnamefont
  {Ye}}, \bibinfo {author} {\bibfnamefont {T.}~\bibnamefont {Cao}}, \bibinfo
  {author} {\bibfnamefont {K.}~\bibnamefont {O'Brien}}, \bibinfo {author}
  {\bibfnamefont {H.}~\bibnamefont {Zhu}}, \bibinfo {author} {\bibfnamefont
  {X.}~\bibnamefont {Yin}}, \bibinfo {author} {\bibfnamefont {Y.}~\bibnamefont
  {Wang}}, \bibinfo {author} {\bibfnamefont {S.~G.}\ \bibnamefont {Louie}}, \
  and\ \bibinfo {author} {\bibfnamefont {X.}~\bibnamefont {Zhang}},\ }\href
  {\doibase 10.1038/nature13734} {\bibfield  {journal} {\bibinfo  {journal}
  {Nature}\ }\textbf {\bibinfo {volume} {513}},\ \bibinfo {pages} {214}
  (\bibinfo {year} {2014})}\BibitemShut {NoStop}%
\bibitem [{\citenamefont {Zhang}\ and\ \citenamefont
  {Ma}(2019)}]{Zhang:JPhysCondMat31-2019}%
  \BibitemOpen
  \bibfield  {author} {\bibinfo {author} {\bibfnamefont {J.-Z.}\ \bibnamefont
  {Zhang}}\ and\ \bibinfo {author} {\bibfnamefont {J.-Z.}\ \bibnamefont {Ma}},\
  }\href {\doibase 10.1088/1361-648x/aaf8c5} {\bibfield  {journal} {\bibinfo
  {journal} {J. Phys. Condens. Matter}\ }\textbf {\bibinfo {volume} {31}},\
  \bibinfo {pages} {105702} (\bibinfo {year} {2019})}\BibitemShut {NoStop}%
\bibitem [{\citenamefont {Kress-Rogers}\ \emph {et~al.}(1982)\citenamefont
  {Kress-Rogers}, \citenamefont {Nicholas}, \citenamefont {Portal},\ and\
  \citenamefont {Chevy}}]{cyclotron}%
  \BibitemOpen
  \bibfield  {author} {\bibinfo {author} {\bibfnamefont {E.}~\bibnamefont
  {Kress-Rogers}}, \bibinfo {author} {\bibfnamefont {R.}~\bibnamefont
  {Nicholas}}, \bibinfo {author} {\bibfnamefont {J.}~\bibnamefont {Portal}}, \
  and\ \bibinfo {author} {\bibfnamefont {A.}~\bibnamefont {Chevy}},\
  }\href@noop {} {\bibfield  {journal} {\bibinfo  {journal} {Solid State
  Communications}\ }\textbf {\bibinfo {volume} {44}},\ \bibinfo {pages} {379}
  (\bibinfo {year} {1982})}\BibitemShut {NoStop}%
\bibitem [{\citenamefont {Kuroda}\ \emph {et~al.}(1979)\citenamefont {Kuroda},
  \citenamefont {Nishina}, \citenamefont {Hori},\ and\ \citenamefont
  {Date}}]{Kuroda:JPSJ47-1979}%
  \BibitemOpen
  \bibfield  {author} {\bibinfo {author} {\bibfnamefont {N.}~\bibnamefont
  {Kuroda}}, \bibinfo {author} {\bibfnamefont {Y.}~\bibnamefont {Nishina}},
  \bibinfo {author} {\bibfnamefont {H.}~\bibnamefont {Hori}}, \ and\ \bibinfo
  {author} {\bibfnamefont {M.}~\bibnamefont {Date}},\ }\href {\doibase
  10.1143/JPSJ.47.1373} {\bibfield  {journal} {\bibinfo  {journal} {Journal of
  the Physical Society of Japan}\ }\textbf {\bibinfo {volume} {47}},\ \bibinfo
  {pages} {1373} (\bibinfo {year} {1979})},\ \Eprint
  {http://arxiv.org/abs/https://doi.org/10.1143/JPSJ.47.1373}
  {https://doi.org/10.1143/JPSJ.47.1373} \BibitemShut {NoStop}%
\bibitem [{\citenamefont {Schindlmayr}(1997)}]{Schindlmayr:EurJPhys18-1997}%
  \BibitemOpen
  \bibfield  {author} {\bibinfo {author} {\bibfnamefont {A.}~\bibnamefont
  {Schindlmayr}},\ }\href {\doibase 10.1088/0143-0807/18/5/011} {\bibfield
  {journal} {\bibinfo  {journal} {European Journal of Physics}\ }\textbf
  {\bibinfo {volume} {18}},\ \bibinfo {pages} {374} (\bibinfo {year}
  {1997})}\BibitemShut {NoStop}%
\bibitem [{\citenamefont {Fang}\ \emph {et~al.}(2014)\citenamefont {Fang},
  \citenamefont {Battaglia}, \citenamefont {Carraro}, \citenamefont {Nemsak},
  \citenamefont {Ozdol}, \citenamefont {Kang}, \citenamefont {Bechtel},
  \citenamefont {Desai}, \citenamefont {Kronast}, \citenamefont {Unal},
  \citenamefont {Conti}, \citenamefont {Conlon}, \citenamefont {Palsson},
  \citenamefont {Martin}, \citenamefont {Minor}, \citenamefont {Fadley},
  \citenamefont {Yablonovitch}, \citenamefont {Maboudian},\ and\ \citenamefont
  {Javey}}]{Fang:PNAS111-2014}%
  \BibitemOpen
  \bibfield  {author} {\bibinfo {author} {\bibfnamefont {H.}~\bibnamefont
  {Fang}}, \bibinfo {author} {\bibfnamefont {C.}~\bibnamefont {Battaglia}},
  \bibinfo {author} {\bibfnamefont {C.}~\bibnamefont {Carraro}}, \bibinfo
  {author} {\bibfnamefont {S.}~\bibnamefont {Nemsak}}, \bibinfo {author}
  {\bibfnamefont {B.}~\bibnamefont {Ozdol}}, \bibinfo {author} {\bibfnamefont
  {J.~S.}\ \bibnamefont {Kang}}, \bibinfo {author} {\bibfnamefont {H.~A.}\
  \bibnamefont {Bechtel}}, \bibinfo {author} {\bibfnamefont {S.~B.}\
  \bibnamefont {Desai}}, \bibinfo {author} {\bibfnamefont {F.}~\bibnamefont
  {Kronast}}, \bibinfo {author} {\bibfnamefont {A.~A.}\ \bibnamefont {Unal}},
  \bibinfo {author} {\bibfnamefont {G.}~\bibnamefont {Conti}}, \bibinfo
  {author} {\bibfnamefont {C.}~\bibnamefont {Conlon}}, \bibinfo {author}
  {\bibfnamefont {G.~K.}\ \bibnamefont {Palsson}}, \bibinfo {author}
  {\bibfnamefont {M.~C.}\ \bibnamefont {Martin}}, \bibinfo {author}
  {\bibfnamefont {A.~M.}\ \bibnamefont {Minor}}, \bibinfo {author}
  {\bibfnamefont {C.~S.}\ \bibnamefont {Fadley}}, \bibinfo {author}
  {\bibfnamefont {E.}~\bibnamefont {Yablonovitch}}, \bibinfo {author}
  {\bibfnamefont {R.}~\bibnamefont {Maboudian}}, \ and\ \bibinfo {author}
  {\bibfnamefont {A.}~\bibnamefont {Javey}},\ }\href {\doibase
  10.1073/pnas.1405435111} {\bibfield  {journal} {\bibinfo  {journal}
  {Proceedings of the National Academy of Sciences}\ }\textbf {\bibinfo
  {volume} {111}},\ \bibinfo {pages} {6198} (\bibinfo {year}
  {2014})}\BibitemShut {NoStop}%
\bibitem [{\citenamefont {Rivera}\ \emph {et~al.}(2015)\citenamefont {Rivera},
  \citenamefont {Schaibley}, \citenamefont {Jones}, \citenamefont {Ross},
  \citenamefont {Wu}, \citenamefont {Aivazian}, \citenamefont {Klement},
  \citenamefont {Seyler}, \citenamefont {Clark}, \citenamefont {Ghimire},
  \citenamefont {Yan}, \citenamefont {Mandrus}, \citenamefont {Yao},\ and\
  \citenamefont {Xu}}]{Rivera:NatComm6-2015}%
  \BibitemOpen
  \bibfield  {author} {\bibinfo {author} {\bibfnamefont {P.}~\bibnamefont
  {Rivera}}, \bibinfo {author} {\bibfnamefont {J.~R.}\ \bibnamefont
  {Schaibley}}, \bibinfo {author} {\bibfnamefont {A.~M.}\ \bibnamefont
  {Jones}}, \bibinfo {author} {\bibfnamefont {J.~S.}\ \bibnamefont {Ross}},
  \bibinfo {author} {\bibfnamefont {S.}~\bibnamefont {Wu}}, \bibinfo {author}
  {\bibfnamefont {G.}~\bibnamefont {Aivazian}}, \bibinfo {author}
  {\bibfnamefont {P.}~\bibnamefont {Klement}}, \bibinfo {author} {\bibfnamefont
  {K.}~\bibnamefont {Seyler}}, \bibinfo {author} {\bibfnamefont
  {G.}~\bibnamefont {Clark}}, \bibinfo {author} {\bibfnamefont {N.~J.}\
  \bibnamefont {Ghimire}}, \bibinfo {author} {\bibfnamefont {J.}~\bibnamefont
  {Yan}}, \bibinfo {author} {\bibfnamefont {D.~G.}\ \bibnamefont {Mandrus}},
  \bibinfo {author} {\bibfnamefont {W.}~\bibnamefont {Yao}}, \ and\ \bibinfo
  {author} {\bibfnamefont {X.}~\bibnamefont {Xu}},\ }\href {\doibase
  10.1038/ncomms7242} {\bibfield  {journal} {\bibinfo  {journal} {Nature
  Communications}\ }\textbf {\bibinfo {volume} {6}} (\bibinfo {year} {2015}),\
  10.1038/ncomms7242}\BibitemShut {NoStop}%
\bibitem [{\citenamefont {Rivera}\ \emph {et~al.}(2016)\citenamefont {Rivera},
  \citenamefont {Seyler}, \citenamefont {Yu}, \citenamefont {Schaibley},
  \citenamefont {Yan}, \citenamefont {Mandrus}, \citenamefont {Yao},\ and\
  \citenamefont {Xu}}]{Rivera:Science351-2016}%
  \BibitemOpen
  \bibfield  {author} {\bibinfo {author} {\bibfnamefont {P.}~\bibnamefont
  {Rivera}}, \bibinfo {author} {\bibfnamefont {K.~L.}\ \bibnamefont {Seyler}},
  \bibinfo {author} {\bibfnamefont {H.}~\bibnamefont {Yu}}, \bibinfo {author}
  {\bibfnamefont {J.~R.}\ \bibnamefont {Schaibley}}, \bibinfo {author}
  {\bibfnamefont {J.}~\bibnamefont {Yan}}, \bibinfo {author} {\bibfnamefont
  {D.~G.}\ \bibnamefont {Mandrus}}, \bibinfo {author} {\bibfnamefont
  {W.}~\bibnamefont {Yao}}, \ and\ \bibinfo {author} {\bibfnamefont
  {X.}~\bibnamefont {Xu}},\ }\href {\doibase 10.1126/science.aac7820}
  {\bibfield  {journal} {\bibinfo  {journal} {Science}\ }\textbf {\bibinfo
  {volume} {351}},\ \bibinfo {pages} {688} (\bibinfo {year}
  {2016})}\BibitemShut {NoStop}%
\bibitem [{\citenamefont {Wang}\ \emph {et~al.}(2017)\citenamefont {Wang},
  \citenamefont {Chiu}, \citenamefont {Honz}, \citenamefont {Mak},\ and\
  \citenamefont {Shan}}]{Wang:NanoLett18-2017}%
  \BibitemOpen
  \bibfield  {author} {\bibinfo {author} {\bibfnamefont {Z.}~\bibnamefont
  {Wang}}, \bibinfo {author} {\bibfnamefont {Y.-H.}\ \bibnamefont {Chiu}},
  \bibinfo {author} {\bibfnamefont {K.}~\bibnamefont {Honz}}, \bibinfo {author}
  {\bibfnamefont {K.~F.}\ \bibnamefont {Mak}}, \ and\ \bibinfo {author}
  {\bibfnamefont {J.}~\bibnamefont {Shan}},\ }\href {\doibase
  10.1021/acs.nanolett.7b03667} {\bibfield  {journal} {\bibinfo  {journal}
  {Nano Letters}\ }\textbf {\bibinfo {volume} {18}},\ \bibinfo {pages} {137}
  (\bibinfo {year} {2017})}\BibitemShut {NoStop}%
\bibitem [{\citenamefont {Ciarrocchi}\ \emph {et~al.}(2018)\citenamefont
  {Ciarrocchi}, \citenamefont {Unuchek}, \citenamefont {Avsar}, \citenamefont
  {Watanabe}, \citenamefont {Taniguchi},\ and\ \citenamefont
  {Kis}}]{Ciarrocchi:NatPhoto13-2018}%
  \BibitemOpen
  \bibfield  {author} {\bibinfo {author} {\bibfnamefont {A.}~\bibnamefont
  {Ciarrocchi}}, \bibinfo {author} {\bibfnamefont {D.}~\bibnamefont {Unuchek}},
  \bibinfo {author} {\bibfnamefont {A.}~\bibnamefont {Avsar}}, \bibinfo
  {author} {\bibfnamefont {K.}~\bibnamefont {Watanabe}}, \bibinfo {author}
  {\bibfnamefont {T.}~\bibnamefont {Taniguchi}}, \ and\ \bibinfo {author}
  {\bibfnamefont {A.}~\bibnamefont {Kis}},\ }\href {\doibase
  10.1038/s41566-018-0325-y} {\bibfield  {journal} {\bibinfo  {journal} {Nature
  Photonics}\ }\textbf {\bibinfo {volume} {13}},\ \bibinfo {pages} {131}
  (\bibinfo {year} {2018})}\BibitemShut {NoStop}%
\bibitem [{\citenamefont {Calman}\ \emph {et~al.}(2018)\citenamefont {Calman},
  \citenamefont {Fogler}, \citenamefont {Butov}, \citenamefont {Hu},
  \citenamefont {Mishchenko},\ and\ \citenamefont
  {Geim}}]{Calman:NatComm9-2018}%
  \BibitemOpen
  \bibfield  {author} {\bibinfo {author} {\bibfnamefont {E.~V.}\ \bibnamefont
  {Calman}}, \bibinfo {author} {\bibfnamefont {M.~M.}\ \bibnamefont {Fogler}},
  \bibinfo {author} {\bibfnamefont {L.~V.}\ \bibnamefont {Butov}}, \bibinfo
  {author} {\bibfnamefont {S.}~\bibnamefont {Hu}}, \bibinfo {author}
  {\bibfnamefont {A.}~\bibnamefont {Mishchenko}}, \ and\ \bibinfo {author}
  {\bibfnamefont {A.~K.}\ \bibnamefont {Geim}},\ }\href {\doibase
  10.1038/s41467-018-04293-7} {\bibfield  {journal} {\bibinfo  {journal}
  {Nature Communications}\ }\textbf {\bibinfo {volume} {9}} (\bibinfo {year}
  {2018}),\ 10.1038/s41467-018-04293-7}\BibitemShut {NoStop}%
\bibitem [{\citenamefont {Ubrig}\ \emph {et~al.}(2020)\citenamefont {Ubrig},
  \citenamefont {Ponomarev}, \citenamefont {Zultak}, \citenamefont
  {Domaretskiy}, \citenamefont {Z{\'{o}}lyomi}, \citenamefont {Terry},
  \citenamefont {Howarth}, \citenamefont {Guti{\'{e}}rrez-Lezama},
  \citenamefont {Zhukov}, \citenamefont {Kudrynskyi}, \citenamefont {Kovalyuk},
  \citenamefont {Patan{\'{e}}}, \citenamefont {Taniguchi}, \citenamefont
  {Watanabe}, \citenamefont {Gorbachev}, \citenamefont {Fal'ko},\ and\
  \citenamefont {Morpurgo}}]{Ubrig:NatMat-2020}%
  \BibitemOpen
  \bibfield  {author} {\bibinfo {author} {\bibfnamefont {N.}~\bibnamefont
  {Ubrig}}, \bibinfo {author} {\bibfnamefont {E.}~\bibnamefont {Ponomarev}},
  \bibinfo {author} {\bibfnamefont {J.}~\bibnamefont {Zultak}}, \bibinfo
  {author} {\bibfnamefont {D.}~\bibnamefont {Domaretskiy}}, \bibinfo {author}
  {\bibfnamefont {V.}~\bibnamefont {Z{\'{o}}lyomi}}, \bibinfo {author}
  {\bibfnamefont {D.}~\bibnamefont {Terry}}, \bibinfo {author} {\bibfnamefont
  {J.}~\bibnamefont {Howarth}}, \bibinfo {author} {\bibfnamefont
  {I.}~\bibnamefont {Guti{\'{e}}rrez-Lezama}}, \bibinfo {author} {\bibfnamefont
  {A.}~\bibnamefont {Zhukov}}, \bibinfo {author} {\bibfnamefont {Z.~R.}\
  \bibnamefont {Kudrynskyi}}, \bibinfo {author} {\bibfnamefont {Z.~D.}\
  \bibnamefont {Kovalyuk}}, \bibinfo {author} {\bibfnamefont {A.}~\bibnamefont
  {Patan{\'{e}}}}, \bibinfo {author} {\bibfnamefont {T.}~\bibnamefont
  {Taniguchi}}, \bibinfo {author} {\bibfnamefont {K.}~\bibnamefont {Watanabe}},
  \bibinfo {author} {\bibfnamefont {R.~V.}\ \bibnamefont {Gorbachev}}, \bibinfo
  {author} {\bibfnamefont {V.~I.}\ \bibnamefont {Fal'ko}}, \ and\ \bibinfo
  {author} {\bibfnamefont {A.~F.}\ \bibnamefont {Morpurgo}},\ }\href {\doibase
  10.1038/s41563-019-0601-3} {\bibfield  {journal} {\bibinfo  {journal} {Nature
  Materials}\ } (\bibinfo {year} {2020}),\
  10.1038/s41563-019-0601-3}\BibitemShut {NoStop}%
\bibitem [{\citenamefont {Mak}\ and\ \citenamefont
  {Shan}(2018)}]{Mak:NatNano13-2018}%
  \BibitemOpen
  \bibfield  {author} {\bibinfo {author} {\bibfnamefont {K.~F.}\ \bibnamefont
  {Mak}}\ and\ \bibinfo {author} {\bibfnamefont {J.}~\bibnamefont {Shan}},\
  }\href {\doibase 10.1038/s41565-018-0301-1} {\bibfield  {journal} {\bibinfo
  {journal} {Nature Nanotechnology}\ }\textbf {\bibinfo {volume} {13}},\
  \bibinfo {pages} {974} (\bibinfo {year} {2018})}\BibitemShut {NoStop}%
\bibitem [{\citenamefont {{Fogler}}\ \emph {et~al.}(2014)\citenamefont
  {{Fogler}}, \citenamefont {{Butov}},\ and\ \citenamefont
  {{Novoselov}}}]{KostyaSuperfluid}%
  \BibitemOpen
  \bibfield  {author} {\bibinfo {author} {\bibfnamefont {M.~M.}\ \bibnamefont
  {{Fogler}}}, \bibinfo {author} {\bibfnamefont {L.~V.}\ \bibnamefont
  {{Butov}}}, \ and\ \bibinfo {author} {\bibfnamefont {K.~S.}\ \bibnamefont
  {{Novoselov}}},\ }\href {\doibase 10.1038/ncomms5555} {\bibfield  {journal}
  {\bibinfo  {journal} {Nature Communications}\ }\textbf {\bibinfo {volume}
  {5}},\ \bibinfo {eid} {4555} (\bibinfo {year} {2014})},\ \Eprint
  {http://arxiv.org/abs/1404.1418} {arXiv:1404.1418 [cond-mat.mes-hall]}
  \BibitemShut {NoStop}%
\bibitem [{\citenamefont {Li}\ and\ \citenamefont
  {Appelbaum}(2015)}]{Appelbaum}%
  \BibitemOpen
  \bibfield  {author} {\bibinfo {author} {\bibfnamefont {P.}~\bibnamefont
  {Li}}\ and\ \bibinfo {author} {\bibfnamefont {I.}~\bibnamefont {Appelbaum}},\
  }\href {\doibase 10.1103/PhysRevB.92.195129} {\bibfield  {journal} {\bibinfo
  {journal} {Phys. Rev. B}\ }\textbf {\bibinfo {volume} {92}},\ \bibinfo
  {pages} {195129} (\bibinfo {year} {2015})}\BibitemShut {NoStop}%
\bibitem [{\citenamefont {Ruiz-Tijerina}\ \emph {et~al.}(2018)\citenamefont
  {Ruiz-Tijerina}, \citenamefont {Danovich}, \citenamefont {Yelgel},
  \citenamefont {Z\'olyomi},\ and\ \citenamefont {Fal'ko}}]{tmd_isb}%
  \BibitemOpen
  \bibfield  {author} {\bibinfo {author} {\bibfnamefont {D.~A.}\ \bibnamefont
  {Ruiz-Tijerina}}, \bibinfo {author} {\bibfnamefont {M.}~\bibnamefont
  {Danovich}}, \bibinfo {author} {\bibfnamefont {C.}~\bibnamefont {Yelgel}},
  \bibinfo {author} {\bibfnamefont {V.}~\bibnamefont {Z\'olyomi}}, \ and\
  \bibinfo {author} {\bibfnamefont {V.~I.}\ \bibnamefont {Fal'ko}},\ }\href
  {\doibase 10.1103/PhysRevB.98.035411} {\bibfield  {journal} {\bibinfo
  {journal} {Phys. Rev. B}\ }\textbf {\bibinfo {volume} {98}},\ \bibinfo
  {pages} {035411} (\bibinfo {year} {2018})}\BibitemShut {NoStop}%
\bibitem [{\citenamefont {Fiorentini}\ and\ \citenamefont
  {Baldereschi}(1995)}]{sc3}%
  \BibitemOpen
  \bibfield  {author} {\bibinfo {author} {\bibfnamefont {V.}~\bibnamefont
  {Fiorentini}}\ and\ \bibinfo {author} {\bibfnamefont {A.}~\bibnamefont
  {Baldereschi}},\ }\href@noop {} {\bibfield  {journal} {\bibinfo  {journal}
  {Phys. Rev. B}\ }\textbf {\bibinfo {volume} {51}},\ \bibinfo {pages} {17196}
  (\bibinfo {year} {1995})}\BibitemShut {NoStop}%
\bibitem [{\citenamefont {Johnson}\ and\ \citenamefont {Ashcroft}(1998)}]{sc4}%
  \BibitemOpen
  \bibfield  {author} {\bibinfo {author} {\bibfnamefont {K.~A.}\ \bibnamefont
  {Johnson}}\ and\ \bibinfo {author} {\bibfnamefont {N.~W.}\ \bibnamefont
  {Ashcroft}},\ }\href@noop {} {\bibfield  {journal} {\bibinfo  {journal}
  {Phys. Rev. B}\ }\textbf {\bibinfo {volume} {58}},\ \bibinfo {pages} {15548}
  (\bibinfo {year} {1998})}\BibitemShut {NoStop}%
\bibitem [{\citenamefont {Bernstein}\ \emph {et~al.}(2002)\citenamefont
  {Bernstein}, \citenamefont {Mehl},\ and\ \citenamefont
  {Papaconstantopoulos}}]{sc5}%
  \BibitemOpen
  \bibfield  {author} {\bibinfo {author} {\bibfnamefont {N.}~\bibnamefont
  {Bernstein}}, \bibinfo {author} {\bibfnamefont {M.~J.}\ \bibnamefont {Mehl}},
  \ and\ \bibinfo {author} {\bibfnamefont {D.~A.}\ \bibnamefont
  {Papaconstantopoulos}},\ }\href@noop {} {\bibfield  {journal} {\bibinfo
  {journal} {Phys. Rev. B}\ }\textbf {\bibinfo {volume} {66}},\ \bibinfo
  {pages} {075212} (\bibinfo {year} {2002})}\BibitemShut {NoStop}%
\bibitem [{\citenamefont {Kotani}\ \emph {et~al.}(2007)\citenamefont {Kotani},
  \citenamefont {van Schilfgaarde},\ and\ \citenamefont
  {Faleev}}]{kotani2007quasiparticle}%
  \BibitemOpen
  \bibfield  {author} {\bibinfo {author} {\bibfnamefont {T.}~\bibnamefont
  {Kotani}}, \bibinfo {author} {\bibfnamefont {M.}~\bibnamefont {van
  Schilfgaarde}}, \ and\ \bibinfo {author} {\bibfnamefont {S.~V.}\ \bibnamefont
  {Faleev}},\ }\href@noop {} {\bibfield  {journal} {\bibinfo  {journal}
  {Physical Review B}\ }\textbf {\bibinfo {volume} {76}},\ \bibinfo {pages}
  {165106} (\bibinfo {year} {2007})}\BibitemShut {NoStop}%
\bibitem [{\citenamefont {Pashov}\ \emph {et~al.}(2020)\citenamefont {Pashov},
  \citenamefont {Acharya}, \citenamefont {Lambrecht}, \citenamefont {Jackson},
  \citenamefont {Belashchenko}, \citenamefont {Chantis}, \citenamefont
  {Jamet},\ and\ \citenamefont {van
  Schilfgaarde}}]{Pashov:CompPhysComm149-2020}%
  \BibitemOpen
  \bibfield  {author} {\bibinfo {author} {\bibfnamefont {D.}~\bibnamefont
  {Pashov}}, \bibinfo {author} {\bibfnamefont {S.}~\bibnamefont {Acharya}},
  \bibinfo {author} {\bibfnamefont {W.~R.}\ \bibnamefont {Lambrecht}}, \bibinfo
  {author} {\bibfnamefont {J.}~\bibnamefont {Jackson}}, \bibinfo {author}
  {\bibfnamefont {K.~D.}\ \bibnamefont {Belashchenko}}, \bibinfo {author}
  {\bibfnamefont {A.}~\bibnamefont {Chantis}}, \bibinfo {author} {\bibfnamefont
  {F.}~\bibnamefont {Jamet}}, \ and\ \bibinfo {author} {\bibfnamefont
  {M.}~\bibnamefont {van Schilfgaarde}},\ }\href {\doibase
  10.1016/j.cpc.2019.107065} {\bibfield  {journal} {\bibinfo  {journal}
  {Computer Physics Communications}\ }\textbf {\bibinfo {volume} {249}},\
  \bibinfo {pages} {107065} (\bibinfo {year} {2020})}\BibitemShut {NoStop}%
\bibitem [{\citenamefont {Li}\ \emph {et~al.}(2018)\citenamefont {Li},
  \citenamefont {Wang}, \citenamefont {Wu}, \citenamefont {Cao}, \citenamefont
  {Chen}, \citenamefont {Sankar}, \citenamefont {Ulaganathan}, \citenamefont
  {Chou}, \citenamefont {Wetzel}, \citenamefont {Xu}, \citenamefont {Louie},\
  and\ \citenamefont {Shi}}]{Li2DMaterials2018}%
  \BibitemOpen
  \bibfield  {author} {\bibinfo {author} {\bibfnamefont {Y.}~\bibnamefont
  {Li}}, \bibinfo {author} {\bibfnamefont {T.}~\bibnamefont {Wang}}, \bibinfo
  {author} {\bibfnamefont {M.}~\bibnamefont {Wu}}, \bibinfo {author}
  {\bibfnamefont {T.}~\bibnamefont {Cao}}, \bibinfo {author} {\bibfnamefont
  {Y.}~\bibnamefont {Chen}}, \bibinfo {author} {\bibfnamefont {R.}~\bibnamefont
  {Sankar}}, \bibinfo {author} {\bibfnamefont {R.~K.}\ \bibnamefont
  {Ulaganathan}}, \bibinfo {author} {\bibfnamefont {F.}~\bibnamefont {Chou}},
  \bibinfo {author} {\bibfnamefont {C.}~\bibnamefont {Wetzel}}, \bibinfo
  {author} {\bibfnamefont {C.-Y.}\ \bibnamefont {Xu}}, \bibinfo {author}
  {\bibfnamefont {S.~G.}\ \bibnamefont {Louie}}, \ and\ \bibinfo {author}
  {\bibfnamefont {S.-F.}\ \bibnamefont {Shi}},\ }\href {\doibase
  10.1088/2053-1583/aaa6eb} {\bibfield  {journal} {\bibinfo  {journal} {2D
  Materials}\ }\textbf {\bibinfo {volume} {5}},\ \bibinfo {pages} {021002}
  (\bibinfo {year} {2018})}\BibitemShut {NoStop}%
\bibitem [{\citenamefont {Song}\ \emph {et~al.}(2018)\citenamefont {Song},
  \citenamefont {Fan}, \citenamefont {Xuan}, \citenamefont {Huang},
  \citenamefont {Zhang}, \citenamefont {Wang}, \citenamefont {Sun},
  \citenamefont {Wu},\ and\ \citenamefont {Yan}}]{Song2018}%
  \BibitemOpen
  \bibfield  {author} {\bibinfo {author} {\bibfnamefont {C.}~\bibnamefont
  {Song}}, \bibinfo {author} {\bibfnamefont {F.}~\bibnamefont {Fan}}, \bibinfo
  {author} {\bibfnamefont {N.}~\bibnamefont {Xuan}}, \bibinfo {author}
  {\bibfnamefont {S.}~\bibnamefont {Huang}}, \bibinfo {author} {\bibfnamefont
  {G.}~\bibnamefont {Zhang}}, \bibinfo {author} {\bibfnamefont
  {C.}~\bibnamefont {Wang}}, \bibinfo {author} {\bibfnamefont {Z.}~\bibnamefont
  {Sun}}, \bibinfo {author} {\bibfnamefont {H.}~\bibnamefont {Wu}}, \ and\
  \bibinfo {author} {\bibfnamefont {H.}~\bibnamefont {Yan}},\ }\href {\doibase
  10.1021/acsami.7b17247} {\bibfield  {journal} {\bibinfo  {journal} {{ACS}
  Applied Materials {\&} Interfaces}\ }\textbf {\bibinfo {volume} {10}},\
  \bibinfo {pages} {3994} (\bibinfo {year} {2018})}\BibitemShut {NoStop}%
\bibitem [{\citenamefont {Rigoult}\ \emph {et~al.}(1980)\citenamefont
  {Rigoult}, \citenamefont {Rimsky},\ and\ \citenamefont {Kuhn}}]{iucr_bulk}%
  \BibitemOpen
  \bibfield  {author} {\bibinfo {author} {\bibfnamefont {J.}~\bibnamefont
  {Rigoult}}, \bibinfo {author} {\bibfnamefont {A.}~\bibnamefont {Rimsky}}, \
  and\ \bibinfo {author} {\bibfnamefont {A.}~\bibnamefont {Kuhn}},\ }\href
  {\doibase 10.1107/s0567740880004840} {\bibfield  {journal} {\bibinfo
  {journal} {Acta Crystallographica B}\ }\textbf {\bibinfo {volume} {36}},\
  \bibinfo {pages} {916} (\bibinfo {year} {1980})}\BibitemShut {NoStop}%
\bibitem [{\citenamefont {Camassel}\ \emph {et~al.}(1978)\citenamefont
  {Camassel}, \citenamefont {Merle}, \citenamefont {Mathieu},\ and\
  \citenamefont {Chevy}}]{Camassel:PRB17-1978}%
  \BibitemOpen
  \bibfield  {author} {\bibinfo {author} {\bibfnamefont {J.}~\bibnamefont
  {Camassel}}, \bibinfo {author} {\bibfnamefont {P.}~\bibnamefont {Merle}},
  \bibinfo {author} {\bibfnamefont {H.}~\bibnamefont {Mathieu}}, \ and\
  \bibinfo {author} {\bibfnamefont {A.}~\bibnamefont {Chevy}},\ }\href
  {\doibase 10.1103/physrevb.17.4718} {\bibfield  {journal} {\bibinfo
  {journal} {Physical Review B}\ }\textbf {\bibinfo {volume} {17}},\ \bibinfo
  {pages} {4718} (\bibinfo {year} {1978})}\BibitemShut {NoStop}%
\bibitem [{\citenamefont {Chadi}\ and\ \citenamefont
  {Cohen}(1973)}]{Chadi:PRB8-1973}%
  \BibitemOpen
  \bibfield  {author} {\bibinfo {author} {\bibfnamefont {D.~J.}\ \bibnamefont
  {Chadi}}\ and\ \bibinfo {author} {\bibfnamefont {M.~L.}\ \bibnamefont
  {Cohen}},\ }\href {\doibase 10.1103/PhysRevB.8.5747} {\bibfield  {journal}
  {\bibinfo  {journal} {Phys. Rev. B}\ }\textbf {\bibinfo {volume} {8}},\
  \bibinfo {pages} {5747} (\bibinfo {year} {1973})}\BibitemShut {NoStop}%
\bibitem [{\citenamefont {Cunningham}(1974)}]{Cunningham:PRB10-1974}%
  \BibitemOpen
  \bibfield  {author} {\bibinfo {author} {\bibfnamefont {S.~L.}\ \bibnamefont
  {Cunningham}},\ }\href {\doibase 10.1103/PhysRevB.10.4988} {\bibfield
  {journal} {\bibinfo  {journal} {Phys. Rev. B}\ }\textbf {\bibinfo {volume}
  {10}},\ \bibinfo {pages} {4988} (\bibinfo {year} {1974})}\BibitemShut
  {NoStop}%
\bibitem [{\citenamefont {Monkhorst}\ and\ \citenamefont
  {Pack}(1976)}]{Monkhorst:PRB13-1976}%
  \BibitemOpen
  \bibfield  {author} {\bibinfo {author} {\bibfnamefont {H.~J.}\ \bibnamefont
  {Monkhorst}}\ and\ \bibinfo {author} {\bibfnamefont {J.~D.}\ \bibnamefont
  {Pack}},\ }\href {\doibase 10.1103/PhysRevB.13.5188} {\bibfield  {journal}
  {\bibinfo  {journal} {Phys. Rev. B}\ }\textbf {\bibinfo {volume} {13}},\
  \bibinfo {pages} {5188} (\bibinfo {year} {1976})}\BibitemShut {NoStop}%
\bibitem [{\citenamefont {Magalhães}(2014)}]{Magalhaes:JChemEdu91-2014}%
  \BibitemOpen
  \bibfield  {author} {\bibinfo {author} {\bibfnamefont {A.~L.}\ \bibnamefont
  {Magalhães}},\ }\href {\doibase 10.1021/ed500437a} {\bibfield  {journal}
  {\bibinfo  {journal} {Journal of Chemical Education}\ }\textbf {\bibinfo
  {volume} {91}},\ \bibinfo {pages} {2124} (\bibinfo {year} {2014})},\ \Eprint
  {http://arxiv.org/abs/https://doi.org/10.1021/ed500437a}
  {https://doi.org/10.1021/ed500437a} \BibitemShut {NoStop}%
\end{thebibliography}%

\end{document}